\documentclass[%
 reprint,
 amsmath,amssymb,
 aps,
pra,
]{revtex4-2}

\usepackage{xcolor}
\usepackage{graphicx}
\usepackage{dcolumn}
\usepackage{bm}

\begin{document}

\preprint{APS/123-QED}

\title{\boldmath Long-term neutrino emission from a core-collapse supernova with axion-photon coupling}

\author{Masamitsu Mori}
 \altaffiliation[Also at ]{ National Astronomical Observatory of Japan}
 \email{masamitsu.mori.astro@gmail.com}
 \email{masamitsu.mori@numazu-ct.ac.jp}
\affiliation{%
 National Institute of Technology, Numazu College, Numazu, Shizuoka 410-8501, Japan
}%
 
\author{Kanji Mori}%
 
\affiliation{%
 National Astronomical Observatory of Japan,\\ 2-21-1 Osawa, Mitaka, Tokyo 181-8588, Japan
}%

\date{\today}

\begin{abstract}
We perform long-term general-relativistic neutrino-radiation hydrodynamic simulations for core-collapse supernovae (CCSNe) which include the cooling effect induced by the coupling between axion-like particles (ALPs) and photons. We take into account the photon coalescence and the Primakoff effect, and investigate ALPs with the mass of 10\,MeV and the coupling constant $g_{a\gamma}$ of $1.0\times10^{-9}{\rm \,GeV^{-1}}$ to $7.0\times10^{-9}{\rm \,GeV^{-1}}$. It is found that the effects of the ALP cooling emerge in the late phase rather than the early phase and the ALP luminosities are always lower than the neutrino luminosity in our simulations. We estimate the number of neutrino events for Super-Kamiokande assuming a 10\,kpc CCSN. We conclude that signatures of ALPs could be found in the long-term neutrino signals from a nearby CCSN event in the future, even if $g_{a\gamma}$ is below an upper limit based on the conventional energy-loss argument.
\end{abstract}

\maketitle
\section{Introduction}\label{sec:intro}
Axion-like particles (ALPs) are a type of the beyond-standard-model particle and pseudo-Nambu-Goldstone bosons predicted by some string theories \cite{2020PhR...870....1D,2021ARNPS..71..225C}. ALPs could interact with photons through the coupling term below~\cite{1988PhRvD..37.1237R}:
\begin{align}
    \mathcal{L} = -\frac{1}{4}g_{a\gamma}F_{\mu\nu}\tilde{F}^{\mu\nu}a,
\end{align}
where $a$ is the ALP field,  $g_{a\gamma}$ is the coupling constant, $F_{\mu\nu}$ is the electromagnetic tensor and $\tilde{F}^{\mu\nu}$ is its dual.

Axion was first introduced to solve the strong CP problem and is a candidate of dark matter. It has been searched for with various methods including collider experiments \cite{2016PhLB..753..482J,2019JHEP...05..213D,2017JHEP...12..094D}, which have excluded $g_{a\gamma} \gtrsim 10^{-7}{\rm \,GeV^{-1}}$. Astronomical observations and  cosmological arguments including Big Bang nucleosynthesis and the cosmic microwave background have also provided constraints on the ALP parameters \cite{2012JCAP...02..032C,2020JCAP...05..009D}. 

Astrophysical objects such as stars and supernovae can produce ALPs. 
Comparison of theoretical models including these effects and  astronomical observation provides various constraints on ALP parameters. For instance, ALPs enhance energy loss in stars and affect the late phase of stellar evolution of stars with initial masses less than 8$M_\odot$, stellar populations, and the structure of asymptotic giant branch stars \cite{2014PhRvL.113s1302A, 2020PhLB..80935709C,Dolan:2022kul}. The energy loss leads to a different initial-final mass relation of white dwarfs ~\cite{2021JCAP...09..010D}.

Among astrophysical objects, core-collapse supernovae (CCSNe) could be prolific sources of ALPs. ALPs formed in the supernova core extract energy from the star, thereby accelerating its cooling as a result. 
In addition, ALPs can assist collapsing stars in exploding if ALPs decay in the gain region and deposit heat into the surrounding matter.
From the observation of SN~1987A, which is humanity's only detection of neutrinos from a CCSN so far and the first example of the multi-messenger astronomy, the ALP parameters have been constrained \cite{2018arXiv180810136L,2019PhRvD..99l1305S,2020JCAP...12..008L,2022PhRvL.128v1103C}. Moreover, heavy ALPs decay into gamma rays while traveling to the earth. The fact that no gamma ray has been detected provides a constraint~\cite{2023JCAP...03..054H} and the coupling constant between approximately $10^{-10}-10^{-9} {\rm \,GeV^{-1}}$ has been excluded. In 2023, SN~2023ixf occurred at 6.9\,Mpc, relatively near the earth. A Fermi-LAT observation of $\gamma$-rays from this object provided another constraint \cite{2024PhRvD.109b3018R}. In the future, photons from nearby supernovae including type Ia supernovae and hypernovae may provide additional constraints~\cite{2011JCAP...01..015G,2018PhRvD..98e5032J,2021PASJ...73.1382M, 2021PhRvL.127r1102C}. 

In the context of simulations of ALP emission from CCSNe, there are two kinds of simulations: post-process simulations and self-consistent simulations with the ALP feedback. Although in both simulations ALP production is investigated~\cite{2020JCAP...12..008L,2022PhRvD.105c5022C,2023JCAP...07..056M,2023EPJP..138..836C,2019JCAP...10..016C}, we need self-consistent simulations for discussion on modification of neutrino emissions and hydrodynamics~\cite{2016PhRvD..94h5012F,2021PhRvD.104j3012F,2022PhRvD.105f3009M}. Table~\ref{tab:my_label} summarizes previous self-consistent simulations. The previous works~\cite{2016PhRvD..94h5012F,2021PhRvD.104j3012F,2022PhRvD.106f3019B} considered axion-nucleon coupling for milli-electron-volt (meV) scale ALPs. On the other hand, Refs.~\cite{2023EPJP..138..836C,2024PhRvD.110b3031M,2025arXiv250309005T} conducted simulations with axion-photon couplings of mega-electron-volt (MeV) scale ALPs. We note that there are simulations that solve not hydrodynamics but the hydrostatic equilibrium of proto-neutron stars to calculate the neutrino and axion emission~\cite{1989PhRvD..39.1020B, 1997PhRvD..56.2419K}.

In this study, we focus on MeV scale ALPs, which we refer to as heavy ALPs. In Refs.~\cite{2016PhRvD..94h5012F,2021PhRvD.104j3012F,2022PhRvD.106f3019B}, they estimated neutrino signals modified by ALPs. In Ref.~\cite{2024PhRvD.110b3031M}, 2D simulations have been conducted and multi-messenger signals including gravitational waves have been estimated. The studies~\cite{2023EPJP..138..836C, 2025arXiv250309005T} have discussed in the explodability of CCSNe. In this paper, we conduct long-term self-consistent 1D simulations with axion-photon coupling with 10MeV ALPs and make a discussion on modification and observation of neutrinos.

In this paper, Section~\ref{sec:method} describes the method of our calculation and Section~\ref{sec:results} shows the calculation results: fluid, neutrino and ALP properties. In Section~\ref{sec:discussion}, we discuss neutrino events and a possibility to find ALPs or constraint on ALP parameters via a Galactic CCSN.

\begin{table*}[]
        \caption{Summary of supernova simulations including ALPs.}
    \scalebox{1.0}{ 
    \hspace{-1cm}
    \begin{tabular}{ccccccccc}
    \hline\hline
        Authors & Year & Progenitor & Gravity & Neutrino & Axion mass & Axion interacting & Dimension & Length [s] \\
        & & mass[$M_\odot$] & & & & matter& & \\ \hline
        
         M. Mori \& K. Mori (This work) & 2025 & 9.6 & G.R. & M1 & 10\,MeV & Photon & 1D & 20 \\
         T. Fischer et al.~\cite{2016PhRvD..94h5012F} & 2016 & 18, 11.2 & G.R. & Boltzmann &  30\,meV&  Nucleon & 1D & 40 \\
         T. Fischer et al.~\cite{2021PhRvD.104j3012F} & 2021 & 18 & G.R. & Boltzmann & 15\,meV & Nucleon, Pion & 1D & 10 \\
        A. Betranhandy et al.~\cite{2022PhRvD.106f3019B} & 2022 & 18 & Effective & M1 & 11-114\,meV & Nucleon & 1D & 4.5 \\
        &  & 20 & & &57-114\,meV & & 2D & 0.8  \\
         K. Mori et al.~\cite{2023EPJP..138..836C} & 2023 & 20 & Effective & IDSA & 50-800\,MeV & Photon & 1D & 0.5\\
         & & 11.2 && & 100\,MeV & &\\
        K. Mori et al.~\cite{2024PhRvD.110b3031M} & 2024 & 20 & Effective & IDSA & 100,200\,MeV & Photon& 2D & 0.4 \\
        T. Tsurugi et al.~\cite{2025arXiv250309005T} & 2025 & 11.2,20,25 & Effective & IDSA & 50-800\,MeV & Photon & 1D & 0.5 \\

    \hline\hline
    \end{tabular}
    }

    \label{tab:my_label}
\end{table*}
\section{Method}\label{sec:method}
\subsection{CCSN simulation}
We employ GR1D as a CCSN simulator. GR1D is an open-source general relativistic fluid simulator with neutrino radiation transport in one dimension~\cite{2010CQGra..27k4103O,2015ApJS..219...24O}. GR1D is implemented with the truncated moment scheme (M1 scheme)~\cite{2011PThPh.125.1255S,2013PhRvD..87j3004C} for neutrino transport. We modify the original version of GR1D for long-term simulation and axion cooling. 
The settings of the CCSN simulation without the ALP and the modifications for the long-term simulation are described in Ref.~\cite{2021PTEP.2021b3E01M}. 

In this study, axion cooling is implemented in GR1D. The ALP production is calculated at every time step according to density, temperature and electron fraction of matter at that time. The heat taken away by ALPs is treated as  feedback to the internal energy in the next time step. It is assumed that the ALP escapes from proto-neutron stars before they decay and proto-neutron stars cool down in the aftermath of the escape of ALPs. We use a 9.6$M_\odot$ progenitor (provided by A.~Heger 2016 private communication), which is reported to explode in 1D~\cite{2021PTEP.2021b3E01M,2015ApJ...801L..24M,2017ApJ...850...43R}, and to form 1.36$M_\odot$ after the explosion. The DD2 equation of state (EoS)~\cite{2005PhRvC..71f4301T,2010NuPhA.837..210H,2010PhRvC..81a5803T} is employed as an EoS table, which is based on the density-dependent mean field theory.

\subsection{Axion-Like Particle (ALP)}

In this study, we consider two channels of ALP production: the photon coalescence and the Primakoff effect. Our treatment for the ALP production processes follows Refs.~\cite{2023EPJP..138..836C,2024PhRvD.110b3031M,2020JCAP...12..008L}.
We fix the mass of ALPs to 10\,MeV and choose $1\times 10^{-9}\rm{\,GeV^{-1}}$ to $7\times 10^{-9}\rm{\,GeV}^{-1}$ as the coupling constants $g_{a\gamma}$, because such ALPs evade constraints from the Sun \cite{2008PhLB..668...93I,2015JCAP...10..015V,2017NatPh..13..584A}. In Ref.~\cite{2020JCAP...12..008L}, the same processes are calculated in post process and then the coupling constants above or equal to $6\times 10^{-9}\rm{\,GeV^{-1}}$ are excluded via neutrinos of SN~1987A.
It should be noted that we do not consider ALP heating because 10\,MeV ALPs are not expected to heat the gain region in the stellar cores~\cite{2023EPJP..138..836C}. Thus we ignore ALP decay and its heating.

In our calculation, we turn the axion processes on from the beginning of core collapse for two simulations with $g_{a\gamma}=7.0\times 10^{-9}\rm{\,GeV^{-1}}$ and $6.0\times 10^{-9}{\rm \,GeV^{-1}}$. For smaller coupling constants, we begin calculating the ALP cooling after the shock wave escapes from the calculation region in order to save computational resources. We compare shock wave propagation for simulations with $g_{a\gamma}=7.0\times 10^{-9}\rm{\,GeV^{-1}}$ and $6.0\times 10^{-9}{\rm \,GeV^{-1}}$ with that without ALP.

For convenience, we define a dimensionless constant $g_{10}$:
\begin{equation}
    g_{10} = g_{a\gamma}\times 10^{10}{\,\rm GeV}.
\end{equation}
Hereafter we employ $g_{10}$ instead of $g_{a\gamma}$.

\section{Results}\label{sec:results}

\subsection{Shock wave propagation}\label{subsec:shockwave}
\begin{figure}
    \centering
    \includegraphics[width=1.0\linewidth]{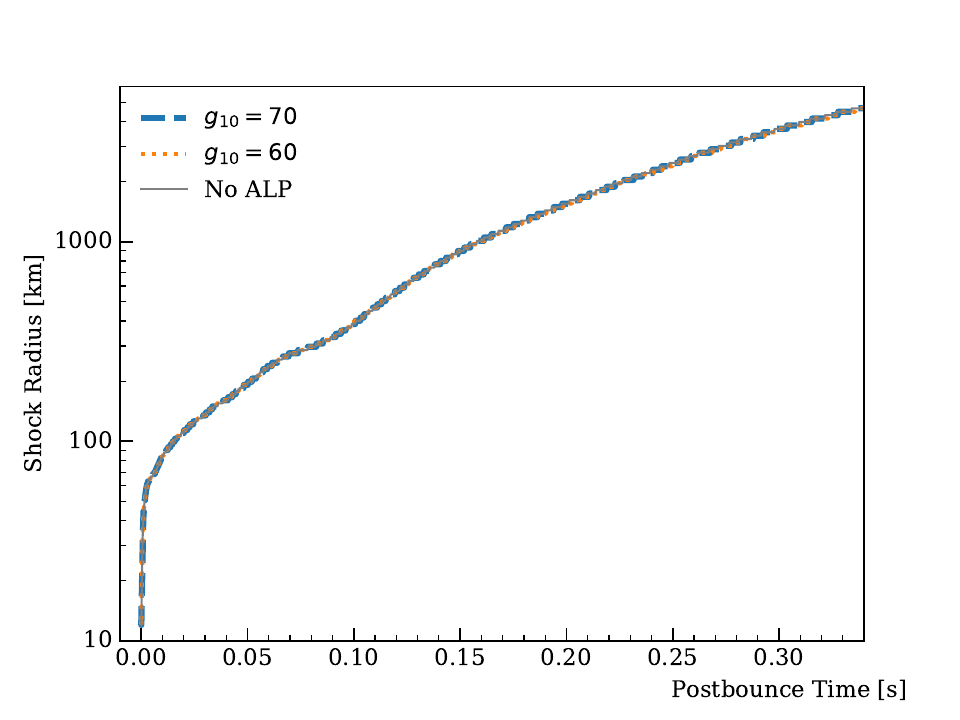}
    \caption{Radius of shock wave with time. There are three curves and they completely overlap.}
   \label{fig:shock_wave}
\end{figure}

Figure~\ref{fig:shock_wave} shows radii of shock waves of the simulation without ALP (the reference model) and the simulations with $g_{10}=70$ and $60$, which are the largest coupling constants in our simulations. Figure~\ref{fig:shock_wave} indicates that the three curves completely overlap and the ALP cooling does not affect fluid dynamics and explodability in our simulations at all. It also agrees with the result in a previous work~\cite{2023EPJP..138..836C}. Thus, we turn the ALP cooling on after the shock wave escapes from the calculation region in order to save computational costs.

\subsection{Temperature profiles}
\begin{figure*}
\begin{minipage}[b]{0.98\columnwidth}
\centering
        \includegraphics[width=1.0\linewidth]{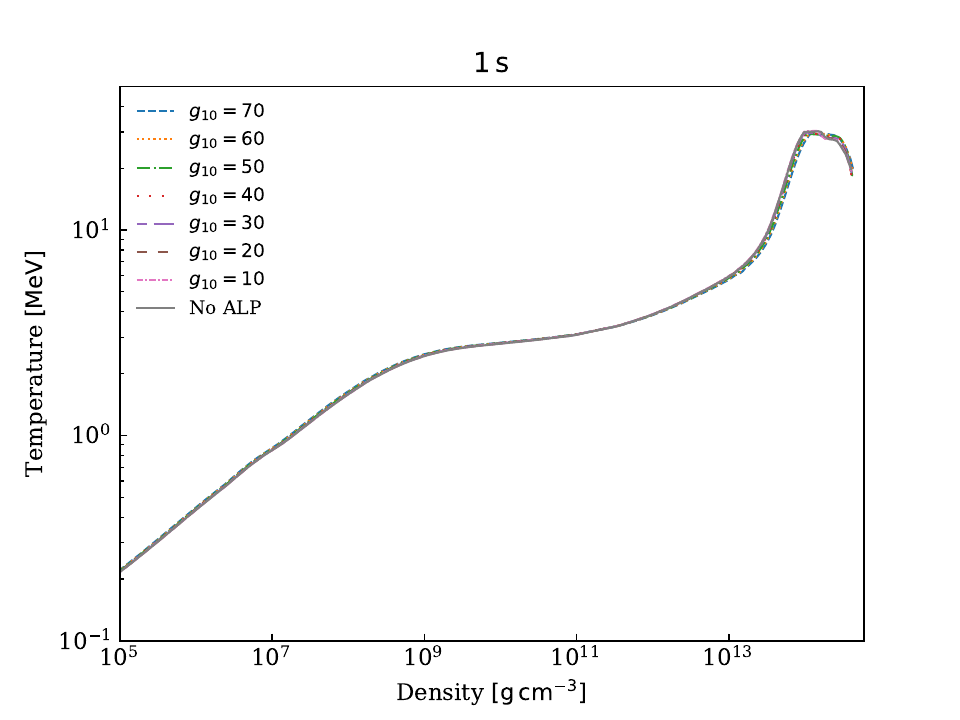}
        \includegraphics[width=1.0\linewidth]{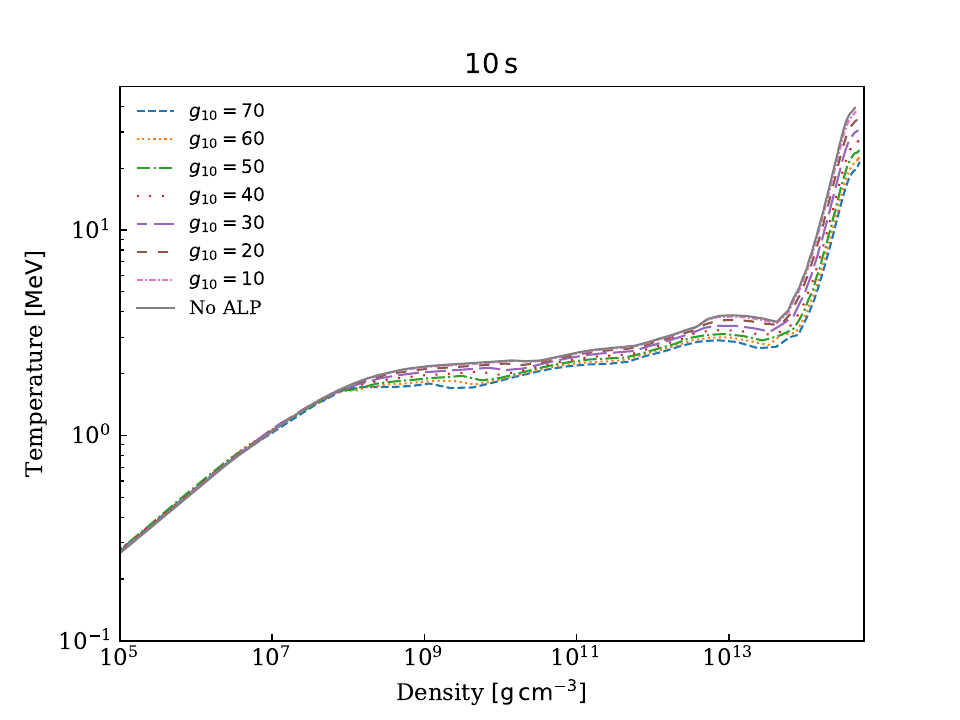}
\end{minipage}
\begin{minipage}[b]{0.98\columnwidth}
\centering
        \includegraphics[width=1.0\linewidth]{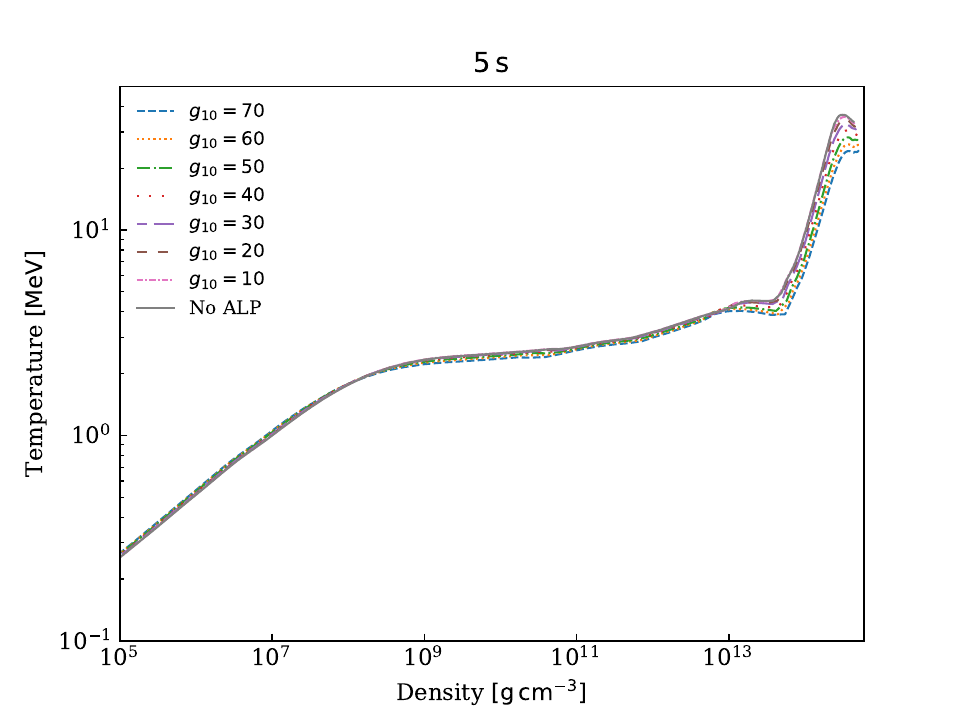}
                \includegraphics[width=1.0\linewidth]{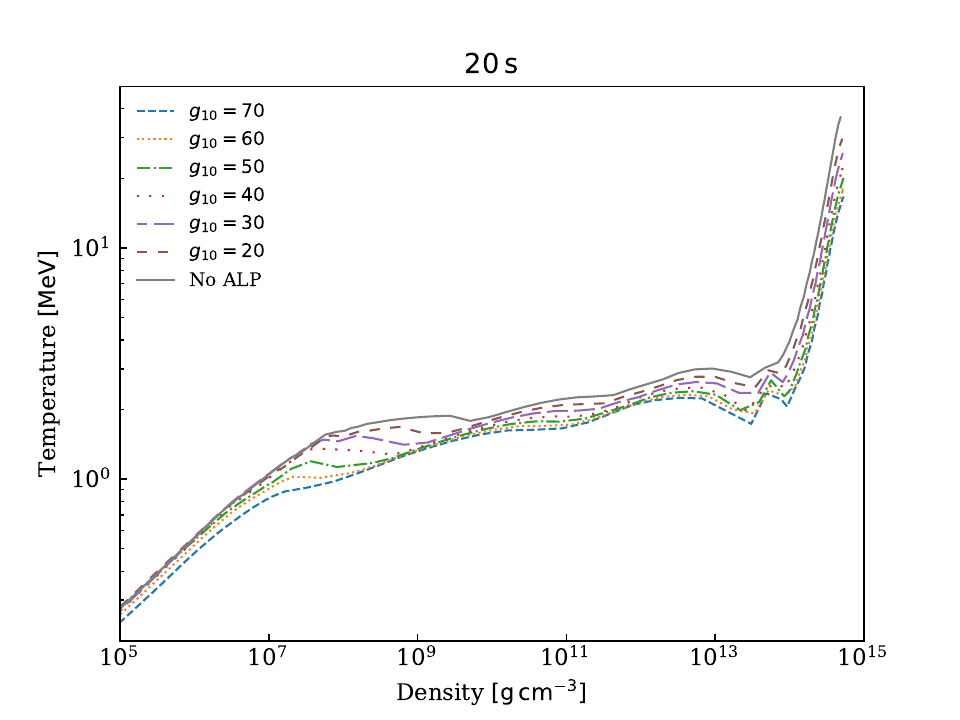}
\end{minipage}
    
    \caption{Temperature versus density. The upper-left plot corresponds to curves at 1\,s after the bounce: the upper-right plot corresponds to 5\,s, the lower-left plot to 10\,s and the lower-right plot to at 20\,s. Regarding the color assignment: the blue corresponds to $g_{a\gamma}=70\,{\rm \,GeV^{-1}}$, the orange to 60, the green to 50, the red to 40, the purple to 30, the brown to 20, the pink to 10 and the gray to the no ALP case.}
    \label{fig:rho_temp}
\end{figure*}

Figure~\ref{fig:rho_temp} shows temperature versus density plots at 1\,s, 5\,s, 10\,s, and 20\,s after the bounce. Hereafter, we use the following color scheme: orange corresponds to 60, green to 50, red to 40, purple to 30, brown to 20, pink to 10, and grey represents the no ALP case (the reference model). At 1\,s after the bounce, there is little difference among the simulations. By 5\,s, the temperature declines above $10^{13}{\rm g\,cm^{-3}}$. At 10\,s, the differences continue to grow and expand down to $10^{8}{\rm g\,cm^{-3}}$. At last by 20\,s, the outer layers have cooled due to the presence of ALPs. The hierarchy of the temperature profiles follows the order of the coupling constants $g_{a\gamma}$. These results show that PNS cooling is enhanced by ALP emission, particularly in the late phase.

\subsection{ALP emission}
\begin{figure*}
    \begin{minipage}[b]{0.48\linewidth}
    \centering
    \includegraphics[width=\linewidth]{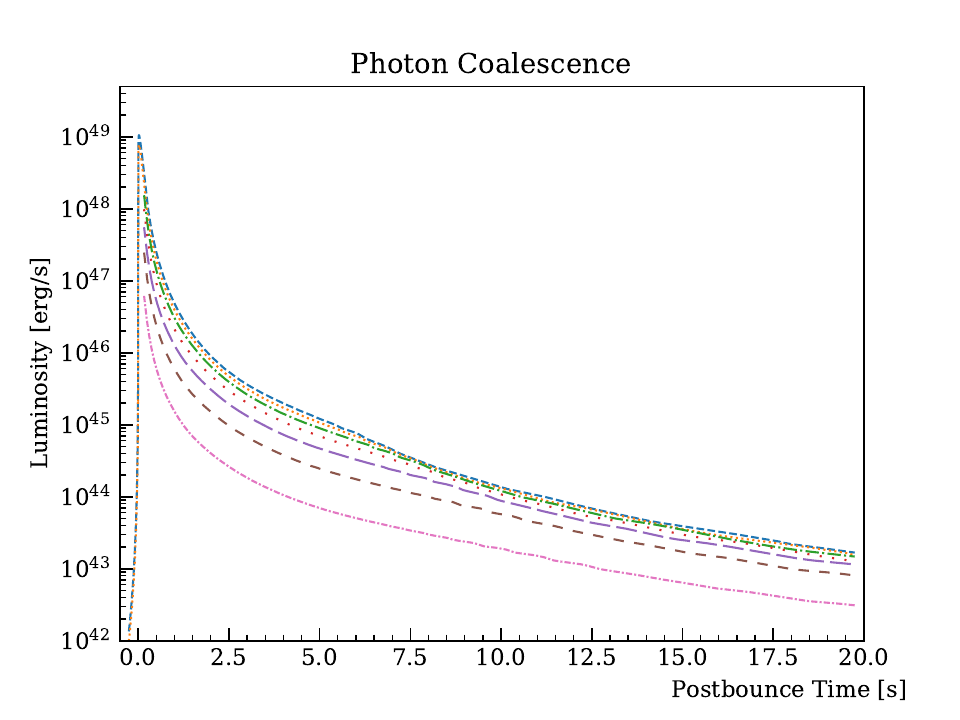}
    \end{minipage}
    \begin{minipage}[b]{0.48\linewidth}
    \centering
    \includegraphics[width=\linewidth]{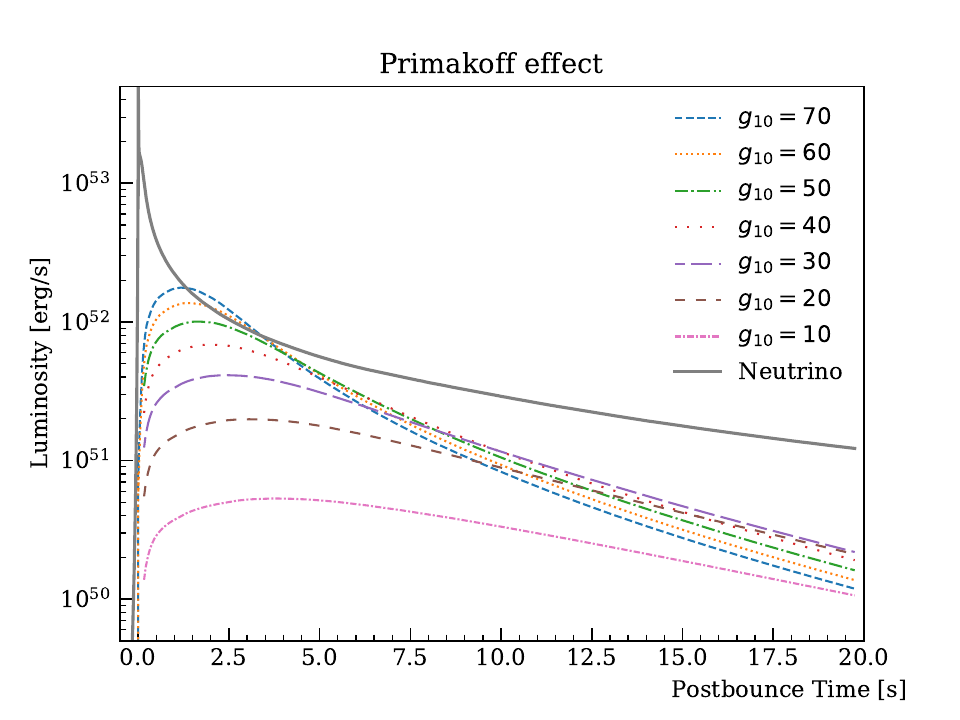}
    \end{minipage}
    
    \caption{ALP luminosities. The left panel represents that via the photon coalescence and the right panel represents that via the Primakoff effect and the total neutrino luminosity of the reference model. The color assignment follows Figure~\ref{fig:rho_temp}.}
    \label{fig:axion_lum}
\end{figure*}

Figure~\ref{fig:axion_lum} shows the comparison of ALP luminosities. The right panel shows emissions via the photon coalescence and the left panel shows those from the Primakoff effect and neutrino luminosity including all flavors as the reference. The luminosities via the photon coalescence are from $10^{49} {\rm erg/s}$ to  $10^{47} {\rm erg/s}$ at the bounce and they decrease to between about $10^{43} {\rm erg/s}$ to  $10^{44} {\rm erg/s}$. The hierarchy follows the order of the coupling constants and is kept until 20\,s. We attribute the kink around 7.5\,s in the curve of the $g_{10} = 70$ model to insufficient resolution, which results from the strong cooling rate and rapid contraction. The luminosities via the Primakoff effect increase more slowly and reach peaks after 1\,s. The order of peak heights also corresponds to the order of the coupling constants. Then, the luminosities gradually decay. After 5\,s in post-bounce time, the luminosities begin to cross each other. The decay rate of luminosity of the model with $g_{10} = 70$ is the highest and the second lowest luminosity at 20\,s after the bounce. The third lowest luminosity corresponds to $g_{10} = 60$ and has the second highest peak at 2\,s. However, the model with $g_{10} = 10$ always maintains the lowest luminosity. Comparing to the neutrino luminosity, the luminosities with $g_{10}=70$ and $g_{10}=60$ exceed the neutrino luminosity between 1.5\,s to 4.0\,s, which is consistent with the result in Ref.~\cite{2020PhLB..80935709C}. They obtained a constraint of $g_{10}\gtrsim 60$ based on the criterion that ALP luminosities must not exceed the total neutrino luminosity at 1\,s after the bounce.  By contrast, the other ALP luminosities invariably are lower than that of neutrino. 

Comparing the two panels proves that the Primakoff effect is considerably dominant. The ALP luminosities via the Primakoff effect is four to six orders of magnitude higher than those via the photon coalescence. Thus, the Primakoff effect determines luminosities in our simulations.

\subsection{Neutrino emission}
\begin{figure*}
\begin{minipage}[b]{0.64\columnwidth}
        \centering
    \includegraphics[width=1.0\linewidth]{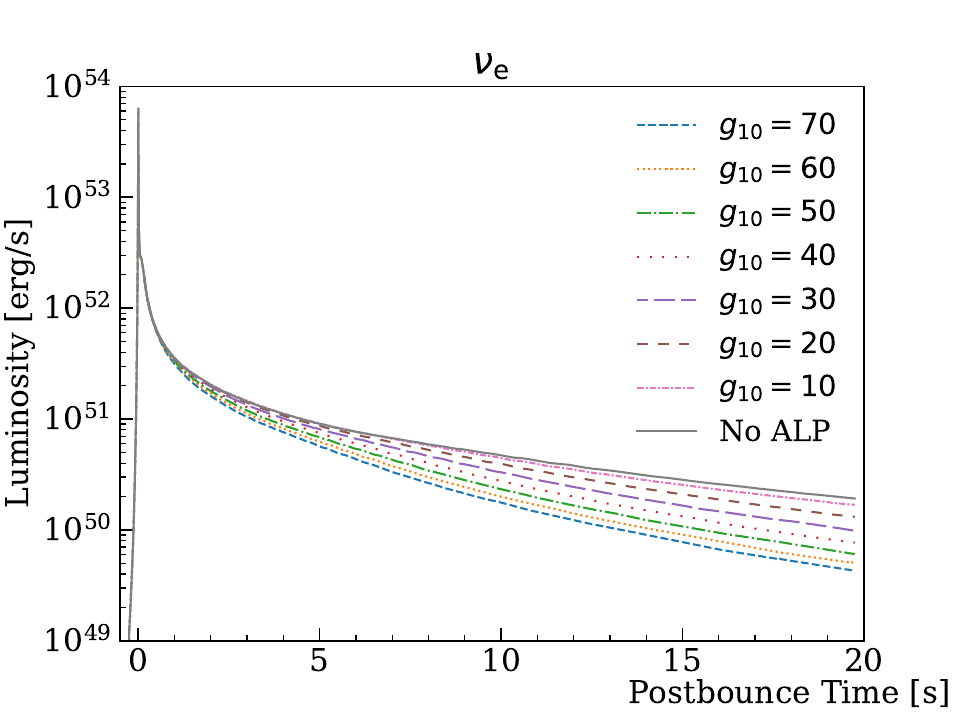}
\end{minipage}
\begin{minipage}[b]{0.64\columnwidth}
        \centering
    \includegraphics[width=1.0\linewidth]{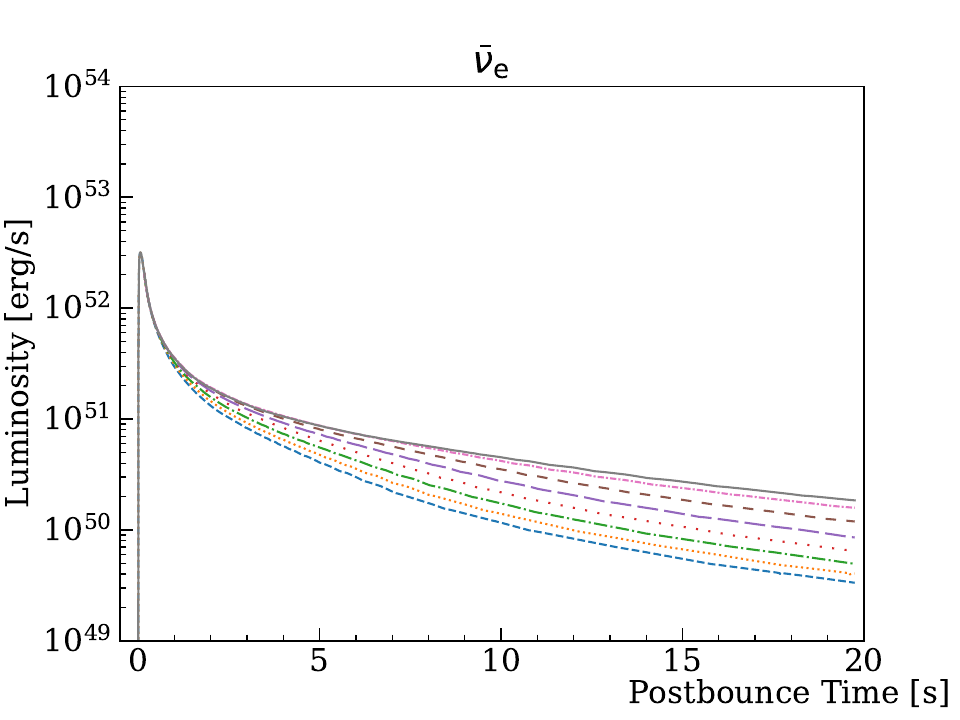}
\end{minipage}
\begin{minipage}[b]{0.64\columnwidth}
        \centering
    \includegraphics[width=1.0\linewidth]{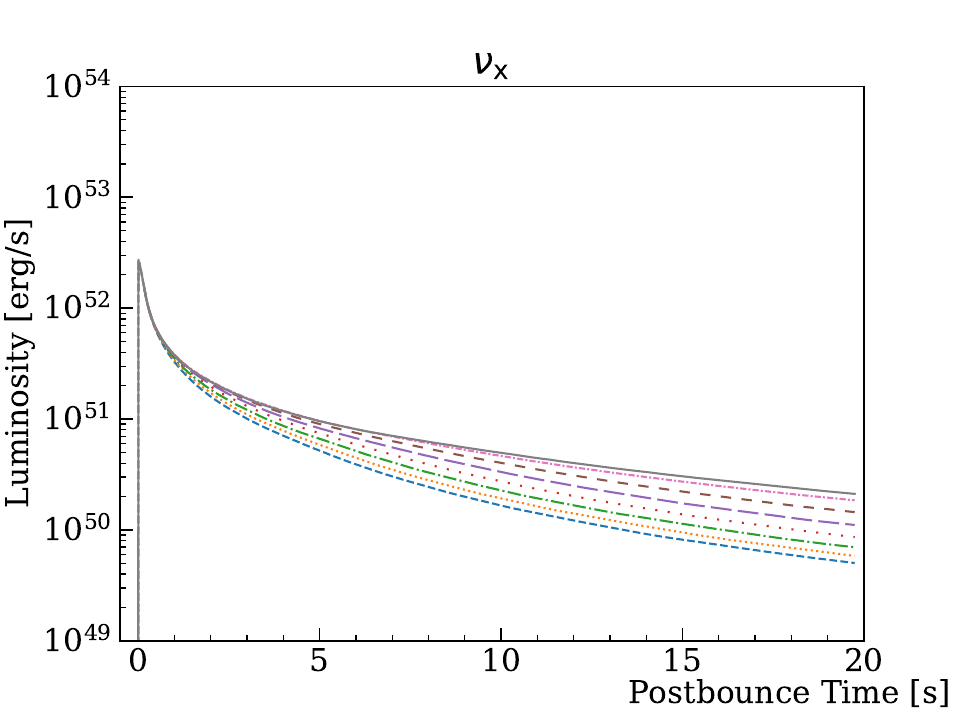}
\end{minipage}

    \caption{Neutrino luminosities. The left panel is luminosities of electron neutrino $\nu_{\rm e}$ and the middle panel is those of anti-electron neutrino $\bar{\nu}_{\rm e}$ and the right panel is those of heavy-lepton neutrino $\nu_{\rm x}$. The color assignment follows Figure~\ref{fig:rho_temp}.
    \label{fig:nu_lum}}
\end{figure*}

Figure~\ref{fig:nu_lum} shows comparison of neutrino luminosities of electron neutrino $\nu_{\rm e}$, anti-electron neutrino $\bar{\nu}_{\rm e}$ and heavy-lepton neutrino $\nu_{\rm x}$. We compare luminosities of each neutrino flavor across models. For all of the three flavors, the luminosities take their peaks at the bounce and monotonically decrease. The luminosities decrease more largely for the larger coupling constants in the late phase. Comparing  the model with $g_{10}=70$ to the no ALP model, the former luminosities are about one-tenth smaller than those of the latter for all flavors. The luminosities of the model with $g_{10}=10$ are almost the same as those of the reference model.

\begin{figure*}
\begin{minipage}[b]{0.64\columnwidth}
        \centering
    \includegraphics[width=1.0\linewidth]{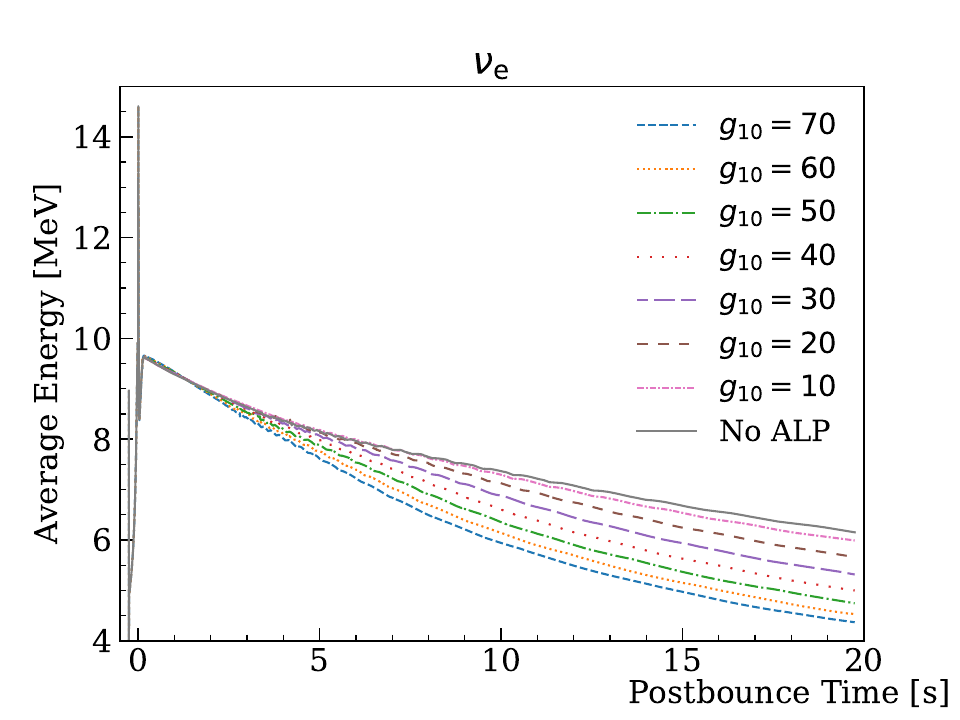}
\end{minipage}
\begin{minipage}[b]{0.64\columnwidth}
        \centering
    \includegraphics[width=1.0\linewidth]{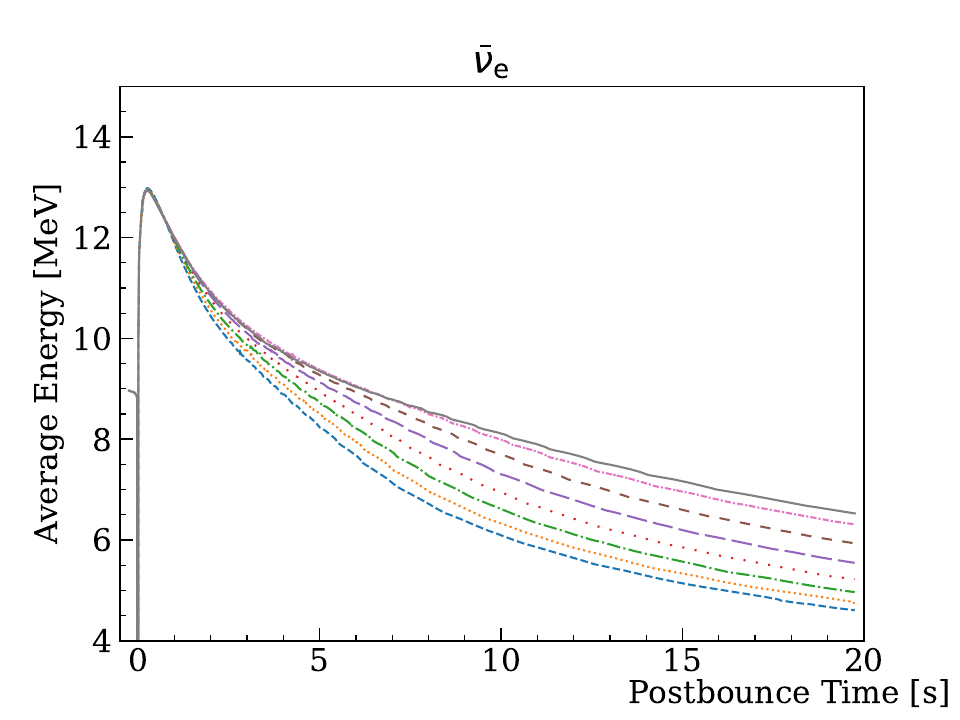}
\end{minipage}
\begin{minipage}[b]{0.64\columnwidth}
        \centering
    \includegraphics[width=1.0\linewidth]{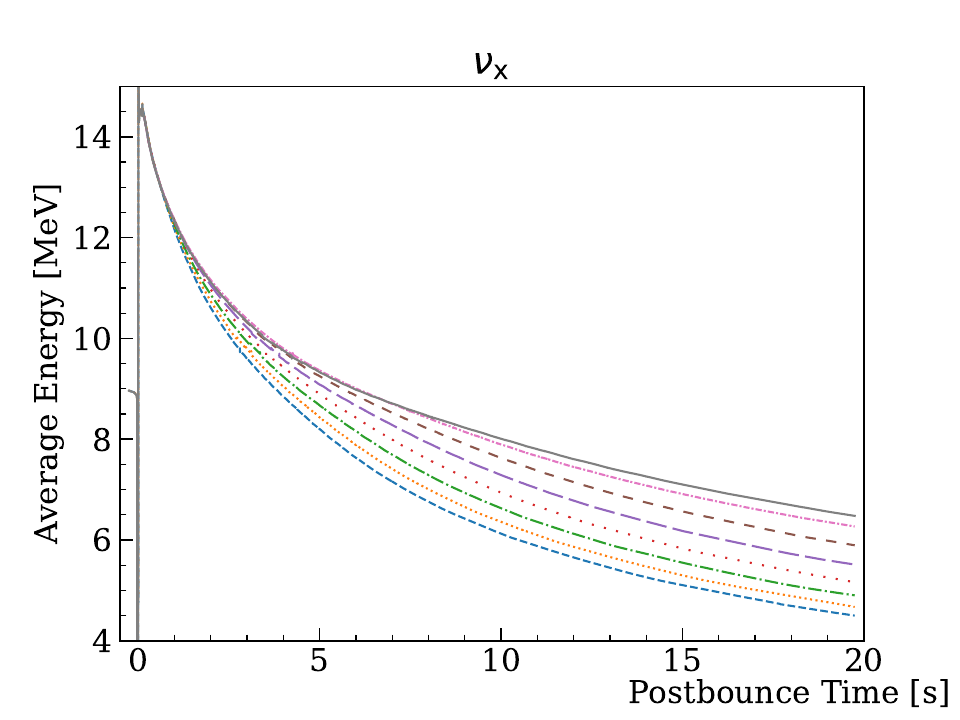}
\end{minipage}

    \caption{Neutrino average energies. The left panel is average energies of electron neutrino $\nu_{\rm e}$ and the middle panel is those of anti-electron neutrino $\bar{\nu}_{\rm e}$ and the right panel is those of heavy-lepton neutrino $\nu_{\rm x}$.
    \label{fig:nu_ave}}
\end{figure*}

Figure~\ref{fig:nu_ave} shows the time evolution of average energies of neutrinos. Our definition of the average energy is 
\begin{equation}
    \left\langle E_{\nu} \right\rangle = \frac{\int Ef(E)dE}{\int f(E)dE},
\end{equation}
with $f(E)$ the spectrum energy density.
The tendency of influence of ALPs is the same as that in luminosities. The larger coupling constant ALPs more largely decrease neutrino average energies. The average energies of the model with $g_{a\gamma}=7.0\times10^{-9} {\rm \,GeV^{-1}}$ are lower by 3\,MeV than those of the reference model. In contract, the difference between average energies of the model with $g_{a\gamma}=1.0\times10^{-9} {\rm \,GeV^{-1}}$ 
and those of the reference model are about 0.1\,MeV at 20\,s.  

\begin{figure*}
\begin{minipage}[b]{0.64\columnwidth}
        \centering
    \includegraphics[width=1.0\linewidth]{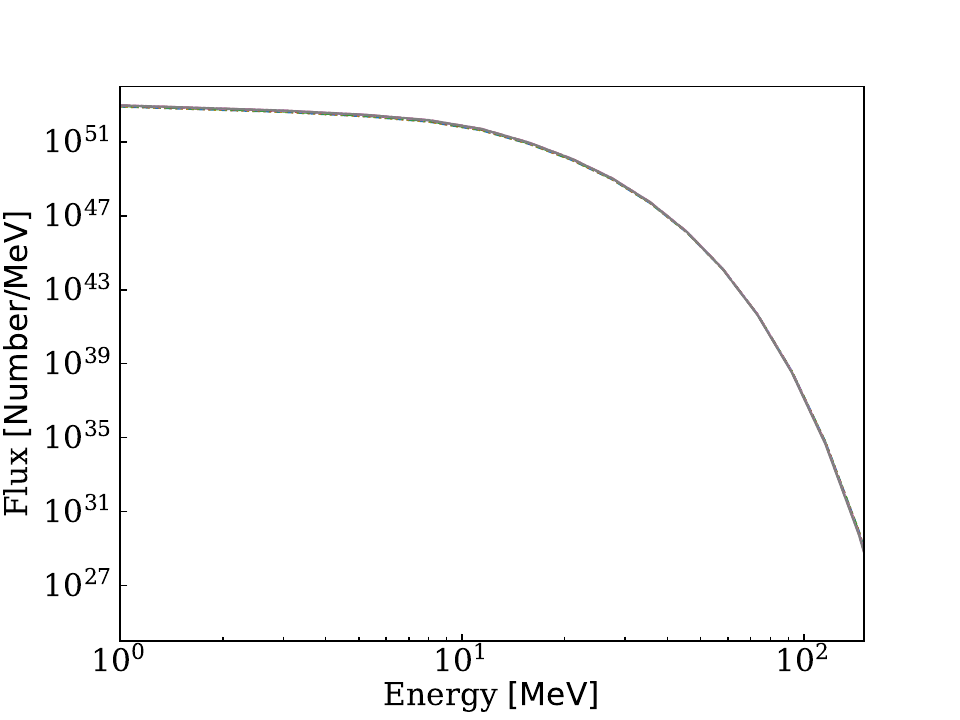}
        \includegraphics[width=1.0\linewidth]{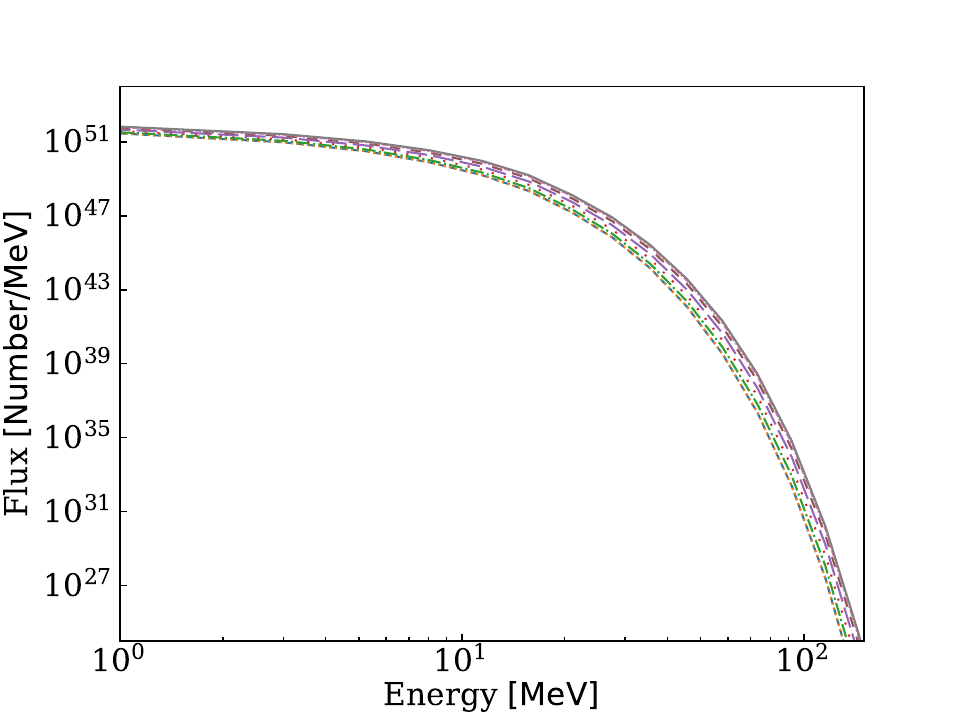}
            \includegraphics[width=1.0\linewidth]{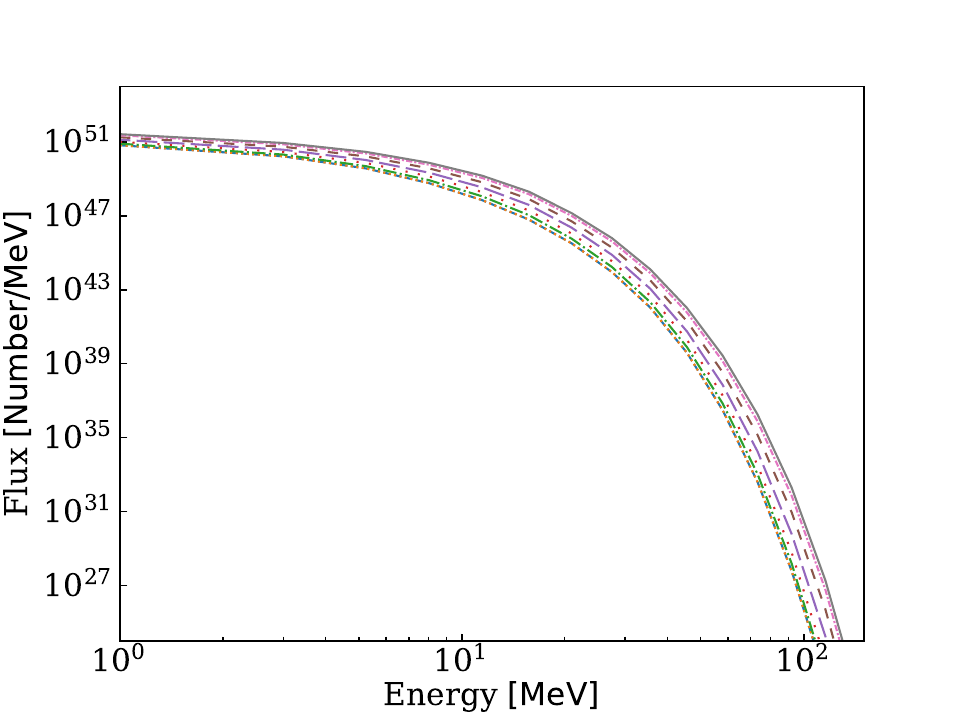}
\end{minipage}
\begin{minipage}[b]{0.64\columnwidth}
        \centering
    \includegraphics[width=1.0\linewidth]{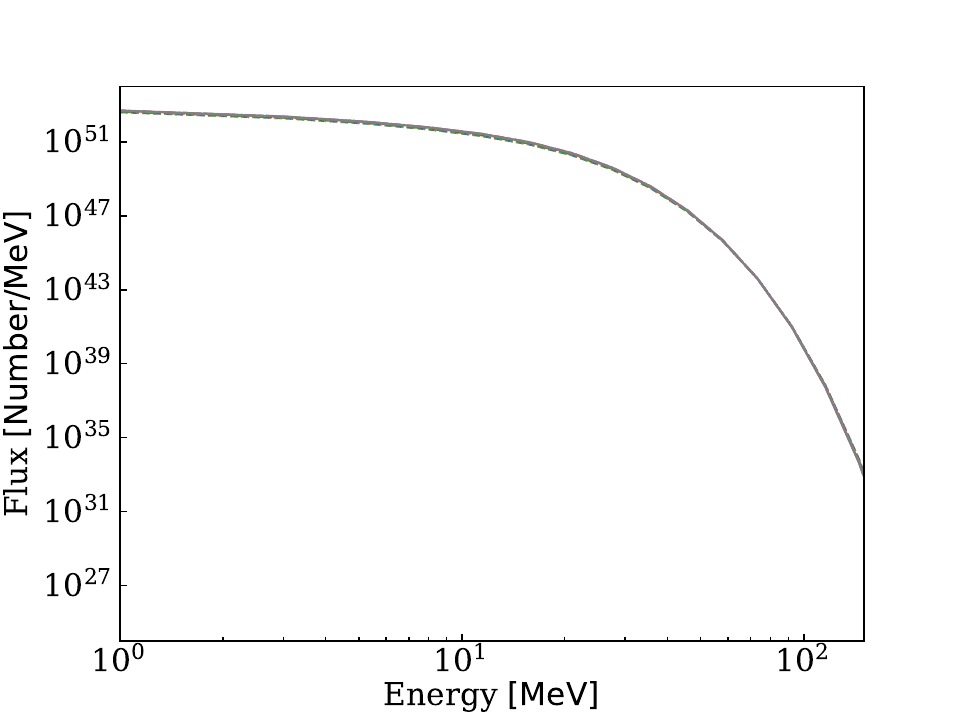}
        \includegraphics[width=1.0\linewidth]{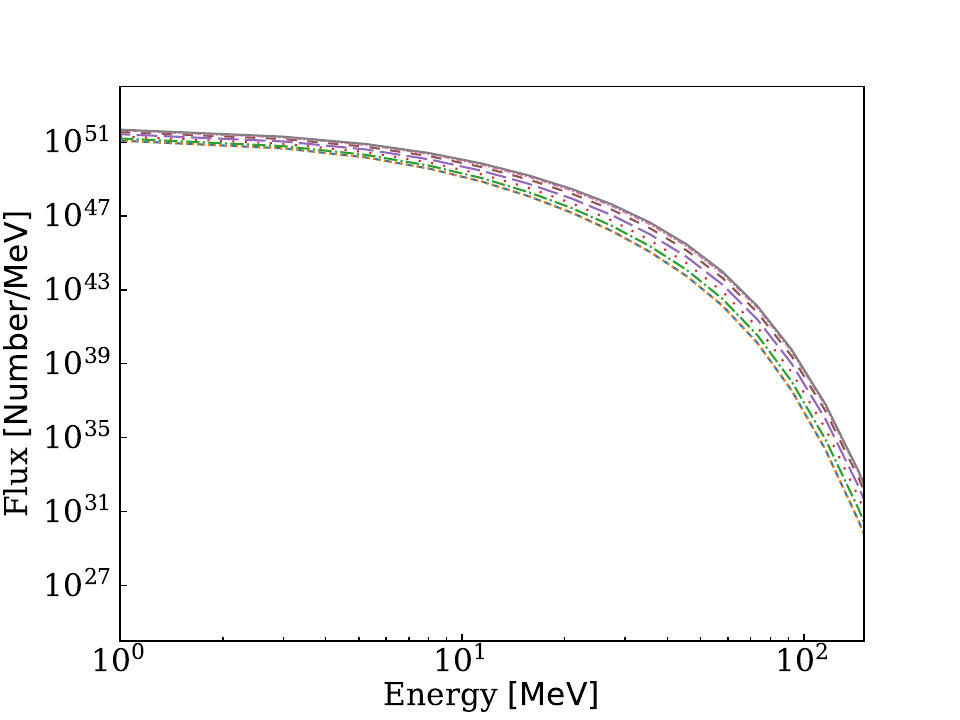}
            \includegraphics[width=1.0\linewidth]{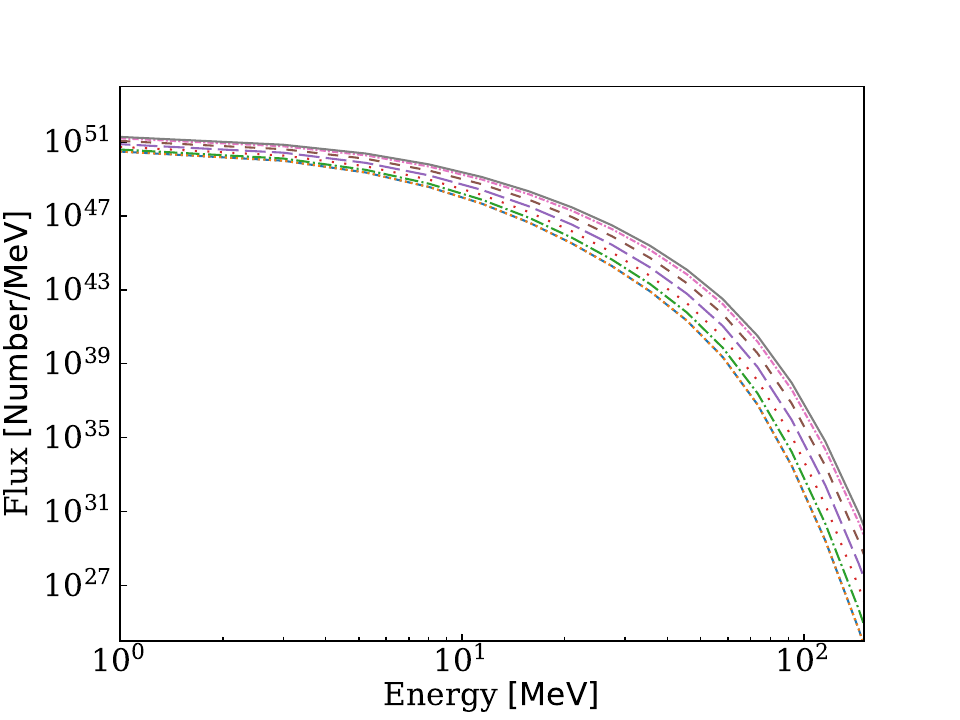}
\end{minipage}
\begin{minipage}[b]{0.64\columnwidth}
        \centering
    \includegraphics[width=1.0\linewidth]{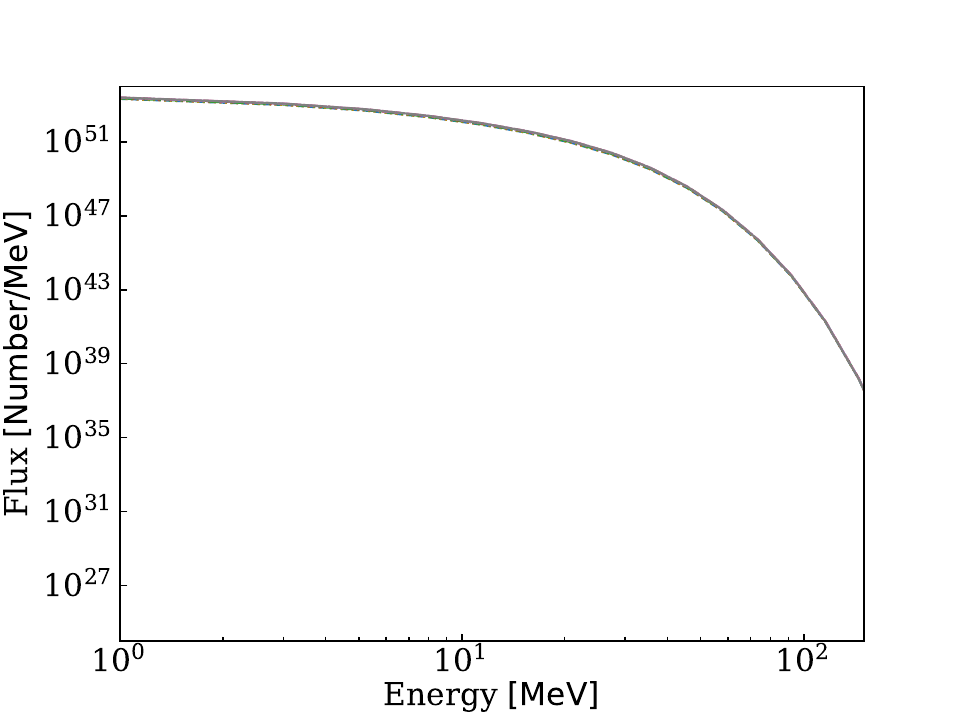}
        \includegraphics[width=1.0\linewidth]{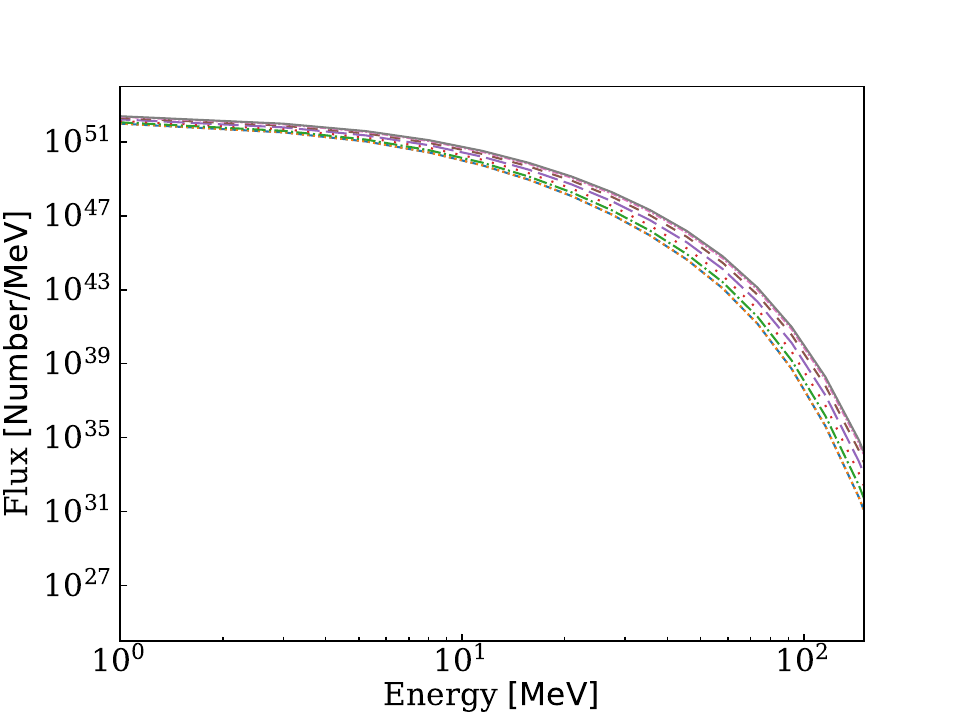}
            \includegraphics[width=1.0\linewidth]{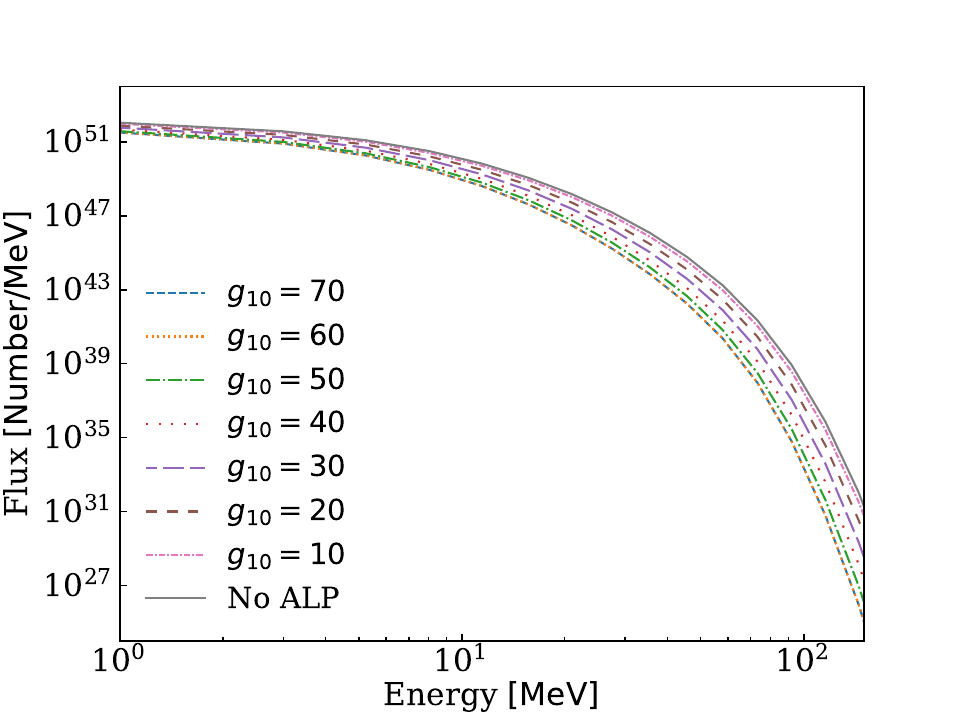}
\end{minipage}

    \caption{Neutrino number spectra. The horizontal axis is neutrino energy in MeV and the vertical axis is number flux per MeV. The panels are arranged vertically in time and horizontally in flavors. The left column is 1\,s after the bounce, the middle column is 10\,s after the bounce and the right column is 20\,s after the bounce. The top row is $\nu_{\rm e}$, the middle row is $\bar{\nu}_{\rm e}$ and the bottom row is $\nu_{\rm x}$.}
    \label{fig:nu_spectra}
\end{figure*}

Figure~\ref{fig:nu_spectra} shows the number spectra at 1\,s, 10\,s and 20\,s after the bounce for three flavors and each model. At 1\,s, all the spectra are overlapped. At 10\,s, the energy spectra begin to decrease and the differences become more pronounced, especially at higher energies. The energy hierarchy is kept for all flavors until 20\,s. Figures~\ref{fig:nu_lum},~\ref{fig:nu_ave} and~\ref{fig:nu_spectra} show neutrinos reflect the cooling enhanced by the ALPs.

\begin{figure}
    \centering
    \includegraphics[width=1.0\linewidth]{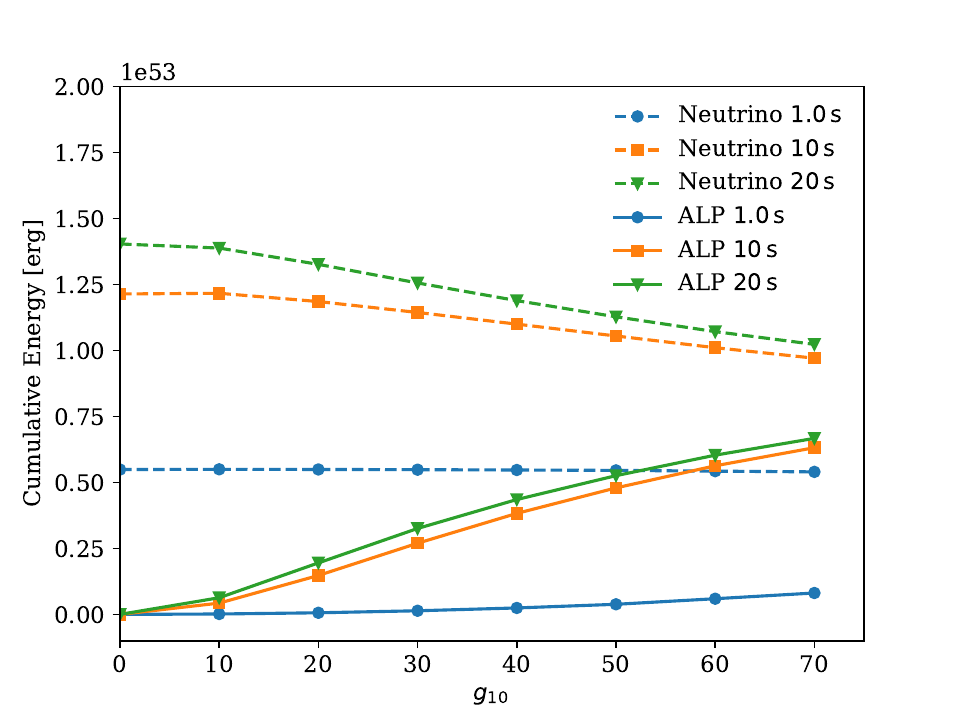}
    \caption{Cumulative energy emitted via neutrinos and ALPs versus coupling constants $g_{10}$. The solid curves mean neutrino and the dashed curves mean ALP. The blue corresponds to 1.0\,s, the orange to 10\,s and the green to 20\,s after the bounce.}
    \label{fig:compare_cum_nu_alp_redshift}
\end{figure}

We show the dependence of cumulative energy emitted via neutrinos and ALPs on the coupling constants $g_{10}$ in Figure~\ref{fig:compare_cum_nu_alp_redshift}. The emitted energy of neutrinos up to 1.0\,s is $5.5\times 10^{52}{\rm\, erg}$ and invariable for $g_{10}$. At 10\,s and 20\,s the cumulative energy emitted as neutrinos decreases from $1.24\times10^{52}{\rm\,erg}$ and $1.40\times10^{52}{\rm \,erg}$  to $1.23\times10^{52}{\rm\,erg}$ and $1.25\times10^{52}{\rm\,erg}$ for as the total energy emitted ALPs increases to $6.2\times10^{52} {\rm\,erg}$.

\section{Discussion}\label{sec:discussion}
This section describes the discussion on neutrino observation and the possibility to find ALP signals from neutrino events. Here we assume Super-Kamiokande as a neutrino detector.
\subsection{Neutrino detection}\label{subsec:neutrino_detection}
Supernova neutrinos can be detected via underground detectors. The Super-Kamiokande (SK) detector~\cite{Super-Kamiokande:2002weg} is a large water-Cherenkov detector in Japan. The SK detector is monitoring supernovae for 24 hours and issues alert to observatories all over the world if it detects supernovae~\cite{2016APh....81...39A}. SK observes mainly supernova neutrinos via the inverse beta decay (IBD)~\cite{2003PhLB..564...42S},
\begin{equation}
    \bar{\nu}_{\rm e} + p^{+} \rightarrow n + e^{+}.
\end{equation}
in which a proton $p$ and an electron antineutrino $\bar{\nu}_{\rm e}$ interact and produce a neutron $n$ and a positron $e^+$.
SK has 100\,\% sensitivity for galactic supernovae~\cite{2007ApJ...669..519I,2022ApJ...938...35M}. SK employs the full of 32.5\,kton of water in the inner detector and some thousands of neutrinos can interact in the SK detector~\cite{2022ApJ...938...35M}. We assume that CCSNe occur at 10\,kpc away from the Earth, neutrino flavors mix following the MSW oscillation~\cite{2000PhRvD..62c3007D, 2015PhRvD..91f5016W} during traveling, mass hierarchy is normal, and neutrinos interact only via the IBD in the SK detector.

\begin{figure*}
    \centering
    \includegraphics[width=0.48\linewidth]{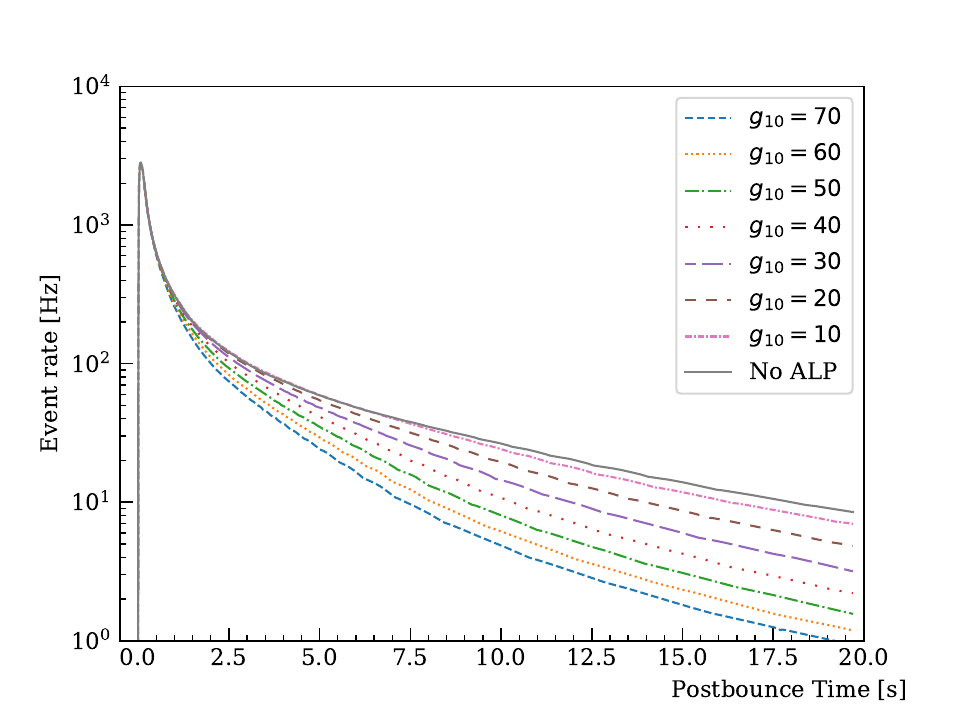}
    \includegraphics[width=0.48\linewidth]{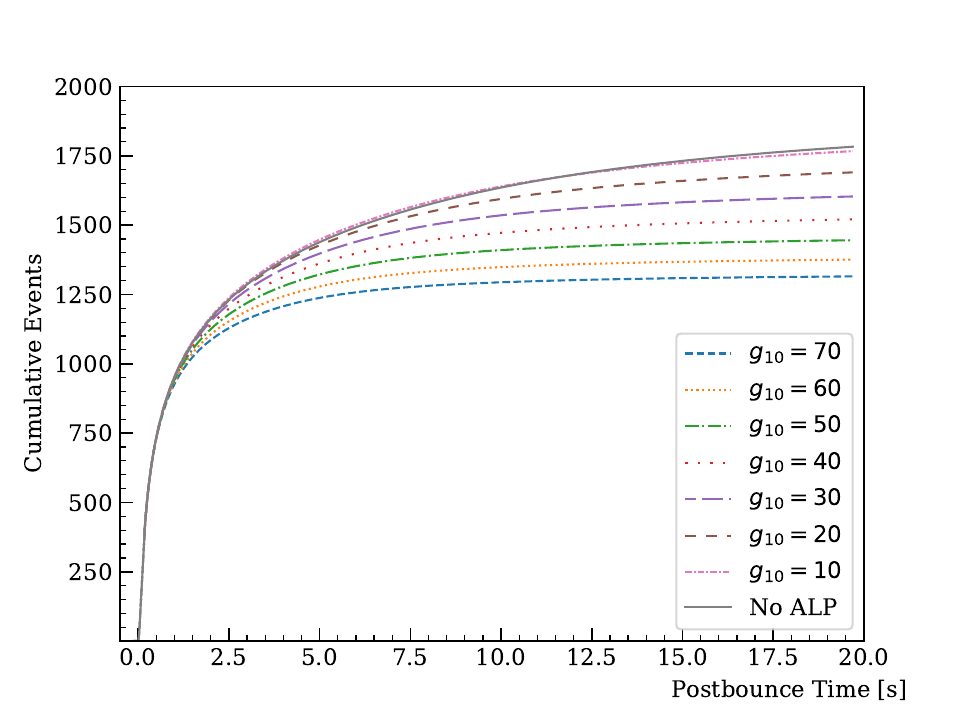}
    \caption{Neutrino event rate (left) and the cumulative number of events (right). It is assumed that CCSNe occurs at 10\,kpc away.}
    \label{fig:nu_event}
\end{figure*}

Figure~\ref{fig:nu_event} shows the event rates and the cumulative event numbers. The event rate and also the cumulative events almost overlap until 1\,s after the bounce. After 2\,s after the bounce, the curves start to separate from each other. At 20\,s after the bounce, the event rate of the model with $g_{10}=70$ decreases below 1\,Hz and smaller than 5\,\% of that of the reference model. This discrepancy arises because event rates depend on both the luminosity and the average energy of neutrinos.

The right panel in Figure~\ref{fig:nu_event} shows cumulative events. The time evolutions of cumulative events also overlap until 1\,s after the bounce and the time evolutions begin to separate from each other. Even after 10\,s, the cumulative events continue to increase. The number of cumulative events of the reference model reaches 1,750 events at 20\,s and on the other hand, that of the model with $g_{10}=70$ reaches 1,250 events.

\begin{figure}
    \centering
    \includegraphics[width=1.0\linewidth]{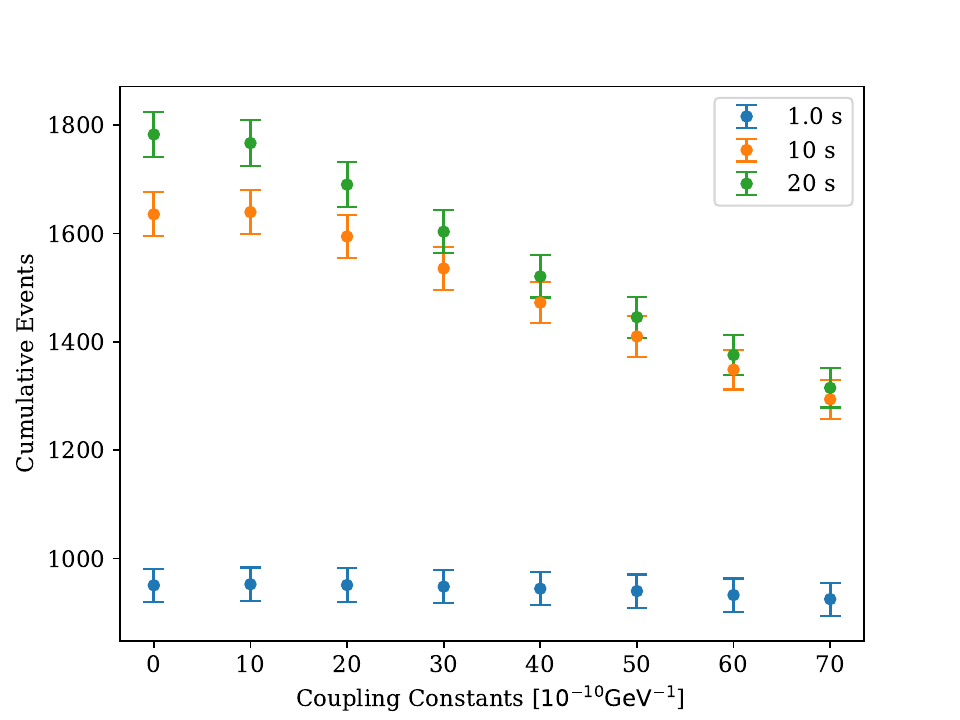}
    \caption{Comparison of the number of cumulative events versus the coupling constants at 1\,s, 10\,s and 20\,s after the bounce. The error bars indicate errors of the Poisson distribution.}
    \label{fig:compare_cum}
\end{figure}

Figure~\ref{fig:compare_cum} plots the numbers of cumulative events at 1\,s, 10\,s and 20\,s after the bounce with error bars. We assume the Poisson distribution and the definition of error bars writes $\sqrt{N(t)}$ with $N(t)$ the number of cumulative events at $t$. We put the errors upward and downward around the center values. At 1\,s after the bounce, we cannot discriminate the models. However, we might discriminate the model with the coupling constants above $g_{10}=30$, which might expand constraint on the ALP parameters.

\subsection{Future plan}

Next, we will conduct simulations of various progenitors. There are a few red supergiants that may be close to undergoing a CCSN such as Betelgeuse. We plan to simulate specific red supergiant progenitors located near Earth~\cite{2019AJ....158...20M,2024MNRAS.529.3630H} and in the local galaxies~\cite{2021ApJ...923..232R}. 

Supernovae can result in black holes instead of neutron stars~\cite{2006AIPC..847..473S,2007ApJ...667..382S,2024arXiv241207831B}. Simulating core collapse leading to black hole formation is more challenging, as it requires fully incorporating general relativity, and involves higher densities and temperatures. These extreme conditions can produce more ALPs than those produced in the neutron star-forming case.  This makes such simulations a worthwhile next step.

In the future, we will construct an analytic formula of neutrino luminosity and event rate and develop a method for estimating existence of ALPs. Analytical late-phase neutrino luminosity and event rate models have been developed in Ref.~\cite{2021PTEP.2021a3E01S} and the methods for estimating PNS mass, radius and so on from neutrinos using the analytic formula have also been proposed~\cite{2022ApJ...934...15S,2023ApJ...954...52H,2025ApJ...980..117S}. Then, an analytic ALP luminosity has been derived~\cite{2023APh...15102855F}. We will incorporate the effect of ALPs as the additional generation of leptons into the formulae in Ref.~\cite{2021PTEP.2021a3E01S}. The modification effectively model the cooling effect of ALPs.


\section{Summary}\label{sec:summary}
In this paper, we described the long-term simulations with the general relativistic neutrino radiation hydrodynamics including axion cooling, in which ALP mass is 10\,MeV and coupling constants are from $10\times10^{-9} {\rm \,GeV^{-1}}$ to $70\times10^{-9}{\rm \,GeV^{-1}}$ and investigated influences of ALPs on fluid and neutrino emission. The ALPs accelerate the PNS cooling. We found that the effects of ALPs do not emerge in the early phase until 1\,s after the bounce but in the late phase on our zero-metallicity 9.6$M_\odot$ progenitor. We assumed that supernovae occur at 10\,kpc away and are observed with SK. The event rate and cumulative events significantly decrease in the late phase. If we detect neutrinos from Galactic supernovae, we might find an indication of an ALP that evade the constraint based on the conventional energy-loss argument.

\acknowledgments

We thank Yudai Suwa, Tomoya Takiwaki, Kohsuke Sumiyoshi and the all members of the nuLC collaboration \footnote{https://sites.google.com/g.ecc.u-tokyo.ac.jp/nulc} for discussions. This work is supported by JSPS KAKENHI Grant Numbers  JP23KJ2147, JP23K13107, JP23KJ2150, and JP25H02194.

\nocite{*}

\bibliography{biblio.bib}

\begin{thebibliography}{65}%
\makeatletter
\providecommand \@ifxundefined [1]{%
 \@ifx{#1\undefined}
}%
\providecommand \@ifnum [1]{%
 \ifnum #1\expandafter \@firstoftwo
 \else \expandafter \@secondoftwo
 \fi
}%
\providecommand \@ifx [1]{%
 \ifx #1\expandafter \@firstoftwo
 \else \expandafter \@secondoftwo
 \fi
}%
\providecommand \natexlab [1]{#1}%
\providecommand \enquote  [1]{``#1''}%
\providecommand \bibnamefont  [1]{#1}%
\providecommand \bibfnamefont [1]{#1}%
\providecommand \citenamefont [1]{#1}%
\providecommand \href@noop [0]{\@secondoftwo}%
\providecommand \href [0]{\begingroup \@sanitize@url \@href}%
\providecommand \@href[1]{\@@startlink{#1}\@@href}%
\providecommand \@@href[1]{\endgroup#1\@@endlink}%
\providecommand \@sanitize@url [0]{\catcode `\\12\catcode `\$12\catcode
  `\&12\catcode `\#12\catcode `\^12\catcode `\_12\catcode `\%12\relax}%
\providecommand \@@startlink[1]{}%
\providecommand \@@endlink[0]{}%
\providecommand \url  [0]{\begingroup\@sanitize@url \@url }%
\providecommand \@url [1]{\endgroup\@href {#1}{\urlprefix }}%
\providecommand \urlprefix  [0]{URL }%
\providecommand \Eprint [0]{\href }%
\providecommand \doibase [0]{https://doi.org/}%
\providecommand \selectlanguage [0]{\@gobble}%
\providecommand \bibinfo  [0]{\@secondoftwo}%
\providecommand \bibfield  [0]{\@secondoftwo}%
\providecommand \translation [1]{[#1]}%
\providecommand \BibitemOpen [0]{}%
\providecommand \bibitemStop [0]{}%
\providecommand \bibitemNoStop [0]{.\EOS\space}%
\providecommand \EOS [0]{\spacefactor3000\relax}%
\providecommand \BibitemShut  [1]{\csname bibitem#1\endcsname}%
\let\auto@bib@innerbib\@empty
\bibitem [{\citenamefont {{Di Luzio}}\ \emph {et~al.}(2020)\citenamefont {{Di
  Luzio}}, \citenamefont {{Giannotti}}, \citenamefont {{Nardi}},\ and\
  \citenamefont {{Visinelli}}}]{2020PhR...870....1D}%
  \BibitemOpen
  \bibfield  {author} {\bibinfo {author} {\bibfnamefont {L.}~\bibnamefont {{Di
  Luzio}}}, \bibinfo {author} {\bibfnamefont {M.}~\bibnamefont {{Giannotti}}},
  \bibinfo {author} {\bibfnamefont {E.}~\bibnamefont {{Nardi}}},\ and\ \bibinfo
  {author} {\bibfnamefont {L.}~\bibnamefont {{Visinelli}}},\ }\bibfield
  {title} {\bibinfo {title} {{Corrigendum to ``The landscape of QCD axion
  models'' [Phys. Rep. 870 (2020) 1-117]}},\ }\href
  {https://doi.org/10.1016/j.physrep.2022.06.006} {\bibfield  {journal}
  {\bibinfo  {journal} {Physics Reports}\ }\textbf {\bibinfo {volume} {870}},\
  \bibinfo {pages} {1} (\bibinfo {year} {2020})},\ \Eprint
  {https://arxiv.org/abs/2003.01100} {arXiv:2003.01100 [hep-ph]} \BibitemShut
  {NoStop}%
\bibitem [{\citenamefont {{Choi}}\ \emph {et~al.}(2021)\citenamefont {{Choi}},
  \citenamefont {{Im}},\ and\ \citenamefont {{Shin}}}]{2021ARNPS..71..225C}%
  \BibitemOpen
  \bibfield  {author} {\bibinfo {author} {\bibfnamefont {K.}~\bibnamefont
  {{Choi}}}, \bibinfo {author} {\bibfnamefont {S.~H.}\ \bibnamefont {{Im}}},\
  and\ \bibinfo {author} {\bibfnamefont {C.~S.}\ \bibnamefont {{Shin}}},\
  }\bibfield  {title} {\bibinfo {title} {{Recent Progress in the Physics of
  Axions and Axion-Like Particles}},\ }\href
  {https://doi.org/10.1146/annurev-nucl-120720-031147} {\bibfield  {journal}
  {\bibinfo  {journal} {Annual Review of Nuclear and Particle Science}\
  }\textbf {\bibinfo {volume} {71}},\ \bibinfo {pages} {225} (\bibinfo {year}
  {2021})},\ \Eprint {https://arxiv.org/abs/2012.05029} {arXiv:2012.05029
  [hep-ph]} \BibitemShut {NoStop}%
\bibitem [{\citenamefont {{Raffelt}}\ and\ \citenamefont
  {{Stodolsky}}(1988)}]{1988PhRvD..37.1237R}%
  \BibitemOpen
  \bibfield  {author} {\bibinfo {author} {\bibfnamefont {G.}~\bibnamefont
  {{Raffelt}}}\ and\ \bibinfo {author} {\bibfnamefont {L.}~\bibnamefont
  {{Stodolsky}}},\ }\bibfield  {title} {\bibinfo {title} {{Mixing of the photon
  with low-mass particles}},\ }\href {https://doi.org/10.1103/PhysRevD.37.1237}
  {\bibfield  {journal} {\bibinfo  {journal} {\prd}\ }\textbf {\bibinfo
  {volume} {37}},\ \bibinfo {pages} {1237} (\bibinfo {year}
  {1988})}\BibitemShut {NoStop}%
\bibitem [{\citenamefont {{Jaeckel}}\ and\ \citenamefont
  {{Spannowsky}}(2016)}]{2016PhLB..753..482J}%
  \BibitemOpen
  \bibfield  {author} {\bibinfo {author} {\bibfnamefont {J.}~\bibnamefont
  {{Jaeckel}}}\ and\ \bibinfo {author} {\bibfnamefont {M.}~\bibnamefont
  {{Spannowsky}}},\ }\bibfield  {title} {\bibinfo {title} {{Probing MeV to 90
  GeV axion-like particles with LEP and LHC}},\ }\href
  {https://doi.org/10.1016/j.physletb.2015.12.037} {\bibfield  {journal}
  {\bibinfo  {journal} {Physics Letters B}\ }\textbf {\bibinfo {volume}
  {753}},\ \bibinfo {pages} {482} (\bibinfo {year} {2016})},\ \Eprint
  {https://arxiv.org/abs/1509.00476} {arXiv:1509.00476 [hep-ph]} \BibitemShut
  {NoStop}%
\bibitem [{\citenamefont {{D{\"o}brich}}\ \emph {et~al.}(2019)\citenamefont
  {{D{\"o}brich}}, \citenamefont {{Jaeckel}},\ and\ \citenamefont
  {{Spadaro}}}]{2019JHEP...05..213D}%
  \BibitemOpen
  \bibfield  {author} {\bibinfo {author} {\bibfnamefont {B.}~\bibnamefont
  {{D{\"o}brich}}}, \bibinfo {author} {\bibfnamefont {J.}~\bibnamefont
  {{Jaeckel}}},\ and\ \bibinfo {author} {\bibfnamefont {T.}~\bibnamefont
  {{Spadaro}}},\ }\bibfield  {title} {\bibinfo {title} {{Light in the beam
  dump. Axion-Like Particle production from decay photons in proton
  beam-dumps}},\ }\href {https://doi.org/10.1007/JHEP05(2019)213} {\bibfield
  {journal} {\bibinfo  {journal} {Journal of High Energy Physics}\ }\textbf
  {\bibinfo {volume} {2019}},\ \bibinfo {eid} {213} (\bibinfo {year}
  {2019})}\BibitemShut {NoStop}%
\bibitem [{\citenamefont {{Dolan}}\ \emph {et~al.}(2017)\citenamefont
  {{Dolan}}, \citenamefont {{Ferber}}, \citenamefont {{Hearty}}, \citenamefont
  {{Kahlhoefer}},\ and\ \citenamefont
  {{Schmidt-Hoberg}}}]{2017JHEP...12..094D}%
  \BibitemOpen
  \bibfield  {author} {\bibinfo {author} {\bibfnamefont {M.~J.}\ \bibnamefont
  {{Dolan}}}, \bibinfo {author} {\bibfnamefont {T.}~\bibnamefont {{Ferber}}},
  \bibinfo {author} {\bibfnamefont {C.}~\bibnamefont {{Hearty}}}, \bibinfo
  {author} {\bibfnamefont {F.}~\bibnamefont {{Kahlhoefer}}},\ and\ \bibinfo
  {author} {\bibfnamefont {K.}~\bibnamefont {{Schmidt-Hoberg}}},\ }\bibfield
  {title} {\bibinfo {title} {{Revised constraints and Belle II sensitivity for
  visible and invisible axion-like particles}},\ }\href
  {https://doi.org/10.1007/JHEP12(2017)094} {\bibfield  {journal} {\bibinfo
  {journal} {Journal of High Energy Physics}\ }\textbf {\bibinfo {volume}
  {2017}},\ \bibinfo {eid} {94} (\bibinfo {year} {2017})},\ \Eprint
  {https://arxiv.org/abs/1709.00009} {arXiv:1709.00009 [hep-ph]} \BibitemShut
  {NoStop}%
\bibitem [{\citenamefont {{Cadamuro}}\ and\ \citenamefont
  {{Redondo}}(2012)}]{2012JCAP...02..032C}%
  \BibitemOpen
  \bibfield  {author} {\bibinfo {author} {\bibfnamefont {D.}~\bibnamefont
  {{Cadamuro}}}\ and\ \bibinfo {author} {\bibfnamefont {J.}~\bibnamefont
  {{Redondo}}},\ }\bibfield  {title} {\bibinfo {title} {{Cosmological bounds on
  pseudo Nambu-Goldstone bosons}},\ }\href
  {https://doi.org/10.1088/1475-7516/2012/02/032} {\bibfield  {journal}
  {\bibinfo  {journal} {Journal of Cosmology and Astroparticle Physics}\
  }\textbf {\bibinfo {volume} {2012}}\bibfield  {number} {\bibinfo  {number} {
  (2)},\ \bibinfo {eid} {032}},\ }\Eprint {https://arxiv.org/abs/1110.2895}
  {arXiv:1110.2895 [hep-ph]} \BibitemShut {NoStop}%
\bibitem [{\citenamefont {{Depta}}\ \emph {et~al.}(2020)\citenamefont
  {{Depta}}, \citenamefont {{Hufnagel}},\ and\ \citenamefont
  {{Schmidt-Hoberg}}}]{2020JCAP...05..009D}%
  \BibitemOpen
  \bibfield  {author} {\bibinfo {author} {\bibfnamefont {P.~F.}\ \bibnamefont
  {{Depta}}}, \bibinfo {author} {\bibfnamefont {M.}~\bibnamefont
  {{Hufnagel}}},\ and\ \bibinfo {author} {\bibfnamefont {K.}~\bibnamefont
  {{Schmidt-Hoberg}}},\ }\bibfield  {title} {\bibinfo {title} {{Robust
  cosmological constraints on axion-like particles}},\ }\href
  {https://doi.org/10.1088/1475-7516/2020/05/009} {\bibfield  {journal}
  {\bibinfo  {journal} {Journal of Cosmology and Astroparticle Physics}\
  }\textbf {\bibinfo {volume} {2020}}\bibfield  {number} {\bibinfo  {number} {
  (5)},\ \bibinfo {eid} {009}},\ }\Eprint {https://arxiv.org/abs/2002.08370}
  {arXiv:2002.08370 [hep-ph]} \BibitemShut {NoStop}%
\bibitem [{\citenamefont {{Carenza}}\ \emph {et~al.}(2020)\citenamefont
  {{Carenza}}, \citenamefont {{Straniero}}, \citenamefont {{D{\"o}brich}},
  \citenamefont {{Giannotti}}, \citenamefont {{Lucente}},\ and\ \citenamefont
  {{Mirizzi}}}]{2020PhLB..80935709C}%
  \BibitemOpen
  \bibfield  {author} {\bibinfo {author} {\bibfnamefont {P.}~\bibnamefont
  {{Carenza}}}, \bibinfo {author} {\bibfnamefont {O.}~\bibnamefont
  {{Straniero}}}, \bibinfo {author} {\bibfnamefont {B.}~\bibnamefont
  {{D{\"o}brich}}}, \bibinfo {author} {\bibfnamefont {M.}~\bibnamefont
  {{Giannotti}}}, \bibinfo {author} {\bibfnamefont {G.}~\bibnamefont
  {{Lucente}}},\ and\ \bibinfo {author} {\bibfnamefont {A.}~\bibnamefont
  {{Mirizzi}}},\ }\bibfield  {title} {\bibinfo {title} {{Constraints on the
  coupling with photons of heavy axion-like-particles from Globular
  Clusters}},\ }\href {https://doi.org/10.1016/j.physletb.2020.135709}
  {\bibfield  {journal} {\bibinfo  {journal} {Physics Letters B}\ }\textbf
  {\bibinfo {volume} {809}},\ \bibinfo {eid} {135709} (\bibinfo {year}
  {2020})},\ \Eprint {https://arxiv.org/abs/2004.08399} {arXiv:2004.08399
  [hep-ph]} \BibitemShut {NoStop}%
\bibitem [{\citenamefont {{Dolan}}\ \emph {et~al.}(2021)\citenamefont
  {{Dolan}}, \citenamefont {{Hiskens}},\ and\ \citenamefont
  {{Volkas}}}]{2021JCAP...09..010D}%
  \BibitemOpen
  \bibfield  {author} {\bibinfo {author} {\bibfnamefont {M.~J.}\ \bibnamefont
  {{Dolan}}}, \bibinfo {author} {\bibfnamefont {F.~J.}\ \bibnamefont
  {{Hiskens}}},\ and\ \bibinfo {author} {\bibfnamefont {R.~R.}\ \bibnamefont
  {{Volkas}}},\ }\bibfield  {title} {\bibinfo {title} {{Constraining axion-like
  particles using the white dwarf initial-final mass relation}},\ }\href
  {https://doi.org/10.1088/1475-7516/2021/09/010} {\bibfield  {journal}
  {\bibinfo  {journal} {Journal of Cosmology and Astroparticle Physics}\
  }\textbf {\bibinfo {volume} {2021}}\bibfield  {number} {\bibinfo  {number} {
  (9)},\ \bibinfo {eid} {010}},\ }\Eprint {https://arxiv.org/abs/2102.00379}
  {arXiv:2102.00379 [hep-ph]} \BibitemShut {NoStop}%
\bibitem [{\citenamefont {{Lee}}(2018)}]{2018arXiv180810136L}%
  \BibitemOpen
  \bibfield  {author} {\bibinfo {author} {\bibfnamefont {J.~S.}\ \bibnamefont
  {{Lee}}},\ }\bibfield  {title} {\bibinfo {title} {{Revisiting Supernova 1987A
  Limits on Axion-Like-Particles}},\ }\href
  {https://doi.org/10.48550/arXiv.1808.10136} {\bibfield  {journal} {\bibinfo
  {journal} {arXiv e-prints}\ ,\ \bibinfo {eid} {arXiv:1808.10136}} (\bibinfo
  {year} {2018})},\ \Eprint {https://arxiv.org/abs/1808.10136}
  {arXiv:1808.10136 [hep-ph]} \BibitemShut {NoStop}%
\bibitem [{\citenamefont {{Lucente}}\ \emph {et~al.}(2020)\citenamefont
  {{Lucente}}, \citenamefont {{Carenza}}, \citenamefont {{Fischer}},
  \citenamefont {{Giannotti}},\ and\ \citenamefont
  {{Mirizzi}}}]{2020JCAP...12..008L}%
  \BibitemOpen
  \bibfield  {author} {\bibinfo {author} {\bibfnamefont {G.}~\bibnamefont
  {{Lucente}}}, \bibinfo {author} {\bibfnamefont {P.}~\bibnamefont
  {{Carenza}}}, \bibinfo {author} {\bibfnamefont {T.}~\bibnamefont
  {{Fischer}}}, \bibinfo {author} {\bibfnamefont {M.}~\bibnamefont
  {{Giannotti}}},\ and\ \bibinfo {author} {\bibfnamefont {A.}~\bibnamefont
  {{Mirizzi}}},\ }\bibfield  {title} {\bibinfo {title} {{Heavy axion-like
  particles and core-collapse supernovae: constraints and impact on the
  explosion mechanism}},\ }\href
  {https://doi.org/10.1088/1475-7516/2020/12/008} {\bibfield  {journal}
  {\bibinfo  {journal} {Journal of Cosmology and Astroparticle Physics}\
  }\textbf {\bibinfo {volume} {2020}}\bibfield  {number} {\bibinfo  {number} {
  (12)},\ \bibinfo {eid} {008}},\ }\Eprint {https://arxiv.org/abs/2008.04918}
  {arXiv:2008.04918 [hep-ph]} \BibitemShut {NoStop}%
\bibitem [{\citenamefont {{Sung}}\ \emph {et~al.}(2019)\citenamefont {{Sung}},
  \citenamefont {{Tu}},\ and\ \citenamefont {{Wu}}}]{2019PhRvD..99l1305S}%
  \BibitemOpen
  \bibfield  {author} {\bibinfo {author} {\bibfnamefont {A.}~\bibnamefont
  {{Sung}}}, \bibinfo {author} {\bibfnamefont {H.}~\bibnamefont {{Tu}}},\ and\
  \bibinfo {author} {\bibfnamefont {M.-R.}\ \bibnamefont {{Wu}}},\ }\bibfield
  {title} {\bibinfo {title} {{New constraint from supernova explosions on light
  particles beyond the Standard Model}},\ }\href
  {https://doi.org/10.1103/PhysRevD.99.121305} {\bibfield  {journal} {\bibinfo
  {journal} {\prd}\ }\textbf {\bibinfo {volume} {99}},\ \bibinfo {eid} {121305}
  (\bibinfo {year} {2019})},\ \Eprint {https://arxiv.org/abs/1903.07923}
  {arXiv:1903.07923 [hep-ph]} \BibitemShut {NoStop}%
\bibitem [{\citenamefont {{Caputo}}\ \emph
  {et~al.}(2022{\natexlab{a}})\citenamefont {{Caputo}}, \citenamefont
  {{Janka}}, \citenamefont {{Raffelt}},\ and\ \citenamefont
  {{Vitagliano}}}]{2022PhRvL.128v1103C}%
  \BibitemOpen
  \bibfield  {author} {\bibinfo {author} {\bibfnamefont {A.}~\bibnamefont
  {{Caputo}}}, \bibinfo {author} {\bibfnamefont {H.-T.}\ \bibnamefont
  {{Janka}}}, \bibinfo {author} {\bibfnamefont {G.}~\bibnamefont {{Raffelt}}},\
  and\ \bibinfo {author} {\bibfnamefont {E.}~\bibnamefont {{Vitagliano}}},\
  }\bibfield  {title} {\bibinfo {title} {{Low-Energy Supernovae Severely
  Constrain Radiative Particle Decays}},\ }\href
  {https://doi.org/10.1103/PhysRevLett.128.221103} {\bibfield  {journal}
  {\bibinfo  {journal} {\prl}\ }\textbf {\bibinfo {volume} {128}},\ \bibinfo
  {eid} {221103} (\bibinfo {year} {2022}{\natexlab{a}})},\ \Eprint
  {https://arxiv.org/abs/2201.09890} {arXiv:2201.09890 [astro-ph.HE]}
  \BibitemShut {NoStop}%
\bibitem [{\citenamefont {{Hoof}}\ and\ \citenamefont
  {{Schulz}}(2023)}]{2023JCAP...03..054H}%
  \BibitemOpen
  \bibfield  {author} {\bibinfo {author} {\bibfnamefont {S.}~\bibnamefont
  {{Hoof}}}\ and\ \bibinfo {author} {\bibfnamefont {L.}~\bibnamefont
  {{Schulz}}},\ }\bibfield  {title} {\bibinfo {title} {{Updated constraints on
  axion-like particles from temporal information in supernova SN1987A gamma-ray
  data}},\ }\href {https://doi.org/10.1088/1475-7516/2023/03/054} {\bibfield
  {journal} {\bibinfo  {journal} {Journal of Cosmology and Astroparticle
  Physics}\ }\textbf {\bibinfo {volume} {2023}}\bibfield  {number} {\bibinfo
  {number} { (3)},\ \bibinfo {eid} {054}},\ }\Eprint
  {https://arxiv.org/abs/2212.09764} {arXiv:2212.09764 [hep-ph]} \BibitemShut
  {NoStop}%
\bibitem [{\citenamefont {{Ravensburg}}\ \emph {et~al.}(2024)\citenamefont
  {{Ravensburg}}, \citenamefont {{Carenza}}, \citenamefont {{Eckner}},\ and\
  \citenamefont {{Goobar}}}]{2024PhRvD.109b3018R}%
  \BibitemOpen
  \bibfield  {author} {\bibinfo {author} {\bibfnamefont {E.}~\bibnamefont
  {{Ravensburg}}}, \bibinfo {author} {\bibfnamefont {P.}~\bibnamefont
  {{Carenza}}}, \bibinfo {author} {\bibfnamefont {C.}~\bibnamefont
  {{Eckner}}},\ and\ \bibinfo {author} {\bibfnamefont {A.}~\bibnamefont
  {{Goobar}}},\ }\bibfield  {title} {\bibinfo {title} {{Constraining MeV-scale
  axionlike particles with Fermi-LAT observations of SN 2023ixf}},\ }\href
  {https://doi.org/10.1103/PhysRevD.109.023018} {\bibfield  {journal} {\bibinfo
   {journal} {\prd}\ }\textbf {\bibinfo {volume} {109}},\ \bibinfo {eid}
  {023018} (\bibinfo {year} {2024})},\ \Eprint
  {https://arxiv.org/abs/2306.16397} {arXiv:2306.16397 [astro-ph.HE]}
  \BibitemShut {NoStop}%
\bibitem [{\citenamefont {{Giannotti}}\ \emph {et~al.}(2011)\citenamefont
  {{Giannotti}}, \citenamefont {{Duffy}},\ and\ \citenamefont
  {{Nita}}}]{2011JCAP...01..015G}%
  \BibitemOpen
  \bibfield  {author} {\bibinfo {author} {\bibfnamefont {M.}~\bibnamefont
  {{Giannotti}}}, \bibinfo {author} {\bibfnamefont {L.~D.}\ \bibnamefont
  {{Duffy}}},\ and\ \bibinfo {author} {\bibfnamefont {R.}~\bibnamefont
  {{Nita}}},\ }\bibfield  {title} {\bibinfo {title} {{New constraints for heavy
  axion-like particles from supernovae}},\ }\href
  {https://doi.org/10.1088/1475-7516/2011/01/015} {\bibfield  {journal}
  {\bibinfo  {journal} {Journal of Cosmology and Astroparticle Physics}\
  }\textbf {\bibinfo {volume} {2011}}\bibfield  {number} {\bibinfo  {number} {
  (1)},\ \bibinfo {eid} {015}},\ }\Eprint {https://arxiv.org/abs/1009.5714}
  {arXiv:1009.5714 [astro-ph.HE]} \BibitemShut {NoStop}%
\bibitem [{\citenamefont {{Jaeckel}}\ \emph {et~al.}(2018)\citenamefont
  {{Jaeckel}}, \citenamefont {{Malta}},\ and\ \citenamefont
  {{Redondo}}}]{2018PhRvD..98e5032J}%
  \BibitemOpen
  \bibfield  {author} {\bibinfo {author} {\bibfnamefont {J.}~\bibnamefont
  {{Jaeckel}}}, \bibinfo {author} {\bibfnamefont {P.~C.}\ \bibnamefont
  {{Malta}}},\ and\ \bibinfo {author} {\bibfnamefont {J.}~\bibnamefont
  {{Redondo}}},\ }\bibfield  {title} {\bibinfo {title} {{Decay photons from the
  axionlike particles burst of type II supernovae}},\ }\href
  {https://doi.org/10.1103/PhysRevD.98.055032} {\bibfield  {journal} {\bibinfo
  {journal} {\prd}\ }\textbf {\bibinfo {volume} {98}},\ \bibinfo {eid} {055032}
  (\bibinfo {year} {2018})},\ \Eprint {https://arxiv.org/abs/1702.02964}
  {arXiv:1702.02964 [hep-ph]} \BibitemShut {NoStop}%
\bibitem [{\citenamefont {{Mori}}(2021)}]{2021PASJ...73.1382M}%
  \BibitemOpen
  \bibfield  {author} {\bibinfo {author} {\bibfnamefont {K.}~\bibnamefont
  {{Mori}}},\ }\bibfield  {title} {\bibinfo {title} {{Heavy axion-like
  particles and MeV decay photons from nearby type Ia supernovae}},\ }\href
  {https://doi.org/10.1093/pasj/psab082} {\bibfield  {journal} {\bibinfo
  {journal} {Publications of the Astronomical Society of Japan}\ }\textbf
  {\bibinfo {volume} {73}},\ \bibinfo {pages} {1382} (\bibinfo {year}
  {2021})},\ \Eprint {https://arxiv.org/abs/2107.09097} {arXiv:2107.09097
  [hep-ph]} \BibitemShut {NoStop}%
\bibitem [{\citenamefont {{Caputo}}\ \emph {et~al.}(2021)\citenamefont
  {{Caputo}}, \citenamefont {{Carenza}}, \citenamefont {{Lucente}},
  \citenamefont {{Vitagliano}}, \citenamefont {{Giannotti}}, \citenamefont
  {{Kotake}}, \citenamefont {{Kuroda}},\ and\ \citenamefont
  {{Mirizzi}}}]{2021PhRvL.127r1102C}%
  \BibitemOpen
  \bibfield  {author} {\bibinfo {author} {\bibfnamefont {A.}~\bibnamefont
  {{Caputo}}}, \bibinfo {author} {\bibfnamefont {P.}~\bibnamefont {{Carenza}}},
  \bibinfo {author} {\bibfnamefont {G.}~\bibnamefont {{Lucente}}}, \bibinfo
  {author} {\bibfnamefont {E.}~\bibnamefont {{Vitagliano}}}, \bibinfo {author}
  {\bibfnamefont {M.}~\bibnamefont {{Giannotti}}}, \bibinfo {author}
  {\bibfnamefont {K.}~\bibnamefont {{Kotake}}}, \bibinfo {author}
  {\bibfnamefont {T.}~\bibnamefont {{Kuroda}}},\ and\ \bibinfo {author}
  {\bibfnamefont {A.}~\bibnamefont {{Mirizzi}}},\ }\bibfield  {title} {\bibinfo
  {title} {{Axionlike Particles from Hypernovae}},\ }\href
  {https://doi.org/10.1103/PhysRevLett.127.181102} {\bibfield  {journal}
  {\bibinfo  {journal} {\prl}\ }\textbf {\bibinfo {volume} {127}},\ \bibinfo
  {eid} {181102} (\bibinfo {year} {2021})},\ \Eprint
  {https://arxiv.org/abs/2104.05727} {arXiv:2104.05727 [hep-ph]} \BibitemShut
  {NoStop}%
\bibitem [{\citenamefont {{Caputo}}\ \emph
  {et~al.}(2022{\natexlab{b}})\citenamefont {{Caputo}}, \citenamefont
  {{Raffelt}},\ and\ \citenamefont {{Vitagliano}}}]{2022PhRvD.105c5022C}%
  \BibitemOpen
  \bibfield  {author} {\bibinfo {author} {\bibfnamefont {A.}~\bibnamefont
  {{Caputo}}}, \bibinfo {author} {\bibfnamefont {G.}~\bibnamefont
  {{Raffelt}}},\ and\ \bibinfo {author} {\bibfnamefont {E.}~\bibnamefont
  {{Vitagliano}}},\ }\bibfield  {title} {\bibinfo {title} {{Muonic boson
  limits: Supernova redux}},\ }\href
  {https://doi.org/10.1103/PhysRevD.105.035022} {\bibfield  {journal} {\bibinfo
   {journal} {\prd}\ }\textbf {\bibinfo {volume} {105}},\ \bibinfo {eid}
  {035022} (\bibinfo {year} {2022}{\natexlab{b}})},\ \Eprint
  {https://arxiv.org/abs/2109.03244} {arXiv:2109.03244 [hep-ph]} \BibitemShut
  {NoStop}%
\bibitem [{\citenamefont {{M{\"u}ller}}\ \emph {et~al.}(2023)\citenamefont
  {{M{\"u}ller}}, \citenamefont {{Calore}}, \citenamefont {{Carenza}},
  \citenamefont {{Eckner}},\ and\ \citenamefont
  {{Marsh}}}]{2023JCAP...07..056M}%
  \BibitemOpen
  \bibfield  {author} {\bibinfo {author} {\bibfnamefont {E.}~\bibnamefont
  {{M{\"u}ller}}}, \bibinfo {author} {\bibfnamefont {F.}~\bibnamefont
  {{Calore}}}, \bibinfo {author} {\bibfnamefont {P.}~\bibnamefont {{Carenza}}},
  \bibinfo {author} {\bibfnamefont {C.}~\bibnamefont {{Eckner}}},\ and\
  \bibinfo {author} {\bibfnamefont {M.~C.~D.}\ \bibnamefont {{Marsh}}},\
  }\bibfield  {title} {\bibinfo {title} {{Investigating the gamma-ray burst
  from decaying MeV-scale axion-like particles produced in supernova
  explosions}},\ }\href {https://doi.org/10.1088/1475-7516/2023/07/056}
  {\bibfield  {journal} {\bibinfo  {journal} {Journal of Cosmology and
  Astroparticle Physics}\ }\textbf {\bibinfo {volume} {2023}}\bibfield
  {number} {\bibinfo  {number} { (7)},\ \bibinfo {eid} {056}},\ }\Eprint
  {https://arxiv.org/abs/2304.01060} {arXiv:2304.01060 [astro-ph.HE]}
  \BibitemShut {NoStop}%
\bibitem [{\citenamefont {{Carenza}}(2023)}]{2023EPJP..138..836C}%
  \BibitemOpen
  \bibfield  {author} {\bibinfo {author} {\bibfnamefont {P.}~\bibnamefont
  {{Carenza}}},\ }\bibfield  {title} {\bibinfo {title} {{Axion emission from
  supernovae: a cheatsheet}},\ }\href
  {https://doi.org/10.1140/epjp/s13360-023-04484-2} {\bibfield  {journal}
  {\bibinfo  {journal} {European Physical Journal Plus}\ }\textbf {\bibinfo
  {volume} {138}},\ \bibinfo {eid} {836} (\bibinfo {year} {2023})},\ \Eprint
  {https://arxiv.org/abs/2309.14798} {arXiv:2309.14798 [hep-ph]} \BibitemShut
  {NoStop}%
\bibitem [{\citenamefont {{Carenza}}\ \emph {et~al.}(2019)\citenamefont
  {{Carenza}}, \citenamefont {{Fischer}}, \citenamefont {{Giannotti}},
  \citenamefont {{Guo}}, \citenamefont {{Mart{\'\i}nez-Pinedo}},\ and\
  \citenamefont {{Mirizzi}}}]{2019JCAP...10..016C}%
  \BibitemOpen
  \bibfield  {author} {\bibinfo {author} {\bibfnamefont {P.}~\bibnamefont
  {{Carenza}}}, \bibinfo {author} {\bibfnamefont {T.}~\bibnamefont
  {{Fischer}}}, \bibinfo {author} {\bibfnamefont {M.}~\bibnamefont
  {{Giannotti}}}, \bibinfo {author} {\bibfnamefont {G.}~\bibnamefont {{Guo}}},
  \bibinfo {author} {\bibfnamefont {G.}~\bibnamefont
  {{Mart{\'\i}nez-Pinedo}}},\ and\ \bibinfo {author} {\bibfnamefont
  {A.}~\bibnamefont {{Mirizzi}}},\ }\bibfield  {title} {\bibinfo {title}
  {{Improved axion emissivity from a supernova via nucleon-nucleon
  bremsstrahlung}},\ }\href {https://doi.org/10.1088/1475-7516/2019/10/016}
  {\bibfield  {journal} {\bibinfo  {journal} {Journal of Cosmology and
  Astroparticle Physics}\ }\textbf {\bibinfo {volume} {2019}}\bibfield
  {number} {\bibinfo  {number} { (10)},\ \bibinfo {eid} {016}},\ }\Eprint
  {https://arxiv.org/abs/1906.11844} {arXiv:1906.11844 [hep-ph]} \BibitemShut
  {NoStop}%
\bibitem [{\citenamefont {{Fischer}}\ \emph {et~al.}(2016)\citenamefont
  {{Fischer}}, \citenamefont {{Chakraborty}}, \citenamefont {{Giannotti}},
  \citenamefont {{Mirizzi}}, \citenamefont {{Payez}},\ and\ \citenamefont
  {{Ringwald}}}]{2016PhRvD..94h5012F}%
  \BibitemOpen
  \bibfield  {author} {\bibinfo {author} {\bibfnamefont {T.}~\bibnamefont
  {{Fischer}}}, \bibinfo {author} {\bibfnamefont {S.}~\bibnamefont
  {{Chakraborty}}}, \bibinfo {author} {\bibfnamefont {M.}~\bibnamefont
  {{Giannotti}}}, \bibinfo {author} {\bibfnamefont {A.}~\bibnamefont
  {{Mirizzi}}}, \bibinfo {author} {\bibfnamefont {A.}~\bibnamefont {{Payez}}},\
  and\ \bibinfo {author} {\bibfnamefont {A.}~\bibnamefont {{Ringwald}}},\
  }\bibfield  {title} {\bibinfo {title} {{Probing axions with the neutrino
  signal from the next Galactic supernova}},\ }\href
  {https://doi.org/10.1103/PhysRevD.94.085012} {\bibfield  {journal} {\bibinfo
  {journal} {\prd}\ }\textbf {\bibinfo {volume} {94}},\ \bibinfo {eid} {085012}
  (\bibinfo {year} {2016})},\ \Eprint {https://arxiv.org/abs/1605.08780}
  {arXiv:1605.08780 [astro-ph.HE]} \BibitemShut {NoStop}%
\bibitem [{\citenamefont {{Fischer}}\ \emph {et~al.}(2021)\citenamefont
  {{Fischer}}, \citenamefont {{Carenza}}, \citenamefont {{Fore}}, \citenamefont
  {{Giannotti}}, \citenamefont {{Mirizzi}},\ and\ \citenamefont
  {{Reddy}}}]{2021PhRvD.104j3012F}%
  \BibitemOpen
  \bibfield  {author} {\bibinfo {author} {\bibfnamefont {T.}~\bibnamefont
  {{Fischer}}}, \bibinfo {author} {\bibfnamefont {P.}~\bibnamefont
  {{Carenza}}}, \bibinfo {author} {\bibfnamefont {B.}~\bibnamefont {{Fore}}},
  \bibinfo {author} {\bibfnamefont {M.}~\bibnamefont {{Giannotti}}}, \bibinfo
  {author} {\bibfnamefont {A.}~\bibnamefont {{Mirizzi}}},\ and\ \bibinfo
  {author} {\bibfnamefont {S.}~\bibnamefont {{Reddy}}},\ }\bibfield  {title}
  {\bibinfo {title} {{Observable signatures of enhanced axion emission from
  protoneutron stars}},\ }\href {https://doi.org/10.1103/PhysRevD.104.103012}
  {\bibfield  {journal} {\bibinfo  {journal} {\prd}\ }\textbf {\bibinfo
  {volume} {104}},\ \bibinfo {eid} {103012} (\bibinfo {year} {2021})},\ \Eprint
  {https://arxiv.org/abs/2108.13726} {arXiv:2108.13726 [hep-ph]} \BibitemShut
  {NoStop}%
\bibitem [{\citenamefont {{Mori}}\ \emph
  {et~al.}(2022{\natexlab{a}})\citenamefont {{Mori}}, \citenamefont
  {{Takiwaki}}, \citenamefont {{Kotake}},\ and\ \citenamefont
  {{Horiuchi}}}]{2022PhRvD.105f3009M}%
  \BibitemOpen
  \bibfield  {author} {\bibinfo {author} {\bibfnamefont {K.}~\bibnamefont
  {{Mori}}}, \bibinfo {author} {\bibfnamefont {T.}~\bibnamefont {{Takiwaki}}},
  \bibinfo {author} {\bibfnamefont {K.}~\bibnamefont {{Kotake}}},\ and\
  \bibinfo {author} {\bibfnamefont {S.}~\bibnamefont {{Horiuchi}}},\ }\bibfield
   {title} {\bibinfo {title} {{Shock revival in core-collapse supernovae
  assisted by heavy axionlike particles}},\ }\href
  {https://doi.org/10.1103/PhysRevD.105.063009} {\bibfield  {journal} {\bibinfo
   {journal} {\prd}\ }\textbf {\bibinfo {volume} {105}},\ \bibinfo {eid}
  {063009} (\bibinfo {year} {2022}{\natexlab{a}})},\ \Eprint
  {https://arxiv.org/abs/2112.03613} {arXiv:2112.03613 [astro-ph.HE]}
  \BibitemShut {NoStop}%
\bibitem [{\citenamefont {{Betranhandy}}\ and\ \citenamefont
  {{O'Connor}}(2022)}]{2022PhRvD.106f3019B}%
  \BibitemOpen
  \bibfield  {author} {\bibinfo {author} {\bibfnamefont {A.}~\bibnamefont
  {{Betranhandy}}}\ and\ \bibinfo {author} {\bibfnamefont {E.}~\bibnamefont
  {{O'Connor}}},\ }\bibfield  {title} {\bibinfo {title} {{Neutrino driven
  explosions aided by axion cooling in multidimensional simulations of
  core-collapse supernovae}},\ }\href
  {https://doi.org/10.1103/PhysRevD.106.063019} {\bibfield  {journal} {\bibinfo
   {journal} {\prd}\ }\textbf {\bibinfo {volume} {106}},\ \bibinfo {eid}
  {063019} (\bibinfo {year} {2022})},\ \Eprint
  {https://arxiv.org/abs/2204.00503} {arXiv:2204.00503 [astro-ph.HE]}
  \BibitemShut {NoStop}%
\bibitem [{\citenamefont {{Mori}}\ \emph {et~al.}(2024)\citenamefont {{Mori}},
  \citenamefont {{Takiwaki}}, \citenamefont {{Kotake}},\ and\ \citenamefont
  {{Horiuchi}}}]{2024PhRvD.110b3031M}%
  \BibitemOpen
  \bibfield  {author} {\bibinfo {author} {\bibfnamefont {K.}~\bibnamefont
  {{Mori}}}, \bibinfo {author} {\bibfnamefont {T.}~\bibnamefont {{Takiwaki}}},
  \bibinfo {author} {\bibfnamefont {K.}~\bibnamefont {{Kotake}}},\ and\
  \bibinfo {author} {\bibfnamefont {S.}~\bibnamefont {{Horiuchi}}},\ }\bibfield
   {title} {\bibinfo {title} {{Two-dimensional models of core-collapse
  supernova explosions assisted by heavy sterile neutrinos}},\ }\href
  {https://doi.org/10.1103/PhysRevD.110.023031} {\bibfield  {journal} {\bibinfo
   {journal} {\prd}\ }\textbf {\bibinfo {volume} {110}},\ \bibinfo {eid}
  {023031} (\bibinfo {year} {2024})},\ \Eprint
  {https://arxiv.org/abs/2402.14333} {arXiv:2402.14333 [astro-ph.HE]}
  \BibitemShut {NoStop}%
\bibitem [{\citenamefont {{Takata}}\ \emph {et~al.}(2025)\citenamefont
  {{Takata}}, \citenamefont {{Mori}}, \citenamefont {{Nakamura}},\ and\
  \citenamefont {{Kotake}}}]{2025arXiv250309005T}%
  \BibitemOpen
  \bibfield  {author} {\bibinfo {author} {\bibfnamefont {T.}~\bibnamefont
  {{Takata}}}, \bibinfo {author} {\bibfnamefont {K.}~\bibnamefont {{Mori}}},
  \bibinfo {author} {\bibfnamefont {K.}~\bibnamefont {{Nakamura}}},\ and\
  \bibinfo {author} {\bibfnamefont {K.}~\bibnamefont {{Kotake}}},\ }\bibfield
  {title} {\bibinfo {title} {{Progenitor Dependence of Neutrino-driven
  Supernova Explosions with the Aid of Heavy Axion-like Particles}},\ }\href
  {https://doi.org/10.48550/arXiv.2503.09005} {\bibfield  {journal} {\bibinfo
  {journal} {arXiv e-prints}\ ,\ \bibinfo {eid} {arXiv:2503.09005}} (\bibinfo
  {year} {2025})},\ \Eprint {https://arxiv.org/abs/2503.09005}
  {arXiv:2503.09005 [astro-ph.HE]} \BibitemShut {NoStop}%
\bibitem [{\citenamefont {{Burrows}}\ \emph {et~al.}(1989)\citenamefont
  {{Burrows}}, \citenamefont {{Turner}},\ and\ \citenamefont
  {{Brinkmann}}}]{1989PhRvD..39.1020B}%
  \BibitemOpen
  \bibfield  {author} {\bibinfo {author} {\bibfnamefont {A.}~\bibnamefont
  {{Burrows}}}, \bibinfo {author} {\bibfnamefont {M.~S.}\ \bibnamefont
  {{Turner}}},\ and\ \bibinfo {author} {\bibfnamefont {R.~P.}\ \bibnamefont
  {{Brinkmann}}},\ }\bibfield  {title} {\bibinfo {title} {{Axions and SN
  1987A}},\ }\href {https://doi.org/10.1103/PhysxvD.39.1020} {\bibfield
  {journal} {\bibinfo  {journal} {\prd}\ }\textbf {\bibinfo {volume} {39}},\
  \bibinfo {pages} {1020} (\bibinfo {year} {1989})}\BibitemShut {NoStop}%
\bibitem [{\citenamefont {{Keil}}\ \emph {et~al.}(1997)\citenamefont {{Keil}},
  \citenamefont {{Janka}}, \citenamefont {{Schramm}}, \citenamefont {{Sigl}},
  \citenamefont {{Turner}},\ and\ \citenamefont
  {{Ellis}}}]{1997PhRvD..56.2419K}%
  \BibitemOpen
  \bibfield  {author} {\bibinfo {author} {\bibfnamefont {W.}~\bibnamefont
  {{Keil}}}, \bibinfo {author} {\bibfnamefont {H.-T.}\ \bibnamefont {{Janka}}},
  \bibinfo {author} {\bibfnamefont {D.~N.}\ \bibnamefont {{Schramm}}}, \bibinfo
  {author} {\bibfnamefont {G.}~\bibnamefont {{Sigl}}}, \bibinfo {author}
  {\bibfnamefont {M.~S.}\ \bibnamefont {{Turner}}},\ and\ \bibinfo {author}
  {\bibfnamefont {J.}~\bibnamefont {{Ellis}}},\ }\bibfield  {title} {\bibinfo
  {title} {{Fresh look at axions and SN 1987A}},\ }\href
  {https://doi.org/10.1103/PhysRevD.56.2419} {\bibfield  {journal} {\bibinfo
  {journal} {\prd}\ }\textbf {\bibinfo {volume} {56}},\ \bibinfo {pages} {2419}
  (\bibinfo {year} {1997})},\ \Eprint {https://arxiv.org/abs/astro-ph/9612222}
  {arXiv:astro-ph/9612222 [astro-ph]} \BibitemShut {NoStop}%
\bibitem [{\citenamefont {{Vinyoles}}\ \emph {et~al.}(2015)\citenamefont
  {{Vinyoles}}, \citenamefont {{Serenelli}}, \citenamefont {{Villante}},
  \citenamefont {{Basu}}, \citenamefont {{Redondo}},\ and\ \citenamefont
  {{Isern}}}]{2015JCAP...10..015V}%
  \BibitemOpen
  \bibfield  {author} {\bibinfo {author} {\bibfnamefont {N.}~\bibnamefont
  {{Vinyoles}}}, \bibinfo {author} {\bibfnamefont {A.}~\bibnamefont
  {{Serenelli}}}, \bibinfo {author} {\bibfnamefont {F.~L.}\ \bibnamefont
  {{Villante}}}, \bibinfo {author} {\bibfnamefont {S.}~\bibnamefont {{Basu}}},
  \bibinfo {author} {\bibfnamefont {J.}~\bibnamefont {{Redondo}}},\ and\
  \bibinfo {author} {\bibfnamefont {J.}~\bibnamefont {{Isern}}},\ }\bibfield
  {title} {\bibinfo {title} {{New axion and hidden photon constraints from a
  solar data global fit}},\ }\href
  {https://doi.org/10.1088/1475-7516/2015/10/015} {\bibfield  {journal}
  {\bibinfo  {journal} {Journal of Cosmology and Astroparticle Physics}\
  }\textbf {\bibinfo {volume} {2015}}\bibfield  {number} {\bibinfo  {number} {
  (10)},\ \bibinfo {pages} {015}},\ }\Eprint {https://arxiv.org/abs/1501.01639}
  {arXiv:1501.01639 [astro-ph.SR]} \BibitemShut {NoStop}%
\bibitem [{\citenamefont {{Ayala}}\ \emph {et~al.}(2014)\citenamefont
  {{Ayala}}, \citenamefont {{Dom{\'\i}nguez}}, \citenamefont {{Giannotti}},
  \citenamefont {{Mirizzi}},\ and\ \citenamefont
  {{Straniero}}}]{2014PhRvL.113s1302A}%
  \BibitemOpen
  \bibfield  {author} {\bibinfo {author} {\bibfnamefont {A.}~\bibnamefont
  {{Ayala}}}, \bibinfo {author} {\bibfnamefont {I.}~\bibnamefont
  {{Dom{\'\i}nguez}}}, \bibinfo {author} {\bibfnamefont {M.}~\bibnamefont
  {{Giannotti}}}, \bibinfo {author} {\bibfnamefont {A.}~\bibnamefont
  {{Mirizzi}}},\ and\ \bibinfo {author} {\bibfnamefont {O.}~\bibnamefont
  {{Straniero}}},\ }\bibfield  {title} {\bibinfo {title} {{Revisiting the Bound
  on Axion-Photon Coupling from Globular Clusters}},\ }\href
  {https://doi.org/10.1103/PhysRevLett.113.191302} {\bibfield  {journal}
  {\bibinfo  {journal} {\prl}\ }\textbf {\bibinfo {volume} {113}},\ \bibinfo
  {eid} {191302} (\bibinfo {year} {2014})},\ \Eprint
  {https://arxiv.org/abs/1406.6053} {arXiv:1406.6053 [astro-ph.SR]}
  \BibitemShut {NoStop}%
\bibitem [{\citenamefont {Dolan}\ \emph {et~al.}(2022)\citenamefont {Dolan},
  \citenamefont {Hiskens},\ and\ \citenamefont {Volkas}}]{Dolan:2022kul}%
  \BibitemOpen
  \bibfield  {author} {\bibinfo {author} {\bibfnamefont {M.~J.}\ \bibnamefont
  {Dolan}}, \bibinfo {author} {\bibfnamefont {F.~J.}\ \bibnamefont {Hiskens}},\
  and\ \bibinfo {author} {\bibfnamefont {R.~R.}\ \bibnamefont {Volkas}},\
  }\bibfield  {title} {\bibinfo {title} {{Advancing globular cluster
  constraints on the axion-photon coupling}},\ }\href
  {https://doi.org/10.1088/1475-7516/2022/10/096} {\bibfield  {journal}
  {\bibinfo  {journal} {JCAP}\ }\textbf {\bibinfo {volume} {10}},\ \bibinfo
  {pages} {096}},\ \Eprint {https://arxiv.org/abs/2207.03102} {arXiv:2207.03102
  [hep-ph]} \BibitemShut {NoStop}%
\bibitem [{\citenamefont {{O'Connor}}\ and\ \citenamefont
  {{Ott}}(2010)}]{2010CQGra..27k4103O}%
  \BibitemOpen
  \bibfield  {author} {\bibinfo {author} {\bibfnamefont {E.}~\bibnamefont
  {{O'Connor}}}\ and\ \bibinfo {author} {\bibfnamefont {C.~D.}\ \bibnamefont
  {{Ott}}},\ }\bibfield  {title} {\bibinfo {title} {{A new open-source code for
  spherically symmetric stellar collapse to neutron stars and black holes}},\
  }\href {https://doi.org/10.1088/0264-9381/27/11/114103} {\bibfield  {journal}
  {\bibinfo  {journal} {Classical and Quantum Gravity}\ }\textbf {\bibinfo
  {volume} {27}},\ \bibinfo {eid} {114103} (\bibinfo {year} {2010})},\ \Eprint
  {https://arxiv.org/abs/0912.2393} {arXiv:0912.2393 [astro-ph.HE]}
  \BibitemShut {NoStop}%
\bibitem [{\citenamefont {{O'Connor}}(2015)}]{2015ApJS..219...24O}%
  \BibitemOpen
  \bibfield  {author} {\bibinfo {author} {\bibfnamefont {E.}~\bibnamefont
  {{O'Connor}}},\ }\bibfield  {title} {\bibinfo {title} {{An Open-source
  Neutrino Radiation Hydrodynamics Code for Core-collapse Supernovae}},\ }\href
  {https://doi.org/10.1088/0067-0049/219/2/24} {\bibfield  {journal} {\bibinfo
  {journal} {The Astrophysical Journal Supplement Series}\ }\textbf {\bibinfo
  {volume} {219}},\ \bibinfo {eid} {24} (\bibinfo {year} {2015})},\ \Eprint
  {https://arxiv.org/abs/1411.7058} {arXiv:1411.7058 [astro-ph.HE]}
  \BibitemShut {NoStop}%
\bibitem [{\citenamefont {{Shibata}}\ \emph {et~al.}(2011)\citenamefont
  {{Shibata}}, \citenamefont {{Kiuchi}}, \citenamefont {{Sekiguchi}},\ and\
  \citenamefont {{Suwa}}}]{2011PThPh.125.1255S}%
  \BibitemOpen
  \bibfield  {author} {\bibinfo {author} {\bibfnamefont {M.}~\bibnamefont
  {{Shibata}}}, \bibinfo {author} {\bibfnamefont {K.}~\bibnamefont {{Kiuchi}}},
  \bibinfo {author} {\bibfnamefont {Y.}~\bibnamefont {{Sekiguchi}}},\ and\
  \bibinfo {author} {\bibfnamefont {Y.}~\bibnamefont {{Suwa}}},\ }\bibfield
  {title} {\bibinfo {title} {{Truncated Moment Formalism for Radiation
  Hydrodynamics in Numerical Relativity}},\ }\href
  {https://doi.org/10.1143/PTP.125.1255} {\bibfield  {journal} {\bibinfo
  {journal} {Progress of Theoretical Physics}\ }\textbf {\bibinfo {volume}
  {125}},\ \bibinfo {pages} {1255} (\bibinfo {year} {2011})},\ \Eprint
  {https://arxiv.org/abs/1104.3937} {arXiv:1104.3937 [astro-ph.HE]}
  \BibitemShut {NoStop}%
\bibitem [{\citenamefont {{Cardall}}\ \emph {et~al.}(2013)\citenamefont
  {{Cardall}}, \citenamefont {{Endeve}},\ and\ \citenamefont
  {{Mezzacappa}}}]{2013PhRvD..87j3004C}%
  \BibitemOpen
  \bibfield  {author} {\bibinfo {author} {\bibfnamefont {C.~Y.}\ \bibnamefont
  {{Cardall}}}, \bibinfo {author} {\bibfnamefont {E.}~\bibnamefont
  {{Endeve}}},\ and\ \bibinfo {author} {\bibfnamefont {A.}~\bibnamefont
  {{Mezzacappa}}},\ }\bibfield  {title} {\bibinfo {title} {{Conservative 3+1
  general relativistic variable Eddington tensor radiation transport
  equations}},\ }\href {https://doi.org/10.1103/PhysRevD.87.103004} {\bibfield
  {journal} {\bibinfo  {journal} {\prd}\ }\textbf {\bibinfo {volume} {87}},\
  \bibinfo {eid} {103004} (\bibinfo {year} {2013})},\ \Eprint
  {https://arxiv.org/abs/1209.2151} {arXiv:1209.2151 [astro-ph.HE]}
  \BibitemShut {NoStop}%
\bibitem [{\citenamefont {{Mori}}\ \emph {et~al.}(2021)\citenamefont {{Mori}},
  \citenamefont {{Suwa}}, \citenamefont {{Nakazato}}, \citenamefont
  {{Sumiyoshi}}, \citenamefont {{Harada}}, \citenamefont {{Harada}},
  \citenamefont {{Koshio}},\ and\ \citenamefont
  {{Wendell}}}]{2021PTEP.2021b3E01M}%
  \BibitemOpen
  \bibfield  {author} {\bibinfo {author} {\bibfnamefont {M.}~\bibnamefont
  {{Mori}}}, \bibinfo {author} {\bibfnamefont {Y.}~\bibnamefont {{Suwa}}},
  \bibinfo {author} {\bibfnamefont {K.}~\bibnamefont {{Nakazato}}}, \bibinfo
  {author} {\bibfnamefont {K.}~\bibnamefont {{Sumiyoshi}}}, \bibinfo {author}
  {\bibfnamefont {M.}~\bibnamefont {{Harada}}}, \bibinfo {author}
  {\bibfnamefont {A.}~\bibnamefont {{Harada}}}, \bibinfo {author}
  {\bibfnamefont {Y.}~\bibnamefont {{Koshio}}},\ and\ \bibinfo {author}
  {\bibfnamefont {R.~A.}\ \bibnamefont {{Wendell}}},\ }\bibfield  {title}
  {\bibinfo {title} {{Developing an end-to-end simulation framework of
  supernova neutrino detection}},\ }\href
  {https://doi.org/10.1093/ptep/ptaa185} {\bibfield  {journal} {\bibinfo
  {journal} {Progress of Theoretical and Experimental Physics}\ }\textbf
  {\bibinfo {volume} {2021}},\ \bibinfo {eid} {023E01} (\bibinfo {year}
  {2021})},\ \Eprint {https://arxiv.org/abs/2010.16254} {arXiv:2010.16254
  [astro-ph.HE]} \BibitemShut {NoStop}%
\bibitem [{\citenamefont {{Melson}}\ \emph {et~al.}(2015)\citenamefont
  {{Melson}}, \citenamefont {{Janka}},\ and\ \citenamefont
  {{Marek}}}]{2015ApJ...801L..24M}%
  \BibitemOpen
  \bibfield  {author} {\bibinfo {author} {\bibfnamefont {T.}~\bibnamefont
  {{Melson}}}, \bibinfo {author} {\bibfnamefont {H.-T.}\ \bibnamefont
  {{Janka}}},\ and\ \bibinfo {author} {\bibfnamefont {A.}~\bibnamefont
  {{Marek}}},\ }\bibfield  {title} {\bibinfo {title} {{Neutrino-driven
  Supernova of a Low-mass Iron-core Progenitor Boosted by Three-dimensional
  Turbulent Convection}},\ }\href {https://doi.org/10.1088/2041-8205/801/2/L24}
  {\bibfield  {journal} {\bibinfo  {journal} {The Astrophysical Journal}\
  }\textbf {\bibinfo {volume} {801}},\ \bibinfo {eid} {L24} (\bibinfo {year}
  {2015})},\ \Eprint {https://arxiv.org/abs/1501.01961} {arXiv:1501.01961
  [astro-ph.SR]} \BibitemShut {NoStop}%
\bibitem [{\citenamefont {{Radice}}\ \emph {et~al.}(2017)\citenamefont
  {{Radice}}, \citenamefont {{Burrows}}, \citenamefont {{Vartanyan}},
  \citenamefont {{Skinner}},\ and\ \citenamefont
  {{Dolence}}}]{2017ApJ...850...43R}%
  \BibitemOpen
  \bibfield  {author} {\bibinfo {author} {\bibfnamefont {D.}~\bibnamefont
  {{Radice}}}, \bibinfo {author} {\bibfnamefont {A.}~\bibnamefont {{Burrows}}},
  \bibinfo {author} {\bibfnamefont {D.}~\bibnamefont {{Vartanyan}}}, \bibinfo
  {author} {\bibfnamefont {M.~A.}\ \bibnamefont {{Skinner}}},\ and\ \bibinfo
  {author} {\bibfnamefont {J.~C.}\ \bibnamefont {{Dolence}}},\ }\bibfield
  {title} {\bibinfo {title} {{Electron-capture and Low-mass Iron-core-collapse
  Supernovae: New Neutrino-radiation-hydrodynamics Simulations}},\ }\href
  {https://doi.org/10.3847/1538-4357/aa92c5} {\bibfield  {journal} {\bibinfo
  {journal} {\apj}\ }\textbf {\bibinfo {volume} {850}},\ \bibinfo {eid} {43}
  (\bibinfo {year} {2017})},\ \Eprint {https://arxiv.org/abs/1702.03927}
  {arXiv:1702.03927 [astro-ph.HE]} \BibitemShut {NoStop}%
\bibitem [{\citenamefont {{Typel}}(2005)}]{2005PhRvC..71f4301T}%
  \BibitemOpen
  \bibfield  {author} {\bibinfo {author} {\bibfnamefont {S.}~\bibnamefont
  {{Typel}}},\ }\bibfield  {title} {\bibinfo {title} {{Relativistic model for
  nuclear matter and atomic nuclei with momentum-dependent self-energies}},\
  }\href {https://doi.org/10.1103/PhysRevC.71.064301} {\bibfield  {journal}
  {\bibinfo  {journal} {\prc}\ }\textbf {\bibinfo {volume} {71}},\ \bibinfo
  {eid} {064301} (\bibinfo {year} {2005})},\ \Eprint
  {https://arxiv.org/abs/nucl-th/0501056} {arXiv:nucl-th/0501056 [nucl-th]}
  \BibitemShut {NoStop}%
\bibitem [{\citenamefont {{Hempel}}\ and\ \citenamefont
  {{Schaffner-Bielich}}(2010)}]{2010NuPhA.837..210H}%
  \BibitemOpen
  \bibfield  {author} {\bibinfo {author} {\bibfnamefont {M.}~\bibnamefont
  {{Hempel}}}\ and\ \bibinfo {author} {\bibfnamefont {J.}~\bibnamefont
  {{Schaffner-Bielich}}},\ }\bibfield  {title} {\bibinfo {title} {{A
  statistical model for a complete supernova equation of state}},\ }\href
  {https://doi.org/10.1016/j.nuclphysa.2010.02.010} {\bibfield  {journal}
  {\bibinfo  {journal} {Nuclear Physics A}\ }\textbf {\bibinfo {volume}
  {837}},\ \bibinfo {pages} {210} (\bibinfo {year} {2010})},\ \Eprint
  {https://arxiv.org/abs/0911.4073} {arXiv:0911.4073 [nucl-th]} \BibitemShut
  {NoStop}%
\bibitem [{\citenamefont {{Typel}}\ \emph {et~al.}(2010)\citenamefont
  {{Typel}}, \citenamefont {{R{\"o}pke}}, \citenamefont {{Kl{\"a}hn}},
  \citenamefont {{Blaschke}},\ and\ \citenamefont
  {{Wolter}}}]{2010PhRvC..81a5803T}%
  \BibitemOpen
  \bibfield  {author} {\bibinfo {author} {\bibfnamefont {S.}~\bibnamefont
  {{Typel}}}, \bibinfo {author} {\bibfnamefont {G.}~\bibnamefont
  {{R{\"o}pke}}}, \bibinfo {author} {\bibfnamefont {T.}~\bibnamefont
  {{Kl{\"a}hn}}}, \bibinfo {author} {\bibfnamefont {D.}~\bibnamefont
  {{Blaschke}}},\ and\ \bibinfo {author} {\bibfnamefont {H.~H.}\ \bibnamefont
  {{Wolter}}},\ }\bibfield  {title} {\bibinfo {title} {{Composition and
  thermodynamics of nuclear matter with light clusters}},\ }\href
  {https://doi.org/10.1103/PhysRevC.81.015803} {\bibfield  {journal} {\bibinfo
  {journal} {\prc}\ }\textbf {\bibinfo {volume} {81}},\ \bibinfo {eid} {015803}
  (\bibinfo {year} {2010})},\ \Eprint {https://arxiv.org/abs/0908.2344}
  {arXiv:0908.2344 [nucl-th]} \BibitemShut {NoStop}%
\bibitem [{\citenamefont {Fukuda}\ \emph {et~al.}(2003)\citenamefont {Fukuda}
  \emph {et~al.}}]{Super-Kamiokande:2002weg}%
  \BibitemOpen
  \bibfield  {author} {\bibinfo {author} {\bibfnamefont {Y.}~\bibnamefont
  {Fukuda}} \emph {et~al.} (\bibinfo {collaboration} {Super-Kamiokande}),\
  }\bibfield  {title} {\bibinfo {title} {{The Super-Kamiokande detector}},\
  }\href {https://doi.org/10.1016/S0168-9002(03)00425-X} {\bibfield  {journal}
  {\bibinfo  {journal} {Nucl. Instrum. Meth. A}\ }\textbf {\bibinfo {volume}
  {501}},\ \bibinfo {pages} {418} (\bibinfo {year} {2003})}\BibitemShut
  {NoStop}%
\bibitem [{\citenamefont {{Abe}}\ \emph {et~al.}(2016)\citenamefont {{Abe}},
  \citenamefont {{Haga}}, \citenamefont {{Hayato}}, \citenamefont {{Ikeda}},
  \citenamefont {{Iyogi}}, \citenamefont {{Kameda}}, \citenamefont
  {{Kishimoto}}, \citenamefont {{Miura}}, \citenamefont {{Moriyama}},
  \citenamefont {{Nakahata}}, \citenamefont {{Nakano}}, \citenamefont
  {{Nakayama}}, \citenamefont {{Sekiya}}, \citenamefont {{Shiozawa}},
  \citenamefont {{Suzuki}}, \citenamefont {{Takeda}}, \citenamefont {{Tanaka}},
  \citenamefont {{Tomura}}, \citenamefont {{Ueno}}, \citenamefont {{Wendell}},
  \citenamefont {{Yokozawa}}, \citenamefont {{Irvine}}, \citenamefont
  {{Kajita}}, \citenamefont {{Kametani}}, \citenamefont {{Kaneyuki}},
  \citenamefont {{Lee}}, \citenamefont {{McLachlan}}, \citenamefont
  {{Nishimura}}, \citenamefont {{Richard}}, \citenamefont {{Okumura}},
  \citenamefont {{Labarga}}, \citenamefont {{Fernandez}}, \citenamefont
  {{Berkman}}, \citenamefont {{Tanaka}}, \citenamefont {{Tobayama}},
  \citenamefont {{Gustafson}}, \citenamefont {{Kearns}}, \citenamefont
  {{Raaf}}, \citenamefont {{Stone}}, \citenamefont {{Sulak}}, \citenamefont
  {{Goldhaber}}, \citenamefont {{Carminati}}, \citenamefont {{Kropp}},
  \citenamefont {{Mine}}, \citenamefont {{Weatherly}}, \citenamefont
  {{Renshaw}}, \citenamefont {{Smy}}, \citenamefont {{Sobel}}, \citenamefont
  {{Takhistov}}, \citenamefont {{Ganezer}}, \citenamefont {{Hartfiel}},
  \citenamefont {{Hill}}, \citenamefont {{Keig}}, \citenamefont {{Hong}},
  \citenamefont {{Kim}}, \citenamefont {{Lim}}, \citenamefont {{Akiri}},
  \citenamefont {{Himmel}}, \citenamefont {{Scholberg}}, \citenamefont
  {{Walter}}, \citenamefont {{Wongjirad}}, \citenamefont {{Ishizuka}},
  \citenamefont {{Tasaka}}, \citenamefont {{Jang}}, \citenamefont {{Learned}},
  \citenamefont {{Matsuno}}, \citenamefont {{Smith}}, \citenamefont
  {{Hasegawa}}, \citenamefont {{Ishida}}, \citenamefont {{Ishii}},
  \citenamefont {{Kobayashi}}, \citenamefont {{Nakadaira}}, \citenamefont
  {{Nakamura}}, \citenamefont {{Oyama}}, \citenamefont {{Sakashita}},
  \citenamefont {{Sekiguchi}}, \citenamefont {{Tsukamoto}}, \citenamefont
  {{Suzuki}}, \citenamefont {{Takeuchi}}, \citenamefont {{Bronner}},
  \citenamefont {{Hirota}}, \citenamefont {{Huang}}, \citenamefont {{Ieki}},
  \citenamefont {{Kikawa}}, \citenamefont {{Minamino}}, \citenamefont
  {{Murakami}}, \citenamefont {{Nakaya}}, \citenamefont {{Suzuki}},
  \citenamefont {{Takahashi}}, \citenamefont {{Tateishi}}, \citenamefont
  {{Fukuda}}, \citenamefont {{Choi}}, \citenamefont {{Itow}}, \citenamefont
  {{Mitsuka}}, \citenamefont {{Mijakowski}}, \citenamefont {{Hignight}},
  \citenamefont {{Imber}}, \citenamefont {{Jung}}, \citenamefont
  {{Yanagisawa}}, \citenamefont {{Wilking}}, \citenamefont {{Ishino}},
  \citenamefont {{Kibayashi}}, \citenamefont {{Koshio}}, \citenamefont
  {{Mori}}, \citenamefont {{Sakuda}}, \citenamefont {{Yamaguchi}},
  \citenamefont {{Yano}}, \citenamefont {{Kuno}}, \citenamefont {{Tacik}},
  \citenamefont {{Kim}}, \citenamefont {{Okazawa}}, \citenamefont {{Choi}},
  \citenamefont {{Nishijima}}, \citenamefont {{Koshiba}}, \citenamefont
  {{Suda}}, \citenamefont {{Totsuka}}, \citenamefont {{Yokoyama}},
  \citenamefont {{Martens}}, \citenamefont {{Marti}}, \citenamefont {{Vagins}},
  \citenamefont {{Martin}}, \citenamefont {{de Perio}}, \citenamefont
  {{Konaka}}, \citenamefont {{Chen}}, \citenamefont {{Zhang}}, \citenamefont
  {{Connolly}},\ and\ \citenamefont {{Wilkes}}}]{2016APh....81...39A}%
  \BibitemOpen
  \bibfield  {author} {\bibinfo {author} {\bibfnamefont {K.}~\bibnamefont
  {{Abe}}}, \bibinfo {author} {\bibfnamefont {Y.}~\bibnamefont {{Haga}}},
  \bibinfo {author} {\bibfnamefont {Y.}~\bibnamefont {{Hayato}}}, \bibinfo
  {author} {\bibfnamefont {M.}~\bibnamefont {{Ikeda}}}, \bibinfo {author}
  {\bibfnamefont {K.}~\bibnamefont {{Iyogi}}}, \bibinfo {author} {\bibfnamefont
  {J.}~\bibnamefont {{Kameda}}}, \bibinfo {author} {\bibfnamefont
  {Y.}~\bibnamefont {{Kishimoto}}}, \bibinfo {author} {\bibfnamefont
  {M.}~\bibnamefont {{Miura}}}, \bibinfo {author} {\bibfnamefont
  {S.}~\bibnamefont {{Moriyama}}}, \bibinfo {author} {\bibfnamefont
  {M.}~\bibnamefont {{Nakahata}}}, \bibinfo {author} {\bibfnamefont
  {Y.}~\bibnamefont {{Nakano}}}, \bibinfo {author} {\bibfnamefont
  {S.}~\bibnamefont {{Nakayama}}}, \bibinfo {author} {\bibfnamefont
  {H.}~\bibnamefont {{Sekiya}}}, \bibinfo {author} {\bibfnamefont
  {M.}~\bibnamefont {{Shiozawa}}}, \bibinfo {author} {\bibfnamefont
  {Y.}~\bibnamefont {{Suzuki}}}, \bibinfo {author} {\bibfnamefont
  {A.}~\bibnamefont {{Takeda}}}, \bibinfo {author} {\bibfnamefont
  {H.}~\bibnamefont {{Tanaka}}}, \bibinfo {author} {\bibfnamefont
  {T.}~\bibnamefont {{Tomura}}}, \bibinfo {author} {\bibfnamefont
  {K.}~\bibnamefont {{Ueno}}}, \bibinfo {author} {\bibfnamefont {R.~A.}\
  \bibnamefont {{Wendell}}}, \bibinfo {author} {\bibfnamefont {T.}~\bibnamefont
  {{Yokozawa}}}, \bibinfo {author} {\bibfnamefont {T.}~\bibnamefont
  {{Irvine}}}, \bibinfo {author} {\bibfnamefont {T.}~\bibnamefont {{Kajita}}},
  \bibinfo {author} {\bibfnamefont {I.}~\bibnamefont {{Kametani}}}, \bibinfo
  {author} {\bibfnamefont {K.}~\bibnamefont {{Kaneyuki}}}, \bibinfo {author}
  {\bibfnamefont {K.~P.}\ \bibnamefont {{Lee}}}, \bibinfo {author}
  {\bibfnamefont {T.}~\bibnamefont {{McLachlan}}}, \bibinfo {author}
  {\bibfnamefont {Y.}~\bibnamefont {{Nishimura}}}, \bibinfo {author}
  {\bibfnamefont {E.}~\bibnamefont {{Richard}}}, \bibinfo {author}
  {\bibfnamefont {K.}~\bibnamefont {{Okumura}}}, \bibinfo {author}
  {\bibfnamefont {L.}~\bibnamefont {{Labarga}}}, \bibinfo {author}
  {\bibfnamefont {P.}~\bibnamefont {{Fernandez}}}, \bibinfo {author}
  {\bibfnamefont {S.}~\bibnamefont {{Berkman}}}, \bibinfo {author}
  {\bibfnamefont {H.~A.}\ \bibnamefont {{Tanaka}}}, \bibinfo {author}
  {\bibfnamefont {S.}~\bibnamefont {{Tobayama}}}, \bibinfo {author}
  {\bibfnamefont {J.}~\bibnamefont {{Gustafson}}}, \bibinfo {author}
  {\bibfnamefont {E.}~\bibnamefont {{Kearns}}}, \bibinfo {author}
  {\bibfnamefont {J.~L.}\ \bibnamefont {{Raaf}}}, \bibinfo {author}
  {\bibfnamefont {J.~L.}\ \bibnamefont {{Stone}}}, \bibinfo {author}
  {\bibfnamefont {L.~R.}\ \bibnamefont {{Sulak}}}, \bibinfo {author}
  {\bibfnamefont {M.}~\bibnamefont {{Goldhaber}}}, \bibinfo {author}
  {\bibfnamefont {G.}~\bibnamefont {{Carminati}}}, \bibinfo {author}
  {\bibfnamefont {W.~R.}\ \bibnamefont {{Kropp}}}, \bibinfo {author}
  {\bibfnamefont {S.}~\bibnamefont {{Mine}}}, \bibinfo {author} {\bibfnamefont
  {P.}~\bibnamefont {{Weatherly}}}, \bibinfo {author} {\bibfnamefont
  {A.}~\bibnamefont {{Renshaw}}}, \bibinfo {author} {\bibfnamefont {M.~B.}\
  \bibnamefont {{Smy}}}, \bibinfo {author} {\bibfnamefont {H.~W.}\ \bibnamefont
  {{Sobel}}}, \bibinfo {author} {\bibfnamefont {V.}~\bibnamefont
  {{Takhistov}}}, \bibinfo {author} {\bibfnamefont {K.~S.}\ \bibnamefont
  {{Ganezer}}}, \bibinfo {author} {\bibfnamefont {B.~L.}\ \bibnamefont
  {{Hartfiel}}}, \bibinfo {author} {\bibfnamefont {J.}~\bibnamefont {{Hill}}},
  \bibinfo {author} {\bibfnamefont {W.~E.}\ \bibnamefont {{Keig}}}, \bibinfo
  {author} {\bibfnamefont {N.}~\bibnamefont {{Hong}}}, \bibinfo {author}
  {\bibfnamefont {J.~Y.}\ \bibnamefont {{Kim}}}, \bibinfo {author}
  {\bibfnamefont {I.~T.}\ \bibnamefont {{Lim}}}, \bibinfo {author}
  {\bibfnamefont {T.}~\bibnamefont {{Akiri}}}, \bibinfo {author} {\bibfnamefont
  {A.}~\bibnamefont {{Himmel}}}, \bibinfo {author} {\bibfnamefont
  {K.}~\bibnamefont {{Scholberg}}}, \bibinfo {author} {\bibfnamefont {C.~W.}\
  \bibnamefont {{Walter}}}, \bibinfo {author} {\bibfnamefont {T.}~\bibnamefont
  {{Wongjirad}}}, \bibinfo {author} {\bibfnamefont {T.}~\bibnamefont
  {{Ishizuka}}}, \bibinfo {author} {\bibfnamefont {S.}~\bibnamefont
  {{Tasaka}}}, \bibinfo {author} {\bibfnamefont {J.~S.}\ \bibnamefont
  {{Jang}}}, \bibinfo {author} {\bibfnamefont {J.~G.}\ \bibnamefont
  {{Learned}}}, \bibinfo {author} {\bibfnamefont {S.}~\bibnamefont
  {{Matsuno}}}, \bibinfo {author} {\bibfnamefont {S.~N.}\ \bibnamefont
  {{Smith}}}, \bibinfo {author} {\bibfnamefont {T.}~\bibnamefont {{Hasegawa}}},
  \bibinfo {author} {\bibfnamefont {T.}~\bibnamefont {{Ishida}}}, \bibinfo
  {author} {\bibfnamefont {T.}~\bibnamefont {{Ishii}}}, \bibinfo {author}
  {\bibfnamefont {T.}~\bibnamefont {{Kobayashi}}}, \bibinfo {author}
  {\bibfnamefont {T.}~\bibnamefont {{Nakadaira}}}, \bibinfo {author}
  {\bibfnamefont {K.}~\bibnamefont {{Nakamura}}}, \bibinfo {author}
  {\bibfnamefont {Y.}~\bibnamefont {{Oyama}}}, \bibinfo {author} {\bibfnamefont
  {K.}~\bibnamefont {{Sakashita}}}, \bibinfo {author} {\bibfnamefont
  {T.}~\bibnamefont {{Sekiguchi}}}, \bibinfo {author} {\bibfnamefont
  {T.}~\bibnamefont {{Tsukamoto}}}, \bibinfo {author} {\bibfnamefont {A.~T.}\
  \bibnamefont {{Suzuki}}}, \bibinfo {author} {\bibfnamefont {Y.}~\bibnamefont
  {{Takeuchi}}}, \bibinfo {author} {\bibfnamefont {C.}~\bibnamefont
  {{Bronner}}}, \bibinfo {author} {\bibfnamefont {S.}~\bibnamefont {{Hirota}}},
  \bibinfo {author} {\bibfnamefont {K.}~\bibnamefont {{Huang}}}, \bibinfo
  {author} {\bibfnamefont {K.}~\bibnamefont {{Ieki}}}, \bibinfo {author}
  {\bibfnamefont {T.}~\bibnamefont {{Kikawa}}}, \bibinfo {author}
  {\bibfnamefont {A.}~\bibnamefont {{Minamino}}}, \bibinfo {author}
  {\bibfnamefont {A.}~\bibnamefont {{Murakami}}}, \bibinfo {author}
  {\bibfnamefont {T.}~\bibnamefont {{Nakaya}}}, \bibinfo {author}
  {\bibfnamefont {K.}~\bibnamefont {{Suzuki}}}, \bibinfo {author}
  {\bibfnamefont {S.}~\bibnamefont {{Takahashi}}}, \bibinfo {author}
  {\bibfnamefont {K.}~\bibnamefont {{Tateishi}}}, \bibinfo {author}
  {\bibfnamefont {Y.}~\bibnamefont {{Fukuda}}}, \bibinfo {author}
  {\bibfnamefont {K.}~\bibnamefont {{Choi}}}, \bibinfo {author} {\bibfnamefont
  {Y.}~\bibnamefont {{Itow}}}, \bibinfo {author} {\bibfnamefont
  {G.}~\bibnamefont {{Mitsuka}}}, \bibinfo {author} {\bibfnamefont
  {P.}~\bibnamefont {{Mijakowski}}}, \bibinfo {author} {\bibfnamefont
  {J.}~\bibnamefont {{Hignight}}}, \bibinfo {author} {\bibfnamefont
  {J.}~\bibnamefont {{Imber}}}, \bibinfo {author} {\bibfnamefont {C.~K.}\
  \bibnamefont {{Jung}}}, \bibinfo {author} {\bibfnamefont {C.}~\bibnamefont
  {{Yanagisawa}}}, \bibinfo {author} {\bibfnamefont {M.~J.}\ \bibnamefont
  {{Wilking}}}, \bibinfo {author} {\bibfnamefont {H.}~\bibnamefont {{Ishino}}},
  \bibinfo {author} {\bibfnamefont {A.}~\bibnamefont {{Kibayashi}}}, \bibinfo
  {author} {\bibfnamefont {Y.}~\bibnamefont {{Koshio}}}, \bibinfo {author}
  {\bibfnamefont {T.}~\bibnamefont {{Mori}}}, \bibinfo {author} {\bibfnamefont
  {M.}~\bibnamefont {{Sakuda}}}, \bibinfo {author} {\bibfnamefont
  {R.}~\bibnamefont {{Yamaguchi}}}, \bibinfo {author} {\bibfnamefont
  {T.}~\bibnamefont {{Yano}}}, \bibinfo {author} {\bibfnamefont
  {Y.}~\bibnamefont {{Kuno}}}, \bibinfo {author} {\bibfnamefont
  {R.}~\bibnamefont {{Tacik}}}, \bibinfo {author} {\bibfnamefont {S.~B.}\
  \bibnamefont {{Kim}}}, \bibinfo {author} {\bibfnamefont {H.}~\bibnamefont
  {{Okazawa}}}, \bibinfo {author} {\bibfnamefont {Y.}~\bibnamefont {{Choi}}},
  \bibinfo {author} {\bibfnamefont {K.}~\bibnamefont {{Nishijima}}}, \bibinfo
  {author} {\bibfnamefont {M.}~\bibnamefont {{Koshiba}}}, \bibinfo {author}
  {\bibfnamefont {Y.}~\bibnamefont {{Suda}}}, \bibinfo {author} {\bibfnamefont
  {Y.}~\bibnamefont {{Totsuka}}}, \bibinfo {author} {\bibfnamefont
  {M.}~\bibnamefont {{Yokoyama}}}, \bibinfo {author} {\bibfnamefont
  {K.}~\bibnamefont {{Martens}}}, \bibinfo {author} {\bibfnamefont
  {L.}~\bibnamefont {{Marti}}}, \bibinfo {author} {\bibfnamefont {M.~R.}\
  \bibnamefont {{Vagins}}}, \bibinfo {author} {\bibfnamefont {J.~F.}\
  \bibnamefont {{Martin}}}, \bibinfo {author} {\bibfnamefont {P.}~\bibnamefont
  {{de Perio}}}, \bibinfo {author} {\bibfnamefont {A.}~\bibnamefont
  {{Konaka}}}, \bibinfo {author} {\bibfnamefont {S.}~\bibnamefont {{Chen}}},
  \bibinfo {author} {\bibfnamefont {Y.}~\bibnamefont {{Zhang}}}, \bibinfo
  {author} {\bibfnamefont {K.}~\bibnamefont {{Connolly}}},\ and\ \bibinfo
  {author} {\bibfnamefont {R.~J.}\ \bibnamefont {{Wilkes}}},\ }\bibfield
  {title} {\bibinfo {title} {{Real-time supernova neutrino burst monitor at
  Super-Kamiokande}},\ }\href
  {https://doi.org/10.1016/j.astropartphys.2016.04.003} {\bibfield  {journal}
  {\bibinfo  {journal} {Astroparticle Physics}\ }\textbf {\bibinfo {volume}
  {81}},\ \bibinfo {pages} {39} (\bibinfo {year} {2016})},\ \Eprint
  {https://arxiv.org/abs/1601.04778} {arXiv:1601.04778 [astro-ph.HE]}
  \BibitemShut {NoStop}%
\bibitem [{\citenamefont {{Strumia}}\ and\ \citenamefont
  {{Vissani}}(2003)}]{2003PhLB..564...42S}%
  \BibitemOpen
  \bibfield  {author} {\bibinfo {author} {\bibfnamefont {A.}~\bibnamefont
  {{Strumia}}}\ and\ \bibinfo {author} {\bibfnamefont {F.}~\bibnamefont
  {{Vissani}}},\ }\bibfield  {title} {\bibinfo {title} {{Precise quasielastic
  neutrino/nucleon cross-section}},\ }\href
  {https://doi.org/10.1016/S0370-2693(03)00616-6} {\bibfield  {journal}
  {\bibinfo  {journal} {Physics Letters B}\ }\textbf {\bibinfo {volume}
  {564}},\ \bibinfo {pages} {42} (\bibinfo {year} {2003})},\ \Eprint
  {https://arxiv.org/abs/astro-ph/0302055} {arXiv:astro-ph/0302055 [astro-ph]}
  \BibitemShut {NoStop}%
\bibitem [{\citenamefont {{Ikeda}}\ \emph {et~al.}(2007)\citenamefont
  {{Ikeda}}, \citenamefont {{Takeda}}, \citenamefont {{Fukuda}}, \citenamefont
  {{Vagins}}, \citenamefont {{Abe}}, \citenamefont {{Iida}}, \citenamefont
  {{Ishihara}}, \citenamefont {{Kameda}}, \citenamefont {{Koshio}},
  \citenamefont {{Minamino}}, \citenamefont {{Mitsuda}}, \citenamefont
  {{Miura}}, \citenamefont {{Moriyama}}, \citenamefont {{Nakahata}},
  \citenamefont {{Obayashi}}, \citenamefont {{Ogawa}}, \citenamefont
  {{Sekiya}}, \citenamefont {{Shiozawa}}, \citenamefont {{Suzuki}},
  \citenamefont {{Takeuchi}}, \citenamefont {{Ueshima}}, \citenamefont
  {{Watanabe}}, \citenamefont {{Yamada}}, \citenamefont {{Higuchi}},
  \citenamefont {{Ishihara}}, \citenamefont {{Ishitsuka}}, \citenamefont
  {{Kajita}}, \citenamefont {{Kaneyuki}}, \citenamefont {{Mitsuka}},
  \citenamefont {{Nakayama}}, \citenamefont {{Nishino}}, \citenamefont
  {{Okumura}}, \citenamefont {{Saji}}, \citenamefont {{Takenaga}},
  \citenamefont {{Clark}}, \citenamefont {{Desai}}, \citenamefont {{Dufour}},
  \citenamefont {{Kearns}}, \citenamefont {{Likhoded}}, \citenamefont
  {{Litos}}, \citenamefont {{Raaf}}, \citenamefont {{Stone}}, \citenamefont
  {{Sulak}}, \citenamefont {{Wang}}, \citenamefont {{Goldhaber}}, \citenamefont
  {{Casper}}, \citenamefont {{Cravens}}, \citenamefont {{Dunmore}},
  \citenamefont {{Kropp}}, \citenamefont {{Liu}}, \citenamefont {{Mine}},
  \citenamefont {{Regis}}, \citenamefont {{Smy}}, \citenamefont {{Sobel}},
  \citenamefont {{Ganezer}}, \citenamefont {{Hill}}, \citenamefont {{Keig}},
  \citenamefont {{Jang}}, \citenamefont {{Kim}}, \citenamefont {{Lim}},
  \citenamefont {{Scholberg}}, \citenamefont {{Tanimoto}}, \citenamefont
  {{Walter}}, \citenamefont {{Wendell}}, \citenamefont {{Ellsworth}},
  \citenamefont {{Tasaka}}, \citenamefont {{Guillian}}, \citenamefont
  {{Learned}}, \citenamefont {{Matsuno}}, \citenamefont {{Messier}},
  \citenamefont {{Hayato}}, \citenamefont {{Ichikawa}}, \citenamefont
  {{Ishida}}, \citenamefont {{Ishii}}, \citenamefont {{Iwashita}},
  \citenamefont {{Kobayashi}}, \citenamefont {{Nakadaira}}, \citenamefont
  {{Nakamura}}, \citenamefont {{Nitta}}, \citenamefont {{Oyama}}, \citenamefont
  {{Totsuka}}, \citenamefont {{Suzuki}}, \citenamefont {{Hasegawa}},
  \citenamefont {{Hiraide}}, \citenamefont {{Maesaka}}, \citenamefont
  {{Nakaya}}, \citenamefont {{Nishikawa}}, \citenamefont {{Sasaki}},
  \citenamefont {{Yamamoto}}, \citenamefont {{Yokoyama}}, \citenamefont
  {{Haines}}, \citenamefont {{Dazeley}}, \citenamefont {{Hatakeyama}},
  \citenamefont {{Svoboda}}, \citenamefont {{Sullivan}}, \citenamefont
  {{Turcan}}, \citenamefont {{Habig}}, \citenamefont {{Sato}}, \citenamefont
  {{Itow}}, \citenamefont {{Koike}}, \citenamefont {{Tanaka}}, \citenamefont
  {{Jung}}, \citenamefont {{Kato}}, \citenamefont {{Kobayashi}}, \citenamefont
  {{Malek}}, \citenamefont {{McGrew}}, \citenamefont {{Sarrat}}, \citenamefont
  {{Terri}}, \citenamefont {{Yanagisawa}}, \citenamefont {{Tamura}},
  \citenamefont {{Idehara}}, \citenamefont {{Sakuda}}, \citenamefont
  {{Sugihara}}, \citenamefont {{Kuno}}, \citenamefont {{Yoshida}},
  \citenamefont {{Kim}}, \citenamefont {{Yang}}, \citenamefont {{Yoo}},
  \citenamefont {{Ishizuka}}, \citenamefont {{Okazawa}}, \citenamefont
  {{Choi}}, \citenamefont {{Seo}}, \citenamefont {{Gando}}, \citenamefont
  {{Hasegawa}}, \citenamefont {{Inoue}}, \citenamefont {{Furuse}},
  \citenamefont {{Ishii}}, \citenamefont {{Nishijima}}, \citenamefont
  {{Ishino}}, \citenamefont {{Watanabe}}, \citenamefont {{Koshiba}},
  \citenamefont {{Chen}}, \citenamefont {{Deng}}, \citenamefont {{Liu}},
  \citenamefont {{Kielczewska}}, \citenamefont {{Zalipska}}, \citenamefont
  {{Berns}}, \citenamefont {{Gran}}, \citenamefont {{Shiraishi}}, \citenamefont
  {{Stachyra}}, \citenamefont {{Thrane}}, \citenamefont {{Washburn}},
  \citenamefont {{Wilkes}},\ and\ \citenamefont {{Super-KAMIOKANDE
  Collaboration}}}]{2007ApJ...669..519I}%
  \BibitemOpen
  \bibfield  {author} {\bibinfo {author} {\bibfnamefont {M.}~\bibnamefont
  {{Ikeda}}}, \bibinfo {author} {\bibfnamefont {A.}~\bibnamefont {{Takeda}}},
  \bibinfo {author} {\bibfnamefont {Y.}~\bibnamefont {{Fukuda}}}, \bibinfo
  {author} {\bibfnamefont {M.~R.}\ \bibnamefont {{Vagins}}}, \bibinfo {author}
  {\bibfnamefont {K.}~\bibnamefont {{Abe}}}, \bibinfo {author} {\bibfnamefont
  {T.}~\bibnamefont {{Iida}}}, \bibinfo {author} {\bibfnamefont
  {K.}~\bibnamefont {{Ishihara}}}, \bibinfo {author} {\bibfnamefont
  {J.}~\bibnamefont {{Kameda}}}, \bibinfo {author} {\bibfnamefont
  {Y.}~\bibnamefont {{Koshio}}}, \bibinfo {author} {\bibfnamefont
  {A.}~\bibnamefont {{Minamino}}}, \bibinfo {author} {\bibfnamefont
  {C.}~\bibnamefont {{Mitsuda}}}, \bibinfo {author} {\bibfnamefont
  {M.}~\bibnamefont {{Miura}}}, \bibinfo {author} {\bibfnamefont
  {S.}~\bibnamefont {{Moriyama}}}, \bibinfo {author} {\bibfnamefont
  {M.}~\bibnamefont {{Nakahata}}}, \bibinfo {author} {\bibfnamefont
  {Y.}~\bibnamefont {{Obayashi}}}, \bibinfo {author} {\bibfnamefont
  {H.}~\bibnamefont {{Ogawa}}}, \bibinfo {author} {\bibfnamefont
  {H.}~\bibnamefont {{Sekiya}}}, \bibinfo {author} {\bibfnamefont
  {M.}~\bibnamefont {{Shiozawa}}}, \bibinfo {author} {\bibfnamefont
  {Y.}~\bibnamefont {{Suzuki}}}, \bibinfo {author} {\bibfnamefont
  {Y.}~\bibnamefont {{Takeuchi}}}, \bibinfo {author} {\bibfnamefont
  {K.}~\bibnamefont {{Ueshima}}}, \bibinfo {author} {\bibfnamefont
  {H.}~\bibnamefont {{Watanabe}}}, \bibinfo {author} {\bibfnamefont
  {S.}~\bibnamefont {{Yamada}}}, \bibinfo {author} {\bibfnamefont
  {I.}~\bibnamefont {{Higuchi}}}, \bibinfo {author} {\bibfnamefont
  {C.}~\bibnamefont {{Ishihara}}}, \bibinfo {author} {\bibfnamefont
  {M.}~\bibnamefont {{Ishitsuka}}}, \bibinfo {author} {\bibfnamefont
  {T.}~\bibnamefont {{Kajita}}}, \bibinfo {author} {\bibfnamefont
  {K.}~\bibnamefont {{Kaneyuki}}}, \bibinfo {author} {\bibfnamefont
  {G.}~\bibnamefont {{Mitsuka}}}, \bibinfo {author} {\bibfnamefont
  {S.}~\bibnamefont {{Nakayama}}}, \bibinfo {author} {\bibfnamefont
  {H.}~\bibnamefont {{Nishino}}}, \bibinfo {author} {\bibfnamefont
  {K.}~\bibnamefont {{Okumura}}}, \bibinfo {author} {\bibfnamefont
  {C.}~\bibnamefont {{Saji}}}, \bibinfo {author} {\bibfnamefont
  {Y.}~\bibnamefont {{Takenaga}}}, \bibinfo {author} {\bibfnamefont
  {S.}~\bibnamefont {{Clark}}}, \bibinfo {author} {\bibfnamefont
  {S.}~\bibnamefont {{Desai}}}, \bibinfo {author} {\bibfnamefont
  {F.}~\bibnamefont {{Dufour}}}, \bibinfo {author} {\bibfnamefont
  {E.}~\bibnamefont {{Kearns}}}, \bibinfo {author} {\bibfnamefont
  {S.}~\bibnamefont {{Likhoded}}}, \bibinfo {author} {\bibfnamefont
  {M.}~\bibnamefont {{Litos}}}, \bibinfo {author} {\bibfnamefont {J.~L.}\
  \bibnamefont {{Raaf}}}, \bibinfo {author} {\bibfnamefont {J.~L.}\
  \bibnamefont {{Stone}}}, \bibinfo {author} {\bibfnamefont {L.~R.}\
  \bibnamefont {{Sulak}}}, \bibinfo {author} {\bibfnamefont {W.}~\bibnamefont
  {{Wang}}}, \bibinfo {author} {\bibfnamefont {M.}~\bibnamefont {{Goldhaber}}},
  \bibinfo {author} {\bibfnamefont {D.}~\bibnamefont {{Casper}}}, \bibinfo
  {author} {\bibfnamefont {J.~P.}\ \bibnamefont {{Cravens}}}, \bibinfo {author}
  {\bibfnamefont {J.}~\bibnamefont {{Dunmore}}}, \bibinfo {author}
  {\bibfnamefont {W.~R.}\ \bibnamefont {{Kropp}}}, \bibinfo {author}
  {\bibfnamefont {D.~W.}\ \bibnamefont {{Liu}}}, \bibinfo {author}
  {\bibfnamefont {S.}~\bibnamefont {{Mine}}}, \bibinfo {author} {\bibfnamefont
  {C.}~\bibnamefont {{Regis}}}, \bibinfo {author} {\bibfnamefont {M.~B.}\
  \bibnamefont {{Smy}}}, \bibinfo {author} {\bibfnamefont {H.~W.}\ \bibnamefont
  {{Sobel}}}, \bibinfo {author} {\bibfnamefont {K.~S.}\ \bibnamefont
  {{Ganezer}}}, \bibinfo {author} {\bibfnamefont {J.}~\bibnamefont {{Hill}}},
  \bibinfo {author} {\bibfnamefont {W.~E.}\ \bibnamefont {{Keig}}}, \bibinfo
  {author} {\bibfnamefont {J.~S.}\ \bibnamefont {{Jang}}}, \bibinfo {author}
  {\bibfnamefont {J.~Y.}\ \bibnamefont {{Kim}}}, \bibinfo {author}
  {\bibfnamefont {I.~T.}\ \bibnamefont {{Lim}}}, \bibinfo {author}
  {\bibfnamefont {K.}~\bibnamefont {{Scholberg}}}, \bibinfo {author}
  {\bibfnamefont {N.}~\bibnamefont {{Tanimoto}}}, \bibinfo {author}
  {\bibfnamefont {C.~W.}\ \bibnamefont {{Walter}}}, \bibinfo {author}
  {\bibfnamefont {R.}~\bibnamefont {{Wendell}}}, \bibinfo {author}
  {\bibfnamefont {R.~W.}\ \bibnamefont {{Ellsworth}}}, \bibinfo {author}
  {\bibfnamefont {S.}~\bibnamefont {{Tasaka}}}, \bibinfo {author}
  {\bibfnamefont {G.}~\bibnamefont {{Guillian}}}, \bibinfo {author}
  {\bibfnamefont {J.~G.}\ \bibnamefont {{Learned}}}, \bibinfo {author}
  {\bibfnamefont {S.}~\bibnamefont {{Matsuno}}}, \bibinfo {author}
  {\bibfnamefont {M.~D.}\ \bibnamefont {{Messier}}}, \bibinfo {author}
  {\bibfnamefont {Y.}~\bibnamefont {{Hayato}}}, \bibinfo {author}
  {\bibfnamefont {A.~K.}\ \bibnamefont {{Ichikawa}}}, \bibinfo {author}
  {\bibfnamefont {T.}~\bibnamefont {{Ishida}}}, \bibinfo {author}
  {\bibfnamefont {T.}~\bibnamefont {{Ishii}}}, \bibinfo {author} {\bibfnamefont
  {T.}~\bibnamefont {{Iwashita}}}, \bibinfo {author} {\bibfnamefont
  {T.}~\bibnamefont {{Kobayashi}}}, \bibinfo {author} {\bibfnamefont
  {T.}~\bibnamefont {{Nakadaira}}}, \bibinfo {author} {\bibfnamefont
  {K.}~\bibnamefont {{Nakamura}}}, \bibinfo {author} {\bibfnamefont
  {K.}~\bibnamefont {{Nitta}}}, \bibinfo {author} {\bibfnamefont
  {Y.}~\bibnamefont {{Oyama}}}, \bibinfo {author} {\bibfnamefont
  {Y.}~\bibnamefont {{Totsuka}}}, \bibinfo {author} {\bibfnamefont {A.~T.}\
  \bibnamefont {{Suzuki}}}, \bibinfo {author} {\bibfnamefont {M.}~\bibnamefont
  {{Hasegawa}}}, \bibinfo {author} {\bibfnamefont {K.}~\bibnamefont
  {{Hiraide}}}, \bibinfo {author} {\bibfnamefont {H.}~\bibnamefont
  {{Maesaka}}}, \bibinfo {author} {\bibfnamefont {T.}~\bibnamefont {{Nakaya}}},
  \bibinfo {author} {\bibfnamefont {K.}~\bibnamefont {{Nishikawa}}}, \bibinfo
  {author} {\bibfnamefont {T.}~\bibnamefont {{Sasaki}}}, \bibinfo {author}
  {\bibfnamefont {S.}~\bibnamefont {{Yamamoto}}}, \bibinfo {author}
  {\bibfnamefont {M.}~\bibnamefont {{Yokoyama}}}, \bibinfo {author}
  {\bibfnamefont {T.~J.}\ \bibnamefont {{Haines}}}, \bibinfo {author}
  {\bibfnamefont {S.}~\bibnamefont {{Dazeley}}}, \bibinfo {author}
  {\bibfnamefont {S.}~\bibnamefont {{Hatakeyama}}}, \bibinfo {author}
  {\bibfnamefont {R.}~\bibnamefont {{Svoboda}}}, \bibinfo {author}
  {\bibfnamefont {G.~W.}\ \bibnamefont {{Sullivan}}}, \bibinfo {author}
  {\bibfnamefont {D.}~\bibnamefont {{Turcan}}}, \bibinfo {author}
  {\bibfnamefont {A.}~\bibnamefont {{Habig}}}, \bibinfo {author} {\bibfnamefont
  {T.}~\bibnamefont {{Sato}}}, \bibinfo {author} {\bibfnamefont
  {Y.}~\bibnamefont {{Itow}}}, \bibinfo {author} {\bibfnamefont
  {T.}~\bibnamefont {{Koike}}}, \bibinfo {author} {\bibfnamefont
  {T.}~\bibnamefont {{Tanaka}}}, \bibinfo {author} {\bibfnamefont {C.~K.}\
  \bibnamefont {{Jung}}}, \bibinfo {author} {\bibfnamefont {T.}~\bibnamefont
  {{Kato}}}, \bibinfo {author} {\bibfnamefont {K.}~\bibnamefont {{Kobayashi}}},
  \bibinfo {author} {\bibfnamefont {M.}~\bibnamefont {{Malek}}}, \bibinfo
  {author} {\bibfnamefont {C.}~\bibnamefont {{McGrew}}}, \bibinfo {author}
  {\bibfnamefont {A.}~\bibnamefont {{Sarrat}}}, \bibinfo {author}
  {\bibfnamefont {R.}~\bibnamefont {{Terri}}}, \bibinfo {author} {\bibfnamefont
  {C.}~\bibnamefont {{Yanagisawa}}}, \bibinfo {author} {\bibfnamefont
  {N.}~\bibnamefont {{Tamura}}}, \bibinfo {author} {\bibfnamefont
  {Y.}~\bibnamefont {{Idehara}}}, \bibinfo {author} {\bibfnamefont
  {M.}~\bibnamefont {{Sakuda}}}, \bibinfo {author} {\bibfnamefont
  {M.}~\bibnamefont {{Sugihara}}}, \bibinfo {author} {\bibfnamefont
  {Y.}~\bibnamefont {{Kuno}}}, \bibinfo {author} {\bibfnamefont
  {M.}~\bibnamefont {{Yoshida}}}, \bibinfo {author} {\bibfnamefont {S.~B.}\
  \bibnamefont {{Kim}}}, \bibinfo {author} {\bibfnamefont {B.~S.}\ \bibnamefont
  {{Yang}}}, \bibinfo {author} {\bibfnamefont {J.}~\bibnamefont {{Yoo}}},
  \bibinfo {author} {\bibfnamefont {T.}~\bibnamefont {{Ishizuka}}}, \bibinfo
  {author} {\bibfnamefont {H.}~\bibnamefont {{Okazawa}}}, \bibinfo {author}
  {\bibfnamefont {Y.}~\bibnamefont {{Choi}}}, \bibinfo {author} {\bibfnamefont
  {H.~K.}\ \bibnamefont {{Seo}}}, \bibinfo {author} {\bibfnamefont
  {Y.}~\bibnamefont {{Gando}}}, \bibinfo {author} {\bibfnamefont
  {T.}~\bibnamefont {{Hasegawa}}}, \bibinfo {author} {\bibfnamefont
  {K.}~\bibnamefont {{Inoue}}}, \bibinfo {author} {\bibfnamefont
  {Y.}~\bibnamefont {{Furuse}}}, \bibinfo {author} {\bibfnamefont
  {H.}~\bibnamefont {{Ishii}}}, \bibinfo {author} {\bibfnamefont
  {K.}~\bibnamefont {{Nishijima}}}, \bibinfo {author} {\bibfnamefont
  {H.}~\bibnamefont {{Ishino}}}, \bibinfo {author} {\bibfnamefont
  {Y.}~\bibnamefont {{Watanabe}}}, \bibinfo {author} {\bibfnamefont
  {M.}~\bibnamefont {{Koshiba}}}, \bibinfo {author} {\bibfnamefont
  {S.}~\bibnamefont {{Chen}}}, \bibinfo {author} {\bibfnamefont
  {Z.}~\bibnamefont {{Deng}}}, \bibinfo {author} {\bibfnamefont
  {Y.}~\bibnamefont {{Liu}}}, \bibinfo {author} {\bibfnamefont
  {D.}~\bibnamefont {{Kielczewska}}}, \bibinfo {author} {\bibfnamefont
  {J.}~\bibnamefont {{Zalipska}}}, \bibinfo {author} {\bibfnamefont
  {H.}~\bibnamefont {{Berns}}}, \bibinfo {author} {\bibfnamefont
  {R.}~\bibnamefont {{Gran}}}, \bibinfo {author} {\bibfnamefont {K.~K.}\
  \bibnamefont {{Shiraishi}}}, \bibinfo {author} {\bibfnamefont
  {A.}~\bibnamefont {{Stachyra}}}, \bibinfo {author} {\bibfnamefont
  {E.}~\bibnamefont {{Thrane}}}, \bibinfo {author} {\bibfnamefont
  {K.}~\bibnamefont {{Washburn}}}, \bibinfo {author} {\bibfnamefont {R.~J.}\
  \bibnamefont {{Wilkes}}},\ and\ \bibinfo {author} {\bibnamefont
  {{Super-KAMIOKANDE Collaboration}}},\ }\bibfield  {title} {\bibinfo {title}
  {{Search for Supernova Neutrino Bursts at Super-Kamiokande}},\ }\href
  {https://doi.org/10.1086/521547} {\bibfield  {journal} {\bibinfo  {journal}
  {\apj}\ }\textbf {\bibinfo {volume} {669}},\ \bibinfo {pages} {519} (\bibinfo
  {year} {2007})},\ \Eprint {https://arxiv.org/abs/0706.2283} {arXiv:0706.2283
  [astro-ph]} \BibitemShut {NoStop}%
\bibitem [{\citenamefont {{Mori}}\ \emph
  {et~al.}(2022{\natexlab{b}})\citenamefont {{Mori}}, \citenamefont {{Abe}},
  \citenamefont {{Hayato}}, \citenamefont {{Hiraide}}, \citenamefont {{Ieki}},
  \citenamefont {{Ikeda}}, \citenamefont {{Imaizumi}}, \citenamefont
  {{Kameda}}, \citenamefont {{Kanemura}}, \citenamefont {{Kaneshima}},
  \citenamefont {{Kashiwagi}}, \citenamefont {{Kataoka}}, \citenamefont
  {{Miki}}, \citenamefont {{Mine}}, \citenamefont {{Miura}}, \citenamefont
  {{Moriyama}}, \citenamefont {{Nagao}}, \citenamefont {{Nakahata}},
  \citenamefont {{Nakano}}, \citenamefont {{Nakayama}}, \citenamefont
  {{Noguchi}}, \citenamefont {{Okada}}, \citenamefont {{Okamoto}},
  \citenamefont {{Orii}}, \citenamefont {{Sato}}, \citenamefont {{Sekiya}},
  \citenamefont {{Shiba}}, \citenamefont {{Shimizu}}, \citenamefont
  {{Shiozawa}}, \citenamefont {{Sonoda}}, \citenamefont {{Suzuki}},
  \citenamefont {{Takeda}}, \citenamefont {{Takemoto}}, \citenamefont
  {{Takenaka}}, \citenamefont {{Tanaka}}, \citenamefont {{Tomiya}},
  \citenamefont {{Watanabe}}, \citenamefont {{Yano}}, \citenamefont
  {{Yoshida}}, \citenamefont {{Han}}, \citenamefont {{Kajita}}, \citenamefont
  {{Okumura}}, \citenamefont {{Tashiro}}, \citenamefont {{Wang}}, \citenamefont
  {{Xia}}, \citenamefont {{Megias}}, \citenamefont {{Bravo-Bergu{\~n}o}},
  \citenamefont {{Fernandez}}, \citenamefont {{Labarga}}, \citenamefont
  {{Ospina}}, \citenamefont {{Zaldivar}}, \citenamefont {{Zsoldos}},
  \citenamefont {{Pointon}}, \citenamefont {{Blaszczyk}}, \citenamefont
  {{Kearns}}, \citenamefont {{Raaf}}, \citenamefont {{Stone}}, \citenamefont
  {{Wan}}, \citenamefont {{Wester}}, \citenamefont {{Bian}}, \citenamefont
  {{Griskevich}}, \citenamefont {{Kropp}}, \citenamefont {{Locke}},
  \citenamefont {{Smy}}, \citenamefont {{Sobel}}, \citenamefont {{Takhistov}},
  \citenamefont {{Yankelevich}}, \citenamefont {{Hill}}, \citenamefont {{Kim}},
  \citenamefont {{Lim}}, \citenamefont {{Park}}, \citenamefont {{Bodur}},
  \citenamefont {{Scholberg}}, \citenamefont {{Walter}}, \citenamefont
  {{Bernard}}, \citenamefont {{Coffani}}, \citenamefont {{Drapier}},
  \citenamefont {{El Hedri}}, \citenamefont {{Giampaolo}}, \citenamefont
  {{Mueller}}, \citenamefont {{Paganini}}, \citenamefont {{Quilain}},
  \citenamefont {{Santos}}, \citenamefont {{Ishizuka}}, \citenamefont
  {{Nakamura}}, \citenamefont {{Jang}}, \citenamefont {{Learned}},
  \citenamefont {{Anthony}}, \citenamefont {{Martin}}, \citenamefont {{Scott}},
  \citenamefont {{Sztuc}}, \citenamefont {{Uchida}}, \citenamefont {{Berardi}},
  \citenamefont {{Catanesi}}, \citenamefont {{Radicioni}}, \citenamefont
  {{Calabria}}, \citenamefont {{Machado}}, \citenamefont {{De Rosa}},
  \citenamefont {{Collazuol}}, \citenamefont {{Iacob}}, \citenamefont
  {{Lamoureux}}, \citenamefont {{Mattiazzi}}, \citenamefont {{Ludovici}},
  \citenamefont {{Gonin}}, \citenamefont {{Pronost}}, \citenamefont
  {{Maekawa}}, \citenamefont {{Nishimura}}, \citenamefont {{Fujisawa}},
  \citenamefont {{Friend}}, \citenamefont {{Hasegawa}}, \citenamefont
  {{Ishida}}, \citenamefont {{Kobayashi}}, \citenamefont {{Jakkapu}},
  \citenamefont {{Matsubara}}, \citenamefont {{Nakadaira}}, \citenamefont
  {{Nakamura}}, \citenamefont {{Oyama}}, \citenamefont {{Sakashita}},
  \citenamefont {{Sekiguchi}}, \citenamefont {{Tsukamoto}}, \citenamefont
  {{Ozaki}}, \citenamefont {{Shiozawa}}, \citenamefont {{Suzuki}},
  \citenamefont {{Takeuchi}}, \citenamefont {{Yamamoto}}, \citenamefont
  {{Kotsar}}, \citenamefont {{Ashida}}, \citenamefont {{Bronner}},
  \citenamefont {{Feng}}, \citenamefont {{Hirota}}, \citenamefont {{Kikawa}},
  \citenamefont {{Nakaya}}, \citenamefont {{Wendell}}, \citenamefont
  {{Yasutome}}, \citenamefont {{McCauley}}, \citenamefont {{Mehta}},
  \citenamefont {{Tsui}}, \citenamefont {{Fukuda}}, \citenamefont {{Itow}},
  \citenamefont {{Menjo}}, \citenamefont {{Ninomiya}}, \citenamefont {{Niwa}},
  \citenamefont {{Tsukada}}, \citenamefont {{Lagoda}}, \citenamefont
  {{Lakshmi}}, \citenamefont {{Mijakowski}}, \citenamefont {{Zalipska}},
  \citenamefont {{Mandal}}, \citenamefont {{Prabhu}}, \citenamefont {{Jiang}},
  \citenamefont {{Jung}}, \citenamefont {{Vilela}}, \citenamefont {{Wilking}},
  \citenamefont {{Yanagisawa}}, \citenamefont {{Jia}}, \citenamefont
  {{Hagiwara}}, \citenamefont {{Harada}}, \citenamefont {{Horai}},
  \citenamefont {{Ishino}}, \citenamefont {{Ito}}, \citenamefont {{Kitagawa}},
  \citenamefont {{Koshio}}, \citenamefont {{Ma}}, \citenamefont {{Nakanishi}},
  \citenamefont {{Piplani}}, \citenamefont {{Sakai}}, \citenamefont {{Barr}},
  \citenamefont {{Barrow}}, \citenamefont {{Cook}}, \citenamefont {{Samani}},
  \citenamefont {{Wark}}, \citenamefont {{Nova}}, \citenamefont {{Boschi}},
  \citenamefont {{Gao}}, \citenamefont {{Goldsack}}, \citenamefont {{Katori}},
  \citenamefont {{Di Lodovico}}, \citenamefont {{Migenda}}, \citenamefont
  {{Taani}}, \citenamefont {{Zsoldos}}, \citenamefont {{Yang}}, \citenamefont
  {{Jenkins}}, \citenamefont {{Malek}}, \citenamefont {{McElwee}},
  \citenamefont {{Stone}}, \citenamefont {{Thiesse}}, \citenamefont
  {{Thompson}}, \citenamefont {{Okazawa}}, \citenamefont {{Kim}}, \citenamefont
  {{Seo}}, \citenamefont {{Yu}}, \citenamefont {{Nishijima}}, \citenamefont
  {{Koshiba}}, \citenamefont {{Nakagiri}}, \citenamefont {{Nakajima}},
  \citenamefont {{Iwamoto}}, \citenamefont {{Taniuchi}}, \citenamefont
  {{Yokoyama}}, \citenamefont {{Martens}},\ and\ \citenamefont {{de
  Perio}}}]{2022ApJ...938...35M}%
  \BibitemOpen
  \bibfield  {author} {\bibinfo {author} {\bibfnamefont {M.}~\bibnamefont
  {{Mori}}}, \bibinfo {author} {\bibfnamefont {K.}~\bibnamefont {{Abe}}},
  \bibinfo {author} {\bibfnamefont {Y.}~\bibnamefont {{Hayato}}}, \bibinfo
  {author} {\bibfnamefont {K.}~\bibnamefont {{Hiraide}}}, \bibinfo {author}
  {\bibfnamefont {K.}~\bibnamefont {{Ieki}}}, \bibinfo {author} {\bibfnamefont
  {M.}~\bibnamefont {{Ikeda}}}, \bibinfo {author} {\bibfnamefont
  {S.}~\bibnamefont {{Imaizumi}}}, \bibinfo {author} {\bibfnamefont
  {J.}~\bibnamefont {{Kameda}}}, \bibinfo {author} {\bibfnamefont
  {Y.}~\bibnamefont {{Kanemura}}}, \bibinfo {author} {\bibfnamefont
  {R.}~\bibnamefont {{Kaneshima}}}, \bibinfo {author} {\bibfnamefont
  {Y.}~\bibnamefont {{Kashiwagi}}}, \bibinfo {author} {\bibfnamefont
  {Y.}~\bibnamefont {{Kataoka}}}, \bibinfo {author} {\bibfnamefont
  {S.}~\bibnamefont {{Miki}}}, \bibinfo {author} {\bibfnamefont
  {S.}~\bibnamefont {{Mine}}}, \bibinfo {author} {\bibfnamefont
  {M.}~\bibnamefont {{Miura}}}, \bibinfo {author} {\bibfnamefont
  {S.}~\bibnamefont {{Moriyama}}}, \bibinfo {author} {\bibfnamefont
  {Y.}~\bibnamefont {{Nagao}}}, \bibinfo {author} {\bibfnamefont
  {M.}~\bibnamefont {{Nakahata}}}, \bibinfo {author} {\bibfnamefont
  {Y.}~\bibnamefont {{Nakano}}}, \bibinfo {author} {\bibfnamefont
  {S.}~\bibnamefont {{Nakayama}}}, \bibinfo {author} {\bibfnamefont
  {Y.}~\bibnamefont {{Noguchi}}}, \bibinfo {author} {\bibfnamefont
  {T.}~\bibnamefont {{Okada}}}, \bibinfo {author} {\bibfnamefont
  {K.}~\bibnamefont {{Okamoto}}}, \bibinfo {author} {\bibfnamefont
  {A.}~\bibnamefont {{Orii}}}, \bibinfo {author} {\bibfnamefont
  {K.}~\bibnamefont {{Sato}}}, \bibinfo {author} {\bibfnamefont
  {H.}~\bibnamefont {{Sekiya}}}, \bibinfo {author} {\bibfnamefont
  {H.}~\bibnamefont {{Shiba}}}, \bibinfo {author} {\bibfnamefont
  {K.}~\bibnamefont {{Shimizu}}}, \bibinfo {author} {\bibfnamefont
  {M.}~\bibnamefont {{Shiozawa}}}, \bibinfo {author} {\bibfnamefont
  {Y.}~\bibnamefont {{Sonoda}}}, \bibinfo {author} {\bibfnamefont
  {Y.}~\bibnamefont {{Suzuki}}}, \bibinfo {author} {\bibfnamefont
  {A.}~\bibnamefont {{Takeda}}}, \bibinfo {author} {\bibfnamefont
  {Y.}~\bibnamefont {{Takemoto}}}, \bibinfo {author} {\bibfnamefont
  {A.}~\bibnamefont {{Takenaka}}}, \bibinfo {author} {\bibfnamefont
  {H.}~\bibnamefont {{Tanaka}}}, \bibinfo {author} {\bibfnamefont
  {T.}~\bibnamefont {{Tomiya}}}, \bibinfo {author} {\bibfnamefont
  {S.}~\bibnamefont {{Watanabe}}}, \bibinfo {author} {\bibfnamefont
  {T.}~\bibnamefont {{Yano}}}, \bibinfo {author} {\bibfnamefont
  {S.}~\bibnamefont {{Yoshida}}}, \bibinfo {author} {\bibfnamefont
  {S.}~\bibnamefont {{Han}}}, \bibinfo {author} {\bibfnamefont
  {T.}~\bibnamefont {{Kajita}}}, \bibinfo {author} {\bibfnamefont
  {K.}~\bibnamefont {{Okumura}}}, \bibinfo {author} {\bibfnamefont
  {T.}~\bibnamefont {{Tashiro}}}, \bibinfo {author} {\bibfnamefont
  {X.}~\bibnamefont {{Wang}}}, \bibinfo {author} {\bibfnamefont
  {J.}~\bibnamefont {{Xia}}}, \bibinfo {author} {\bibfnamefont {G.~D.}\
  \bibnamefont {{Megias}}}, \bibinfo {author} {\bibfnamefont {D.}~\bibnamefont
  {{Bravo-Bergu{\~n}o}}}, \bibinfo {author} {\bibfnamefont {P.}~\bibnamefont
  {{Fernandez}}}, \bibinfo {author} {\bibfnamefont {L.}~\bibnamefont
  {{Labarga}}}, \bibinfo {author} {\bibfnamefont {N.}~\bibnamefont {{Ospina}}},
  \bibinfo {author} {\bibfnamefont {B.}~\bibnamefont {{Zaldivar}}}, \bibinfo
  {author} {\bibfnamefont {S.}~\bibnamefont {{Zsoldos}}}, \bibinfo {author}
  {\bibfnamefont {B.~W.}\ \bibnamefont {{Pointon}}}, \bibinfo {author}
  {\bibfnamefont {F.~D.~M.}\ \bibnamefont {{Blaszczyk}}}, \bibinfo {author}
  {\bibfnamefont {E.}~\bibnamefont {{Kearns}}}, \bibinfo {author}
  {\bibfnamefont {J.~L.}\ \bibnamefont {{Raaf}}}, \bibinfo {author}
  {\bibfnamefont {J.~L.}\ \bibnamefont {{Stone}}}, \bibinfo {author}
  {\bibfnamefont {L.}~\bibnamefont {{Wan}}}, \bibinfo {author} {\bibfnamefont
  {T.}~\bibnamefont {{Wester}}}, \bibinfo {author} {\bibfnamefont
  {J.}~\bibnamefont {{Bian}}}, \bibinfo {author} {\bibfnamefont {N.~J.}\
  \bibnamefont {{Griskevich}}}, \bibinfo {author} {\bibfnamefont {W.~R.}\
  \bibnamefont {{Kropp}}}, \bibinfo {author} {\bibfnamefont {S.}~\bibnamefont
  {{Locke}}}, \bibinfo {author} {\bibfnamefont {M.~B.}\ \bibnamefont {{Smy}}},
  \bibinfo {author} {\bibfnamefont {H.~W.}\ \bibnamefont {{Sobel}}}, \bibinfo
  {author} {\bibfnamefont {V.}~\bibnamefont {{Takhistov}}}, \bibinfo {author}
  {\bibfnamefont {A.}~\bibnamefont {{Yankelevich}}}, \bibinfo {author}
  {\bibfnamefont {J.}~\bibnamefont {{Hill}}}, \bibinfo {author} {\bibfnamefont
  {J.~Y.}\ \bibnamefont {{Kim}}}, \bibinfo {author} {\bibfnamefont {I.~T.}\
  \bibnamefont {{Lim}}}, \bibinfo {author} {\bibfnamefont {R.~G.}\ \bibnamefont
  {{Park}}}, \bibinfo {author} {\bibfnamefont {B.}~\bibnamefont {{Bodur}}},
  \bibinfo {author} {\bibfnamefont {K.}~\bibnamefont {{Scholberg}}}, \bibinfo
  {author} {\bibfnamefont {C.~W.}\ \bibnamefont {{Walter}}}, \bibinfo {author}
  {\bibfnamefont {L.}~\bibnamefont {{Bernard}}}, \bibinfo {author}
  {\bibfnamefont {A.}~\bibnamefont {{Coffani}}}, \bibinfo {author}
  {\bibfnamefont {O.}~\bibnamefont {{Drapier}}}, \bibinfo {author}
  {\bibfnamefont {S.}~\bibnamefont {{El Hedri}}}, \bibinfo {author}
  {\bibfnamefont {A.}~\bibnamefont {{Giampaolo}}}, \bibinfo {author}
  {\bibfnamefont {T.~A.}\ \bibnamefont {{Mueller}}}, \bibinfo {author}
  {\bibfnamefont {P.}~\bibnamefont {{Paganini}}}, \bibinfo {author}
  {\bibfnamefont {B.}~\bibnamefont {{Quilain}}}, \bibinfo {author}
  {\bibfnamefont {A.~D.}\ \bibnamefont {{Santos}}}, \bibinfo {author}
  {\bibfnamefont {T.}~\bibnamefont {{Ishizuka}}}, \bibinfo {author}
  {\bibfnamefont {T.}~\bibnamefont {{Nakamura}}}, \bibinfo {author}
  {\bibfnamefont {J.~S.}\ \bibnamefont {{Jang}}}, \bibinfo {author}
  {\bibfnamefont {J.~G.}\ \bibnamefont {{Learned}}}, \bibinfo {author}
  {\bibfnamefont {L.~H.~V.}\ \bibnamefont {{Anthony}}}, \bibinfo {author}
  {\bibfnamefont {D.}~\bibnamefont {{Martin}}}, \bibinfo {author}
  {\bibfnamefont {M.}~\bibnamefont {{Scott}}}, \bibinfo {author} {\bibfnamefont
  {A.~A.}\ \bibnamefont {{Sztuc}}}, \bibinfo {author} {\bibfnamefont
  {Y.}~\bibnamefont {{Uchida}}}, \bibinfo {author} {\bibfnamefont
  {V.}~\bibnamefont {{Berardi}}}, \bibinfo {author} {\bibfnamefont {M.~G.}\
  \bibnamefont {{Catanesi}}}, \bibinfo {author} {\bibfnamefont
  {E.}~\bibnamefont {{Radicioni}}}, \bibinfo {author} {\bibfnamefont {N.~F.}\
  \bibnamefont {{Calabria}}}, \bibinfo {author} {\bibfnamefont {L.~N.}\
  \bibnamefont {{Machado}}}, \bibinfo {author} {\bibfnamefont {G.}~\bibnamefont
  {{De Rosa}}}, \bibinfo {author} {\bibfnamefont {G.}~\bibnamefont
  {{Collazuol}}}, \bibinfo {author} {\bibfnamefont {F.}~\bibnamefont
  {{Iacob}}}, \bibinfo {author} {\bibfnamefont {M.}~\bibnamefont
  {{Lamoureux}}}, \bibinfo {author} {\bibfnamefont {M.}~\bibnamefont
  {{Mattiazzi}}}, \bibinfo {author} {\bibfnamefont {L.}~\bibnamefont
  {{Ludovici}}}, \bibinfo {author} {\bibfnamefont {M.}~\bibnamefont {{Gonin}}},
  \bibinfo {author} {\bibfnamefont {G.}~\bibnamefont {{Pronost}}}, \bibinfo
  {author} {\bibfnamefont {Y.}~\bibnamefont {{Maekawa}}}, \bibinfo {author}
  {\bibfnamefont {Y.}~\bibnamefont {{Nishimura}}}, \bibinfo {author}
  {\bibfnamefont {C.}~\bibnamefont {{Fujisawa}}}, \bibinfo {author}
  {\bibfnamefont {M.}~\bibnamefont {{Friend}}}, \bibinfo {author}
  {\bibfnamefont {T.}~\bibnamefont {{Hasegawa}}}, \bibinfo {author}
  {\bibfnamefont {T.}~\bibnamefont {{Ishida}}}, \bibinfo {author}
  {\bibfnamefont {T.}~\bibnamefont {{Kobayashi}}}, \bibinfo {author}
  {\bibfnamefont {M.}~\bibnamefont {{Jakkapu}}}, \bibinfo {author}
  {\bibfnamefont {T.}~\bibnamefont {{Matsubara}}}, \bibinfo {author}
  {\bibfnamefont {T.}~\bibnamefont {{Nakadaira}}}, \bibinfo {author}
  {\bibfnamefont {K.}~\bibnamefont {{Nakamura}}}, \bibinfo {author}
  {\bibfnamefont {Y.}~\bibnamefont {{Oyama}}}, \bibinfo {author} {\bibfnamefont
  {K.}~\bibnamefont {{Sakashita}}}, \bibinfo {author} {\bibfnamefont
  {T.}~\bibnamefont {{Sekiguchi}}}, \bibinfo {author} {\bibfnamefont
  {T.}~\bibnamefont {{Tsukamoto}}}, \bibinfo {author} {\bibfnamefont
  {H.}~\bibnamefont {{Ozaki}}}, \bibinfo {author} {\bibfnamefont
  {T.}~\bibnamefont {{Shiozawa}}}, \bibinfo {author} {\bibfnamefont {A.~T.}\
  \bibnamefont {{Suzuki}}}, \bibinfo {author} {\bibfnamefont {Y.}~\bibnamefont
  {{Takeuchi}}}, \bibinfo {author} {\bibfnamefont {S.}~\bibnamefont
  {{Yamamoto}}}, \bibinfo {author} {\bibfnamefont {Y.}~\bibnamefont
  {{Kotsar}}}, \bibinfo {author} {\bibfnamefont {Y.}~\bibnamefont {{Ashida}}},
  \bibinfo {author} {\bibfnamefont {C.}~\bibnamefont {{Bronner}}}, \bibinfo
  {author} {\bibfnamefont {J.}~\bibnamefont {{Feng}}}, \bibinfo {author}
  {\bibfnamefont {S.}~\bibnamefont {{Hirota}}}, \bibinfo {author}
  {\bibfnamefont {T.}~\bibnamefont {{Kikawa}}}, \bibinfo {author}
  {\bibfnamefont {T.}~\bibnamefont {{Nakaya}}}, \bibinfo {author}
  {\bibfnamefont {R.~A.}\ \bibnamefont {{Wendell}}}, \bibinfo {author}
  {\bibfnamefont {K.}~\bibnamefont {{Yasutome}}}, \bibinfo {author}
  {\bibfnamefont {N.}~\bibnamefont {{McCauley}}}, \bibinfo {author}
  {\bibfnamefont {P.}~\bibnamefont {{Mehta}}}, \bibinfo {author} {\bibfnamefont
  {K.~M.}\ \bibnamefont {{Tsui}}}, \bibinfo {author} {\bibfnamefont
  {Y.}~\bibnamefont {{Fukuda}}}, \bibinfo {author} {\bibfnamefont
  {Y.}~\bibnamefont {{Itow}}}, \bibinfo {author} {\bibfnamefont
  {H.}~\bibnamefont {{Menjo}}}, \bibinfo {author} {\bibfnamefont
  {K.}~\bibnamefont {{Ninomiya}}}, \bibinfo {author} {\bibfnamefont
  {T.}~\bibnamefont {{Niwa}}}, \bibinfo {author} {\bibfnamefont
  {M.}~\bibnamefont {{Tsukada}}}, \bibinfo {author} {\bibfnamefont
  {J.}~\bibnamefont {{Lagoda}}}, \bibinfo {author} {\bibfnamefont {S.~M.}\
  \bibnamefont {{Lakshmi}}}, \bibinfo {author} {\bibfnamefont {P.}~\bibnamefont
  {{Mijakowski}}}, \bibinfo {author} {\bibfnamefont {J.}~\bibnamefont
  {{Zalipska}}}, \bibinfo {author} {\bibfnamefont {M.}~\bibnamefont
  {{Mandal}}}, \bibinfo {author} {\bibfnamefont {Y.~S.}\ \bibnamefont
  {{Prabhu}}}, \bibinfo {author} {\bibfnamefont {J.}~\bibnamefont {{Jiang}}},
  \bibinfo {author} {\bibfnamefont {C.~K.}\ \bibnamefont {{Jung}}}, \bibinfo
  {author} {\bibfnamefont {C.}~\bibnamefont {{Vilela}}}, \bibinfo {author}
  {\bibfnamefont {M.~J.}\ \bibnamefont {{Wilking}}}, \bibinfo {author}
  {\bibfnamefont {C.}~\bibnamefont {{Yanagisawa}}}, \bibinfo {author}
  {\bibfnamefont {M.}~\bibnamefont {{Jia}}}, \bibinfo {author} {\bibfnamefont
  {K.}~\bibnamefont {{Hagiwara}}}, \bibinfo {author} {\bibfnamefont
  {M.}~\bibnamefont {{Harada}}}, \bibinfo {author} {\bibfnamefont
  {T.}~\bibnamefont {{Horai}}}, \bibinfo {author} {\bibfnamefont
  {H.}~\bibnamefont {{Ishino}}}, \bibinfo {author} {\bibfnamefont
  {S.}~\bibnamefont {{Ito}}}, \bibinfo {author} {\bibfnamefont
  {H.}~\bibnamefont {{Kitagawa}}}, \bibinfo {author} {\bibfnamefont
  {Y.}~\bibnamefont {{Koshio}}}, \bibinfo {author} {\bibfnamefont
  {W.}~\bibnamefont {{Ma}}}, \bibinfo {author} {\bibfnamefont {F.}~\bibnamefont
  {{Nakanishi}}}, \bibinfo {author} {\bibfnamefont {N.}~\bibnamefont
  {{Piplani}}}, \bibinfo {author} {\bibfnamefont {S.}~\bibnamefont {{Sakai}}},
  \bibinfo {author} {\bibfnamefont {G.}~\bibnamefont {{Barr}}}, \bibinfo
  {author} {\bibfnamefont {D.}~\bibnamefont {{Barrow}}}, \bibinfo {author}
  {\bibfnamefont {L.}~\bibnamefont {{Cook}}}, \bibinfo {author} {\bibfnamefont
  {S.}~\bibnamefont {{Samani}}}, \bibinfo {author} {\bibfnamefont
  {D.}~\bibnamefont {{Wark}}}, \bibinfo {author} {\bibfnamefont
  {F.}~\bibnamefont {{Nova}}}, \bibinfo {author} {\bibfnamefont
  {T.}~\bibnamefont {{Boschi}}}, \bibinfo {author} {\bibfnamefont
  {J.}~\bibnamefont {{Gao}}}, \bibinfo {author} {\bibfnamefont
  {A.}~\bibnamefont {{Goldsack}}}, \bibinfo {author} {\bibfnamefont
  {T.}~\bibnamefont {{Katori}}}, \bibinfo {author} {\bibfnamefont
  {F.}~\bibnamefont {{Di Lodovico}}}, \bibinfo {author} {\bibfnamefont
  {J.}~\bibnamefont {{Migenda}}}, \bibinfo {author} {\bibfnamefont
  {M.}~\bibnamefont {{Taani}}}, \bibinfo {author} {\bibfnamefont
  {S.}~\bibnamefont {{Zsoldos}}}, \bibinfo {author} {\bibfnamefont {J.~Y.}\
  \bibnamefont {{Yang}}}, \bibinfo {author} {\bibfnamefont {S.~J.}\
  \bibnamefont {{Jenkins}}}, \bibinfo {author} {\bibfnamefont {M.}~\bibnamefont
  {{Malek}}}, \bibinfo {author} {\bibfnamefont {J.~M.}\ \bibnamefont
  {{McElwee}}}, \bibinfo {author} {\bibfnamefont {O.}~\bibnamefont {{Stone}}},
  \bibinfo {author} {\bibfnamefont {M.~D.}\ \bibnamefont {{Thiesse}}}, \bibinfo
  {author} {\bibfnamefont {L.~F.}\ \bibnamefont {{Thompson}}}, \bibinfo
  {author} {\bibfnamefont {H.}~\bibnamefont {{Okazawa}}}, \bibinfo {author}
  {\bibfnamefont {S.~B.}\ \bibnamefont {{Kim}}}, \bibinfo {author}
  {\bibfnamefont {J.~W.}\ \bibnamefont {{Seo}}}, \bibinfo {author}
  {\bibfnamefont {I.}~\bibnamefont {{Yu}}}, \bibinfo {author} {\bibfnamefont
  {K.}~\bibnamefont {{Nishijima}}}, \bibinfo {author} {\bibfnamefont
  {M.}~\bibnamefont {{Koshiba}}}, \bibinfo {author} {\bibfnamefont
  {K.}~\bibnamefont {{Nakagiri}}}, \bibinfo {author} {\bibfnamefont
  {Y.}~\bibnamefont {{Nakajima}}}, \bibinfo {author} {\bibfnamefont
  {K.}~\bibnamefont {{Iwamoto}}}, \bibinfo {author} {\bibfnamefont
  {N.}~\bibnamefont {{Taniuchi}}}, \bibinfo {author} {\bibfnamefont
  {M.}~\bibnamefont {{Yokoyama}}}, \bibinfo {author} {\bibfnamefont
  {K.}~\bibnamefont {{Martens}}},\ and\ \bibinfo {author} {\bibfnamefont
  {P.}~\bibnamefont {{de Perio}}},\ }\bibfield  {title} {\bibinfo {title}
  {{Searching for Supernova Bursts in Super-Kamiokande IV}},\ }\href
  {https://doi.org/10.3847/1538-4357/ac8f41} {\bibfield  {journal} {\bibinfo
  {journal} {\apj}\ }\textbf {\bibinfo {volume} {938}},\ \bibinfo {eid} {35}
  (\bibinfo {year} {2022}{\natexlab{b}})},\ \Eprint
  {https://arxiv.org/abs/2206.01380} {arXiv:2206.01380 [astro-ph.HE]}
  \BibitemShut {NoStop}%
\bibitem [{\citenamefont {{Dighe}}\ and\ \citenamefont
  {{Smirnov}}(2000)}]{2000PhRvD..62c3007D}%
  \BibitemOpen
  \bibfield  {author} {\bibinfo {author} {\bibfnamefont {A.~S.}\ \bibnamefont
  {{Dighe}}}\ and\ \bibinfo {author} {\bibfnamefont {A.~Y.}\ \bibnamefont
  {{Smirnov}}},\ }\bibfield  {title} {\bibinfo {title} {{Identifying the
  neutrino mass spectrum from a supernova neutrino burst}},\ }\href
  {https://doi.org/10.1103/PhysRevD.62.033007} {\bibfield  {journal} {\bibinfo
  {journal} {\prd}\ }\textbf {\bibinfo {volume} {62}},\ \bibinfo {eid} {033007}
  (\bibinfo {year} {2000})},\ \Eprint {https://arxiv.org/abs/hep-ph/9907423}
  {arXiv:hep-ph/9907423 [hep-ph]} \BibitemShut {NoStop}%
\bibitem [{\citenamefont {{Wu}}\ \emph {et~al.}(2015)\citenamefont {{Wu}},
  \citenamefont {{Qian}}, \citenamefont {{Mart{\'\i}nez-Pinedo}}, \citenamefont
  {{Fischer}},\ and\ \citenamefont {{Huther}}}]{2015PhRvD..91f5016W}%
  \BibitemOpen
  \bibfield  {author} {\bibinfo {author} {\bibfnamefont {M.-R.}\ \bibnamefont
  {{Wu}}}, \bibinfo {author} {\bibfnamefont {Y.-Z.}\ \bibnamefont {{Qian}}},
  \bibinfo {author} {\bibfnamefont {G.}~\bibnamefont {{Mart{\'\i}nez-Pinedo}}},
  \bibinfo {author} {\bibfnamefont {T.}~\bibnamefont {{Fischer}}},\ and\
  \bibinfo {author} {\bibfnamefont {L.}~\bibnamefont {{Huther}}},\ }\bibfield
  {title} {\bibinfo {title} {{Effects of neutrino oscillations on
  nucleosynthesis and neutrino signals for an 18 M$_{{\ensuremath{\odot}}}$
  supernova model}},\ }\href {https://doi.org/10.1103/PhysRevD.91.065016}
  {\bibfield  {journal} {\bibinfo  {journal} {\prd}\ }\textbf {\bibinfo
  {volume} {91}},\ \bibinfo {eid} {065016} (\bibinfo {year} {2015})},\ \Eprint
  {https://arxiv.org/abs/1412.8587} {arXiv:1412.8587 [astro-ph.HE]}
  \BibitemShut {NoStop}%
\bibitem [{\citenamefont {{Messineo}}\ and\ \citenamefont
  {{Brown}}(2019)}]{2019AJ....158...20M}%
  \BibitemOpen
  \bibfield  {author} {\bibinfo {author} {\bibfnamefont {M.}~\bibnamefont
  {{Messineo}}}\ and\ \bibinfo {author} {\bibfnamefont {A.~G.~A.}\ \bibnamefont
  {{Brown}}},\ }\bibfield  {title} {\bibinfo {title} {{A Catalog of Known
  Galactic K-M Stars of Class I Candidate Red Supergiants in Gaia DR2}},\
  }\href {https://doi.org/10.3847/1538-3881/ab1cbd} {\bibfield  {journal}
  {\bibinfo  {journal} {The Astronomical Journal}\ }\textbf {\bibinfo {volume}
  {158}},\ \bibinfo {eid} {20} (\bibinfo {year} {2019})},\ \Eprint
  {https://arxiv.org/abs/1905.03744} {arXiv:1905.03744 [astro-ph.GA]}
  \BibitemShut {NoStop}%
\bibitem [{\citenamefont {{Healy}}\ \emph {et~al.}(2024)\citenamefont
  {{Healy}}, \citenamefont {{Horiuchi}}, \citenamefont {{Colomer Molla}},
  \citenamefont {{Milisavljevic}}, \citenamefont {{Tseng}}, \citenamefont
  {{Bergin}}, \citenamefont {{Weil}}, \citenamefont {{Tanaka}},\ and\
  \citenamefont {{Otero}}}]{2024MNRAS.529.3630H}%
  \BibitemOpen
  \bibfield  {author} {\bibinfo {author} {\bibfnamefont {S.}~\bibnamefont
  {{Healy}}}, \bibinfo {author} {\bibfnamefont {S.}~\bibnamefont {{Horiuchi}}},
  \bibinfo {author} {\bibfnamefont {M.}~\bibnamefont {{Colomer Molla}}},
  \bibinfo {author} {\bibfnamefont {D.}~\bibnamefont {{Milisavljevic}}},
  \bibinfo {author} {\bibfnamefont {J.}~\bibnamefont {{Tseng}}}, \bibinfo
  {author} {\bibfnamefont {F.}~\bibnamefont {{Bergin}}}, \bibinfo {author}
  {\bibfnamefont {K.}~\bibnamefont {{Weil}}}, \bibinfo {author} {\bibfnamefont
  {M.}~\bibnamefont {{Tanaka}}},\ and\ \bibinfo {author} {\bibfnamefont
  {S.}~\bibnamefont {{Otero}}},\ }\bibfield  {title} {\bibinfo {title} {{Red
  supergiant candidates for multimessenger monitoring of the next Galactic
  supernova}},\ }\href {https://doi.org/10.1093/mnras/stae738} {\bibfield
  {journal} {\bibinfo  {journal} {Monthly Notices of the Royal Astronomical
  Society}\ }\textbf {\bibinfo {volume} {529}},\ \bibinfo {pages} {3630}
  (\bibinfo {year} {2024})},\ \Eprint {https://arxiv.org/abs/2307.08785}
  {arXiv:2307.08785 [astro-ph.SR]} \BibitemShut {NoStop}%
\bibitem [{\citenamefont {{Ren}}\ \emph {et~al.}(2021)\citenamefont {{Ren}},
  \citenamefont {{Jiang}}, \citenamefont {{Yang}}, \citenamefont {{Wang}},\
  and\ \citenamefont {{Ren}}}]{2021ApJ...923..232R}%
  \BibitemOpen
  \bibfield  {author} {\bibinfo {author} {\bibfnamefont {Y.}~\bibnamefont
  {{Ren}}}, \bibinfo {author} {\bibfnamefont {B.}~\bibnamefont {{Jiang}}},
  \bibinfo {author} {\bibfnamefont {M.}~\bibnamefont {{Yang}}}, \bibinfo
  {author} {\bibfnamefont {T.}~\bibnamefont {{Wang}}},\ and\ \bibinfo {author}
  {\bibfnamefont {T.}~\bibnamefont {{Ren}}},\ }\bibfield  {title} {\bibinfo
  {title} {{The Sample of Red Supergiants in 12 Low-mass Galaxies of the Local
  Group}},\ }\href {https://doi.org/10.3847/1538-4357/ac307b} {\bibfield
  {journal} {\bibinfo  {journal} {\apj}\ }\textbf {\bibinfo {volume} {923}},\
  \bibinfo {eid} {232} (\bibinfo {year} {2021})},\ \Eprint
  {https://arxiv.org/abs/2110.08793} {arXiv:2110.08793 [astro-ph.GA]}
  \BibitemShut {NoStop}%
\bibitem [{\citenamefont {{Sumiyoshi}}\ \emph {et~al.}(2006)\citenamefont
  {{Sumiyoshi}}, \citenamefont {{Suzuki}},\ and\ \citenamefont
  {{Yamada}}}]{2006AIPC..847..473S}%
  \BibitemOpen
  \bibfield  {author} {\bibinfo {author} {\bibfnamefont {K.}~\bibnamefont
  {{Sumiyoshi}}}, \bibinfo {author} {\bibfnamefont {H.}~\bibnamefont
  {{Suzuki}}},\ and\ \bibinfo {author} {\bibfnamefont {S.}~\bibnamefont
  {{Yamada}}},\ }\bibfield  {title} {\bibinfo {title} {{Fate of core-collapse
  supernovae: formation of neutron star and black hole}},\ }in\ \href
  {https://doi.org/10.1063/1.2234463} {\emph {\bibinfo {booktitle} {Origin of
  Matter and Evolution of Galaxies}}},\ \bibinfo {series} {American Institute
  of Physics Conference Series}, Vol.\ \bibinfo {volume} {847},\ \bibinfo
  {editor} {edited by\ \bibinfo {editor} {\bibfnamefont {S.}~\bibnamefont
  {{Kubono}}}, \bibinfo {editor} {\bibfnamefont {W.}~\bibnamefont {{Aoki}}},
  \bibinfo {editor} {\bibfnamefont {T.}~\bibnamefont {{Kajino}}}, \bibinfo
  {editor} {\bibfnamefont {T.}~\bibnamefont {{Motobayashi}}},\ and\ \bibinfo
  {editor} {\bibfnamefont {K.}~\bibnamefont {{Nomoto}}}}\ (\bibinfo
  {publisher} {AIP},\ \bibinfo {year} {2006})\ pp.\ \bibinfo {pages}
  {473--475}\BibitemShut {NoStop}%
\bibitem [{\citenamefont {{Sumiyoshi}}\ \emph {et~al.}(2007)\citenamefont
  {{Sumiyoshi}}, \citenamefont {{Yamada}},\ and\ \citenamefont
  {{Suzuki}}}]{2007ApJ...667..382S}%
  \BibitemOpen
  \bibfield  {author} {\bibinfo {author} {\bibfnamefont {K.}~\bibnamefont
  {{Sumiyoshi}}}, \bibinfo {author} {\bibfnamefont {S.}~\bibnamefont
  {{Yamada}}},\ and\ \bibinfo {author} {\bibfnamefont {H.}~\bibnamefont
  {{Suzuki}}},\ }\bibfield  {title} {\bibinfo {title} {{Dynamics and Neutrino
  Signal of Black Hole Formation in Nonrotating Failed Supernovae. I. Equation
  of State Dependence}},\ }\href {https://doi.org/10.1086/520876} {\bibfield
  {journal} {\bibinfo  {journal} {\apj}\ }\textbf {\bibinfo {volume} {667}},\
  \bibinfo {pages} {382} (\bibinfo {year} {2007})},\ \Eprint
  {https://arxiv.org/abs/0706.3762} {arXiv:0706.3762 [astro-ph]} \BibitemShut
  {NoStop}%
\bibitem [{\citenamefont {{Burrows}}\ \emph {et~al.}(2024)\citenamefont
  {{Burrows}}, \citenamefont {{Wang}},\ and\ \citenamefont
  {{Vartanyan}}}]{2024arXiv241207831B}%
  \BibitemOpen
  \bibfield  {author} {\bibinfo {author} {\bibfnamefont {A.}~\bibnamefont
  {{Burrows}}}, \bibinfo {author} {\bibfnamefont {T.}~\bibnamefont {{Wang}}},\
  and\ \bibinfo {author} {\bibfnamefont {D.}~\bibnamefont {{Vartanyan}}},\
  }\bibfield  {title} {\bibinfo {title} {{Channels of Stellar-mass Black Hole
  Formation}},\ }\href {https://doi.org/10.48550/arXiv.2412.07831} {\bibfield
  {journal} {\bibinfo  {journal} {arXiv e-prints}\ ,\ \bibinfo {eid}
  {arXiv:2412.07831}} (\bibinfo {year} {2024})},\ \Eprint
  {https://arxiv.org/abs/2412.07831} {arXiv:2412.07831 [astro-ph.SR]}
  \BibitemShut {NoStop}%
\bibitem [{\citenamefont {{Suwa}}\ \emph {et~al.}(2021)\citenamefont {{Suwa}},
  \citenamefont {{Harada}}, \citenamefont {{Nakazato}},\ and\ \citenamefont
  {{Sumiyoshi}}}]{2021PTEP.2021a3E01S}%
  \BibitemOpen
  \bibfield  {author} {\bibinfo {author} {\bibfnamefont {Y.}~\bibnamefont
  {{Suwa}}}, \bibinfo {author} {\bibfnamefont {A.}~\bibnamefont {{Harada}}},
  \bibinfo {author} {\bibfnamefont {K.}~\bibnamefont {{Nakazato}}},\ and\
  \bibinfo {author} {\bibfnamefont {K.}~\bibnamefont {{Sumiyoshi}}},\
  }\bibfield  {title} {\bibinfo {title} {{Analytic solutions for neutrino-light
  curves of core-collapse supernovae}},\ }\href
  {https://doi.org/10.1093/ptep/ptaa154} {\bibfield  {journal} {\bibinfo
  {journal} {Progress of Theoretical and Experimental Physics}\ }\textbf
  {\bibinfo {volume} {2021}},\ \bibinfo {eid} {013E01} (\bibinfo {year}
  {2021})},\ \Eprint {https://arxiv.org/abs/2008.07070} {arXiv:2008.07070
  [astro-ph.HE]} \BibitemShut {NoStop}%
\bibitem [{\citenamefont {{Suwa}}\ \emph {et~al.}(2022)\citenamefont {{Suwa}},
  \citenamefont {{Harada}}, \citenamefont {{Harada}}, \citenamefont {{Koshio}},
  \citenamefont {{Mori}}, \citenamefont {{Nakanishi}}, \citenamefont
  {{Nakazato}}, \citenamefont {{Sumiyoshi}},\ and\ \citenamefont
  {{Wendell}}}]{2022ApJ...934...15S}%
  \BibitemOpen
  \bibfield  {author} {\bibinfo {author} {\bibfnamefont {Y.}~\bibnamefont
  {{Suwa}}}, \bibinfo {author} {\bibfnamefont {A.}~\bibnamefont {{Harada}}},
  \bibinfo {author} {\bibfnamefont {M.}~\bibnamefont {{Harada}}}, \bibinfo
  {author} {\bibfnamefont {Y.}~\bibnamefont {{Koshio}}}, \bibinfo {author}
  {\bibfnamefont {M.}~\bibnamefont {{Mori}}}, \bibinfo {author} {\bibfnamefont
  {F.}~\bibnamefont {{Nakanishi}}}, \bibinfo {author} {\bibfnamefont
  {K.}~\bibnamefont {{Nakazato}}}, \bibinfo {author} {\bibfnamefont
  {K.}~\bibnamefont {{Sumiyoshi}}},\ and\ \bibinfo {author} {\bibfnamefont
  {R.~A.}\ \bibnamefont {{Wendell}}},\ }\bibfield  {title} {\bibinfo {title}
  {{Observing Supernova Neutrino Light Curves with Super-Kamiokande. III.
  Extraction of Mass and Radius of Neutron Stars from Synthetic Data}},\ }\href
  {https://doi.org/10.3847/1538-4357/ac795e} {\bibfield  {journal} {\bibinfo
  {journal} {\apj}\ }\textbf {\bibinfo {volume} {934}},\ \bibinfo {eid} {15}
  (\bibinfo {year} {2022})},\ \Eprint {https://arxiv.org/abs/2204.08363}
  {arXiv:2204.08363 [astro-ph.HE]} \BibitemShut {NoStop}%
\bibitem [{\citenamefont {{Harada}}\ \emph {et~al.}(2023)\citenamefont
  {{Harada}}, \citenamefont {{Suwa}}, \citenamefont {{Harada}}, \citenamefont
  {{Koshio}}, \citenamefont {{Mori}}, \citenamefont {{Nakanishi}},
  \citenamefont {{Nakazato}}, \citenamefont {{Sumiyoshi}},\ and\ \citenamefont
  {{Wendell}}}]{2023ApJ...954...52H}%
  \BibitemOpen
  \bibfield  {author} {\bibinfo {author} {\bibfnamefont {A.}~\bibnamefont
  {{Harada}}}, \bibinfo {author} {\bibfnamefont {Y.}~\bibnamefont {{Suwa}}},
  \bibinfo {author} {\bibfnamefont {M.}~\bibnamefont {{Harada}}}, \bibinfo
  {author} {\bibfnamefont {Y.}~\bibnamefont {{Koshio}}}, \bibinfo {author}
  {\bibfnamefont {M.}~\bibnamefont {{Mori}}}, \bibinfo {author} {\bibfnamefont
  {F.}~\bibnamefont {{Nakanishi}}}, \bibinfo {author} {\bibfnamefont
  {K.}~\bibnamefont {{Nakazato}}}, \bibinfo {author} {\bibfnamefont
  {K.}~\bibnamefont {{Sumiyoshi}}},\ and\ \bibinfo {author} {\bibfnamefont
  {R.~A.}\ \bibnamefont {{Wendell}}},\ }\bibfield  {title} {\bibinfo {title}
  {{Observing Supernova Neutrino Light Curves with Super-Kamiokande. IV.
  Development of SPECIAL BLEND: A New Public Analysis Code for Supernova
  Neutrinos}},\ }\href {https://doi.org/10.3847/1538-4357/ace52e} {\bibfield
  {journal} {\bibinfo  {journal} {\apj}\ }\textbf {\bibinfo {volume} {954}},\
  \bibinfo {eid} {52} (\bibinfo {year} {2023})},\ \Eprint
  {https://arxiv.org/abs/2304.05437} {arXiv:2304.05437 [astro-ph.HE]}
  \BibitemShut {NoStop}%
\bibitem [{\citenamefont {{Suwa}}\ \emph {et~al.}(2025)\citenamefont {{Suwa}},
  \citenamefont {{Harada}}, \citenamefont {{Mori}}, \citenamefont {{Nakazato}},
  \citenamefont {{Akaho}}, \citenamefont {{Harada}}, \citenamefont {{Koshio}},
  \citenamefont {{Nakanishi}}, \citenamefont {{Sumiyoshi}},\ and\ \citenamefont
  {{Wendell}}}]{2025ApJ...980..117S}%
  \BibitemOpen
  \bibfield  {author} {\bibinfo {author} {\bibfnamefont {Y.}~\bibnamefont
  {{Suwa}}}, \bibinfo {author} {\bibfnamefont {A.}~\bibnamefont {{Harada}}},
  \bibinfo {author} {\bibfnamefont {M.}~\bibnamefont {{Mori}}}, \bibinfo
  {author} {\bibfnamefont {K.}~\bibnamefont {{Nakazato}}}, \bibinfo {author}
  {\bibfnamefont {R.}~\bibnamefont {{Akaho}}}, \bibinfo {author} {\bibfnamefont
  {M.}~\bibnamefont {{Harada}}}, \bibinfo {author} {\bibfnamefont
  {Y.}~\bibnamefont {{Koshio}}}, \bibinfo {author} {\bibfnamefont
  {F.}~\bibnamefont {{Nakanishi}}}, \bibinfo {author} {\bibfnamefont
  {K.}~\bibnamefont {{Sumiyoshi}}},\ and\ \bibinfo {author} {\bibfnamefont
  {R.~A.}\ \bibnamefont {{Wendell}}},\ }\bibfield  {title} {\bibinfo {title}
  {{Observing Supernova Neutrino Light Curves with Super-Kamiokande. V.
  Distance Estimation with Neutrinos}},\ }\href
  {https://doi.org/10.3847/1538-4357/adabe2} {\bibfield  {journal} {\bibinfo
  {journal} {\apj}\ }\textbf {\bibinfo {volume} {980}},\ \bibinfo {eid} {117}
  (\bibinfo {year} {2025})},\ \Eprint {https://arxiv.org/abs/2404.18248}
  {arXiv:2404.18248 [astro-ph.HE]} \BibitemShut {NoStop}%
\bibitem [{\citenamefont {{Foguel}}\ and\ \citenamefont
  {{Fraga}}(2023)}]{2023APh...15102855F}%
  \BibitemOpen
  \bibfield  {author} {\bibinfo {author} {\bibfnamefont {A.~L.}\ \bibnamefont
  {{Foguel}}}\ and\ \bibinfo {author} {\bibfnamefont {E.~S.}\ \bibnamefont
  {{Fraga}}},\ }\bibfield  {title} {\bibinfo {title} {{Analytic approach to
  axion-like-particle emission in core-collapse supernovae}},\ }\href
  {https://doi.org/10.1016/j.astropartphys.2023.102855} {\bibfield  {journal}
  {\bibinfo  {journal} {Astroparticle Physics}\ }\textbf {\bibinfo {volume}
  {151}},\ \bibinfo {eid} {102855} (\bibinfo {year} {2023})},\ \Eprint
  {https://arxiv.org/abs/2209.14318} {arXiv:2209.14318 [hep-ph]} \BibitemShut
  {NoStop}%
\bibitem [{Note1()}]{Note1}%
  \BibitemOpen
  \bibinfo {note}
  {Https://sites.google.com/g.ecc.u-tokyo.ac.jp/nulc}\BibitemShut {NoStop}%
\bibitem [{\citenamefont {A.}(YYYY)}]{smaplearticle}%
  \BibitemOpen
  \bibfield  {author} {\bibinfo {author} {\bibfnamefont {F.}~\bibnamefont
  {A.}},\ }\bibfield  {title} {\bibinfo {title} {{Extended Theories of
  Gravity}},\ }\href {https://doi.org/DOI} {\bibfield  {journal} {\bibinfo
  {journal} {JHEP}\ }\textbf {\bibinfo {volume} {VV}},\ \bibinfo {pages}
  {PPP}},\ \Eprint {https://arxiv.org/abs/0000.0000} {arXiv:0000.0000}
  \BibitemShut {NoStop}%
\end{thebibliography}%

\end{document}